\documentclass{aa}  

\usepackage{graphicx}
\usepackage{txfonts}
\usepackage{amsmath}

\usepackage{hyperref}
\hypersetup{
    colorlinks=true,
    linkcolor=blue,
    citecolor=blue,
    filecolor=magenta,      
    urlcolor=cyan,
    }

\graphicspath{{./}{Figs/}}
\defcitealias{Bemis:2023}{Paper I}
\defcitealias{Schneider:2013}{a}

\begin{document} 

\title{Does the HCN/CO ratio trace the star-forming fraction of gas?}

\subtitle{II. Variations in CO and HCN emissivity} 

\author{Ashley R. Bemis\inst{\ref{leiden},\ref{waterloo1},\ref{waterloo2}}
        \and
        Christine D. Wilson\inst{\ref{mcmaster}}
        \and
        Piyush Sharda\inst{\ref{leiden}}
        \and
        Ian D. Roberts\inst{\ref{waterloo1},\ref{waterloo2},\ref{leiden}}
        \and
        Hao He\inst{\ref{bonn}}
        }

\institute{
Leiden Observatory, Leiden University, PO Box 9513, 2300 RA Leiden, The Netherlands \label{leiden}
\and
Department of Physics \& Astronomy, University of Waterloo, Waterloo, ON N2L 3G1, Canada \label{waterloo1}
\and
Waterloo Centre for Astrophysics, University of Waterloo, 200 University Ave W, Waterloo, ON N2L 3G1, Canada, Canada \label{waterloo2}
\and
Department of Physics \& Astronomy, McMaster University, 1280 Main St W, Hamilton ON L8S 4M1, Canada \label{mcmaster}
\and
Argelander-Institut f\"ur Astronomie, Universit\"at Bonn, Auf 
dem H\"ugel 71, 53121 Bonn, Germany\label{bonn}
}

 
\abstract{We modeled emissivities of the HCN and CO $J=1-0$ transitions across a grid of molecular cloud models encapsulating observed properties that span from normal star-forming galaxies to more extreme {merging} systems. These models are compared with archival observations of the HCN and CO $J=1-0$  transitions, in addition to the radio continuum at 93 GHz, for ten nearby galaxies. We combined these model emissivities with the predictions of gravoturbulent models of star formation presented in the first paper in this series. In particular, we explored the impact of excitation and optical depth on CO and HCN emission {and} assess {if} the HCN/CO ratio tracks the fraction of  gravitationally bound dense gas, $f_\mathrm{grav}$, in molecular clouds. We find that our  modeled HCN/CO ratios are consistent with the measurements within our sample, and our modeled HCN and CO emissivities are consistent with the results of observational studies of nearby galaxies and clouds in the Milky Way. CO emission shows a wide range of optical depths across different environments, ranging from optically thick in normal galaxies to moderately optically thin in more extreme systems. HCN appears only moderately optically thick and shows significant subthermal excitation in both normal and extreme galaxies. We find an anticorrelation between HCN/CO and $f_\mathrm{grav}$, which implies that the HCN/CO ratio is not a {reliable} tracer of $f_\mathrm{grav}$. Instead, this ratio appears to best track gas at moderate densities ($n>10^{3.5}\ \mathrm{cm}^{-3}$), which is below the typically assumed dense gas threshold of $n>10^{4.5}\ \mathrm{cm}^{-3}$. We also find that variations in CO emissivity depend strongly on optical depth, which is a product of variations in the dynamics of the cloud gas. HCN emissivity is more strongly dependent on excitation, as opposed to optical depth, and thus does not necessarily track variations in CO emissivity. We further conclude that a single line ratio, such as HCN/CO, will not consistently track the fraction of gravitationally bound, star-forming gas if the critical density for star formation varies in molecular clouds. This work highlights important uncertainties that need to be considered when observationally applying an HCN conversion factor in order to estimate the dense (i.e.,\ $n>10^{4.5}\ \mathrm{cm}^{-3}$) gas content in nearby galaxies.}

\maketitle
%

\section{Introduction} \label{sec:paper3-intro}

The HCN/CO ratio is commonly used to assess the fraction of dense gas ($\gtrsim10^{4}-10^5$ cm$^{-3}$) that might be associated with star formation in external galaxies. The seminal work by \citet{Gao:2004a,Gao:2004b} found a linear scaling (slope of unity) between the logarithm of the HCN luminosity and the star formation rate as traced in the infrared (IR)\footnote{A slope of unity between $\mathrm{log}\,L_\mathrm{IR}$ and $\mathrm{log}\,L_\mathrm{HCN}$ also implies a linear scaling between $L_\mathrm{IR}$ and $L_\mathrm{HCN}$.} for a diverse sample of galaxies, including normal disk galaxies as well as more extreme ultra-luminous and luminous infrared galaxies (U/LIRGs). This correlation suggests that HCN is a useful tracer of star-forming gas for a range of galaxy types. The linear scaling between $L_\mathrm{IR}$ and $L_\mathrm{HCN}$ also implies that the critical density for HCN $J=1-0$ emission, $n_\mathrm{crit,HCN}$, is close to a common mean threshold density, $n_\mathrm{thresh}$, that is important for star formation. Although individual galaxies have scatter in the $L_\mathrm{IR}$ and $L_\mathrm{HCN}$ relationship, a linear scaling implies that the average value of  $L_\mathrm{IR}/L_\mathrm{HCN}$$=900\,L_\odot\,(\mathrm{K\,km\,s^{-s}\,pc^2})$ \citep{Gao:2004b}  is relatively constant over many orders of magnitude. However, systematic deviations from linearity have since been found in U/LIRGs \citep{Gracia-Carpio:2008,Garcia-Burillo:2012}, at subkiloparsec scales in disk galaxies \citep{Usero:2015,Chen:2015,Gallagher:2018,Querejeta:2019,Jimenez_Donaire:2019,Beslic:2021,Neumann:2023}, and at subkiloparsec scales in galaxy mergers \citep{Bigiel:2015,Bemis:2019}. These  nonlinearities do not appear associated with the presence of an active galactic nucleus (AGN), which otherwise could enhance HCN emissivity via infrared pumping \citep{Sakamoto:2010}. In the absence of an AGN, these variations in emissivity can be interpreted as a fundamental difference in the depletion time of dense gas within different systems, which may signal a connection between star formation and environment within galaxies.
\par
Variations are also seen in the star formation efficiency of dense gas in our own Milky Way. Gas in the central molecular zone (CMZ) of the Milky Way is dense ($n\sim10^4$ cm$^{-3}$) and warm ($T\sim65$ K) compared to gas in the disk ($n\sim10^2$ cm$^{-3}$, $T\sim 10$ K, \citealt{Rathborne:2014,Ginsburg:2016}). Despite their abundance of dense gas \citep{Mills:2017}, some clouds in the CMZ display a lack of star formation \citep{Longmore:2013,Kruijssen:2014,Walker:2018}. CMZ clouds   experience high external pressures ($\sim10^8\, \mathrm{K\, cm^{-3}}$), and it is theorized that their lack of star-forming activity may be due to a higher star formation threshold density as a result of higher internal turbulent pressures \citep[][]{Walker:2018}. Additionally, shear from solenoidal turbulence may also suppress the onset of star formation in the CMZ  \citep{Federrath:2016ApJ...832..143F}. A lack of star formation in dense gas traced by HCN is also apparent in the centers of nearby disk galaxies \citep{Gallagher:2018,Querejeta:2019,Jimenez_Donaire:2019,Beslic:2021,Neumann:2023} and the nuclei of the Antennae galaxies (NGC 4038/9, \citealt{Bemis:2019}). Gas density is well-constrained in the CMZ, suggesting that there are true variations in star formation from dense gas in this environment relative to the Milky Way disk. Many studies of external galaxies rely on a single molecular gas tracer, HCN, to estimate the dense gas fraction, and recent work calls into question its ability to consistently trace dense gas in molecular clouds \citep[e.g.,][]{Kauffmann:2017hcn_moderate,Pety:2017,Shimajiri:2017,Barnes:2020,Tafalla:2021,Tafalla:2023,Santa-Maria:2023}. Furthermore, if the star formation threshold density also varies with the local environment within galaxies, a gas fraction estimate from a single line ratio may not reliably track the fraction of gas above this threshold \citep{Bemis:2023,Neumann:2023}.
\par
In \citet[][hereafter Paper I]{Bemis:2023}, we assess the ability of the relative intensity of HCN to CO, $I_\mathrm{HCN}/I_\mathrm{CO}$, to determine the fraction of gravitationally bound gas by comparing the observed star formation properties and HCN/CO ratios in ten galaxies to the predictions of analytical models of star formation \citep{Krumholz:2005,Federrath:2012,Hennebelle:2011,Burkhart:2018}. In \citetalias{Bemis:2023} we find that the trends observed in our sample of dense gas traced by $I_\mathrm{HCN}$, the SFR traced by the radio continuum at 93 GHz, and the total molecular gas traced by $I_\mathrm{CO}$ are best reproduced by gravoturbulent models of star formation with varying star formation thresholds under the assumption that $I_\mathrm{HCN}/I_\mathrm{CO}$ is tracing the fraction of gas above a relatively fixed density, such as $n\gtrsim10^{4.5}\ \mathrm{cm}^{-3}$, but not necessarily the fraction of gas that is gravitationally bound or star-forming. Furthermore, in \citetalias{Bemis:2023} we show that turbulent models of star formation with varying star formation thresholds predict an increase in the depletion time of dense gas at $n\gtrsim10^{4.5}\ \mathrm{cm}^{-3}$ in clouds with higher dense gas fractions due to higher levels of turbulence. This corroborates observations in the CMZ where star formation appears suppressed relative to the amount of dense gas mass, estimates of turbulent pressure ($P_\mathrm{turb}$) appear higher, and the dominant mode of turbulence may be more solenoidal compared to spiral arms \citep{Federrath:2016ApJ...832..143F,Walker:2018}. Similar trends are observed in galaxy centers \citep{Gallagher:2018,Querejeta:2019,Jimenez_Donaire:2019,Beslic:2021} and in the nuclei of the Antennae \citepalias{Bemis:2023}, although estimates of the dominant turbulent mode are unavailable for these studies.
\par 
One key uncertainty in these results is the ratio of the emissivities of HCN and CO (i.e., the conversion of HCN or CO intensity to gas column density) and whether systematic variations in HCN and CO emissivity can occur in such a way that may also contribute to the observed trends. The CO conversion factor, $\alpha_\mathrm{CO}$, is estimated to vary with excitation \citep[e.g.,][]{Narayanan:2012,Narayanan:2014} and metallicity \citep[e.g.,][]{Schruba:2012,Hu:2022}; can be nearly five times lower in U/LIRGs \citep[e.g.,][]{Downes:1993}; and is lower in the centers of disk galaxies \citep[e.g.,][]{Sandstrom:2013}. The HCN conversion factor, $\alpha_\mathrm{HCN}$, is also likely to vary across different systems \citep{Usero:2015}, but may not necessarily track $\alpha_\mathrm{CO}$ \citep{2018MNRAS.479.1702O}. Observations of HCN and  H$^{13}$CN in galaxy centers suggest that HCN is only moderately optically thick \citep{Jimenez_Donaire:2017}, unlike CO which typically has $\tau_\mathrm{CO}>10$. The original estimate of $\alpha_\mathrm{HCN}$ was made under the assumption of optically thick emission \citep{Gao:2004a,Gao:2004b}. Since HCN emission appears only moderately optically thick, variations in the optical depth of HCN emission could impact the accuracy of gas masses using this estimate of the HCN conversion factor. Thus, the HCN/CO intensity ratio may not scale linearly with the fraction of gas $\gtrsim10^4\,\mathrm{cm^{-3}}$, due to variations in excitation and optical depth. As we refine our understanding of star formation in galaxies, it is clear that we must also adopt a more sophisticated approach to estimating masses using molecular line emission, and we must develop a better understanding of the information that these measurements can provide on star formation.
\par 
In this paper we model emissivities of HCN and CO using the non-LTE radiative transfer code RADEX \citep{vanderTak:2007,Leroy:2017b} across a grid of cloud models that encapsulates observed trends in cloud properties that span from from normal star-forming galaxies to U/LIRGs \citep{Sun:2020,Brunetti:2021,Brunetti:2024}. We assess the impact of variations in optical depth and excitation on the emissivities of HCN and CO across this grid, and we compare our modeling results with the star-forming trends observed in the sample of ten galaxies from \citetalias{Bemis:2023} using archival ALMA data of the HCN and CO $J=1\rightarrow0$ transitions and the radio continuum at 93 GHz (see \citealt{Wilson:2023} for details on imaging). This sample includes the dense centers of five disk galaxies and five U/LIRGs (see Table 1 of \citetalias{Bemis:2023}).   In Sect. \ref{sec:model_framework} we describe the model framework that we adopted to derive emissivities using analytical models of star formation, as well as the parameter space we considered. We present the results of our models and compare these results with observations in Sect. \ref{sec:model_results}. Finally, in Sect. \ref{sec:conc} we provide a brief discussion and summary of our main results. Throughout the text  we take  `` HCN/CO ratio'' to mean the ratio of HCN to CO intensities, unless explicitly stated otherwise.

\section{Model framework}
\label{sec:model_framework}

We model molecular line emissivities using the radiative transfer code RADEX \citep{vanderTak:2007}, and we connect these emissivities to gravoturbulent models of star formation. We present several gravoturbulent models of star formation in \citetalias{Bemis:2023}, which predict clouds have gas density distributions that are either purely lognormal (LN, cf. \citealt{Padoan:2011,Krumholz:2005, Federrath:2012}) or a composite lognormal and power-law distribution (LN+PL, cf. \citealt{Burkhart:2017}). We refer to these distributions as the gas density probability density functions (PDFs) for the remainder of the text. We focus on the results of the composite LN+PL models in this analysis.

\subsection{Emissivity}
\label{sec:emiss}

We adopt the following definition of the emissivity of a molecular transition \citep[e.g.,][]{Leroy:2017b}:
\begin{equation} \label{eq:emiss}
    \epsilon_\mathrm{mol} = \frac{I_\mathrm{mol}}{N} = \frac{I_\mathrm{mol}}{N_\mathrm{mol}/x_\mathrm{mol}}
.\end{equation}
\noindent Here $I_\mathrm{mol}$ is the total line intensity of a molecular transition (in units of K km s$^{-1}$), $N$ is the column density of gas that emits $I$, $N_\mathrm{mol}$ is the molecular column density of an observed molecule (both in units of cm$^{-2}$), and $x_\mathrm{mol}$ is the fractional abundance of the molecule relative to the molecular hydrogen, $\mathrm{H_2}$. The ratio of the HCN and CO emissivities is then given by
\begin{equation}\label{eq:hcn_co_emiss}
    \frac{\epsilon_\mathrm{HCN}}{\epsilon_\mathrm{CO}} = \frac{I_\mathrm{HCN}}{I_\mathrm{CO}} \frac{N_\mathrm{CO}}{N_\mathrm{HCN}} \frac{x_\mathrm{HCN}}{x_\mathrm{CO}}
.\end{equation}
Emissivity is analogous to the inverse of molecular conversion factors ($\alpha_\mathrm{mol} = X_\mathrm{mol}/6.3\times10^{19}, $ \citealt{Leroy:2017a}), which are commonly used to estimate the mass traced by a molecular transition. In practice, the relationship between the total emissivity of an observed molecular cloud and an appropriate conversion factor also depends on the beam filling fraction and the uniformity of gas properties within the beam \citep[cf.][]{Bolatto:2013}. 

\subsection{Cloud models} \label{sec:cloud_models}
Under the assumption that we can derive information of the gas density distribution from estimates of gas velocity dispersion, we build our cloud models using gas density distributions predicted by gravoturbulent models of star formation that employ the gas density variance -- mach number relation (cf. Eq. \ref{eq:pdf_nstd}). {There are well-established theories connecting the gas density variance ($\sigma_{n/n_0}^2$ ) in molecular clouds to mach number \citep[cf.][]{Passot:1998,Beetz:2008PhLA..372.3037B, Burkhart:2010ApJ...708.1204B, Padoan:2011, Price:2011ApJ...727L..21P, Konstandin:2012ApJ...761..149K, Molina:2012, Federrath:2015, Nolan:2015MNRAS.451.1380N, Pan:2016ApJ...825...30P, Squire:2017MNRAS.471.3753S, Beattie:2021MNRAS.504.4354B}. In general, these theories predict that the gas density variance increases with mach number, such that \citep[cf.][]{Federrath:2008,Federrath:2010,Molina:2012}
\begin{equation}
    \sigma_{n/n_0}^2 = b^2 \mathcal{M}^2 \frac{\beta}{\beta+1} \label{eq:pdf_nstd}
,\end{equation}
where $b$ is the turbulent forcing parameter which describes the dominant mode(s) of turbulence (i.e., compressive, mixed, or solenoidal) and spans $b=1/3-1$ \citep{Federrath:2008,Federrath:2010}, $\mathcal{M}$ is the sonic mach number, and $\beta$ is the ratio of thermal to magnetic pressure \citep[cf.][]{Molina:2012}. Numerical work shows that a connection is also expected between the 2D gas density variance, $\sigma_{\Sigma/\Sigma_0}$, an observable, and mach number \citep[e.g.,][]{Brunt:2010MNRAS.405L..56B,Brunt:2010MNRAS.403.1507B, Burkhart:2012}. Thus, resolved studies of the gas column density distribution can, in theory, provide information on the Mach number of the initial turbulent velocity field shaping the dynamics of a cloud. Alternatively, lower-resolution studies that are limited to global cloud measurements (as is often the case in extragalactic studies) may also be able to derive information on the gas density distribution using estimates of the gas velocity dispersion. We use this as a basis to build our cloud models and our model parameter space, described in detail in Sect. \ref{sec:param_space}. We also highlight the relevant uncertainties for this approach, both in this section and in Appendix~\ref{ap:large_scale_motions}.}
\par
{We note that there are a number of analytical prescriptions describing the gas density distribution \citep[e.g.,][]{Krumholz:2005,Padoan:2011,Hennebelle:2011, Hopkins:2013MNRAS.430.1653H,Burkhart:2018}. For simplicity}, we focus on the piecewise lognormal and power-law distribution from \citet{Burkhart:2018} {and we note that we do not expect significant changes to our main conclusions if we were to adopt a different prescription. In particular, our models produce gas density distributions with widths and density ranges comparable to those observed in resolved studies of clouds in the Milky Way \citep[cf.][]{Schneider:2022A&A...666A.165S}.} The piecewise volume density PDF is given in \citetalias{Bemis:2023} (Eq. 16) and is originally given in \citet{Burkhart:2018} (Eqs. 2 and 6) in terms of the logarithmic density, $s=\mathrm{ln}(n/n_0)$, where $n$ is gas volume density and $n_0$ is the mean gas volume density. We refer   to \citetalias{Bemis:2023} and \citet{Burkhart:2018} for a thorough description of these models in terms of the logarithmic density. Here we briefly summarise the main components of these models in terms of the linear gas volume density, $n$, which is directly used in our modeling of molecular line emissivities.
\par
\begin{figure*}
    \centering
    \includegraphics[width=0.32\linewidth]{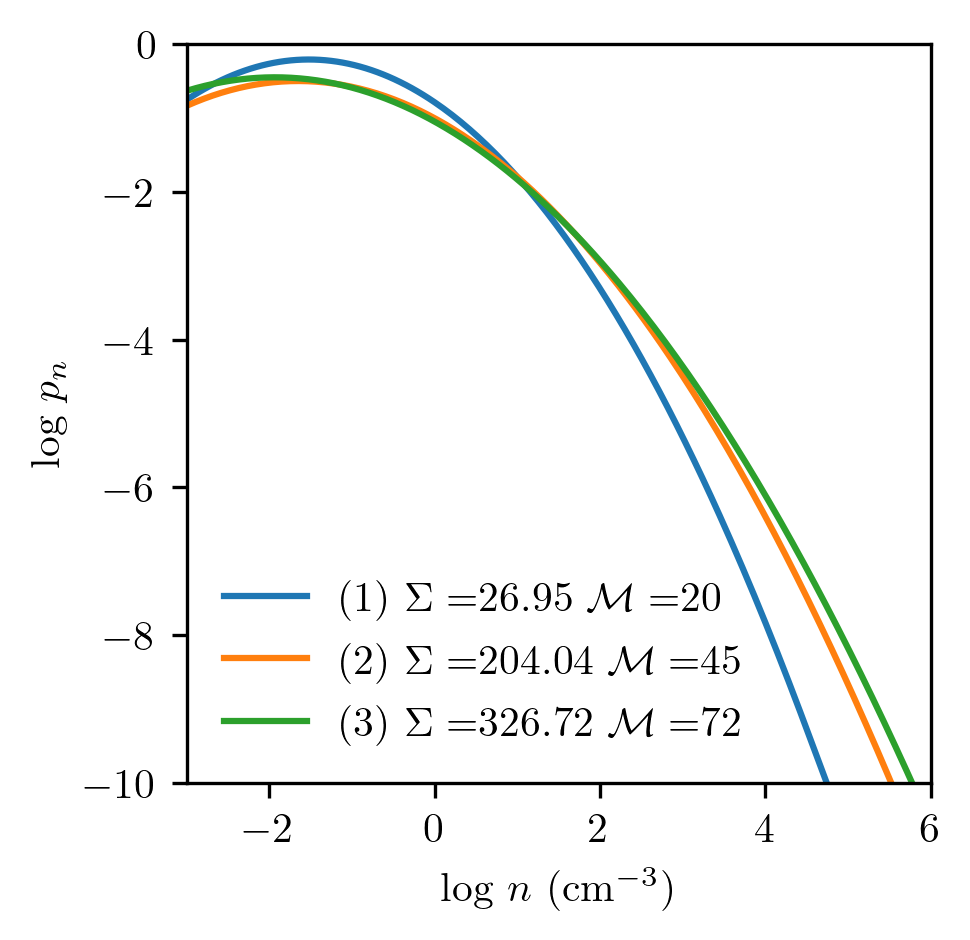}
    \includegraphics[width=0.32\linewidth]{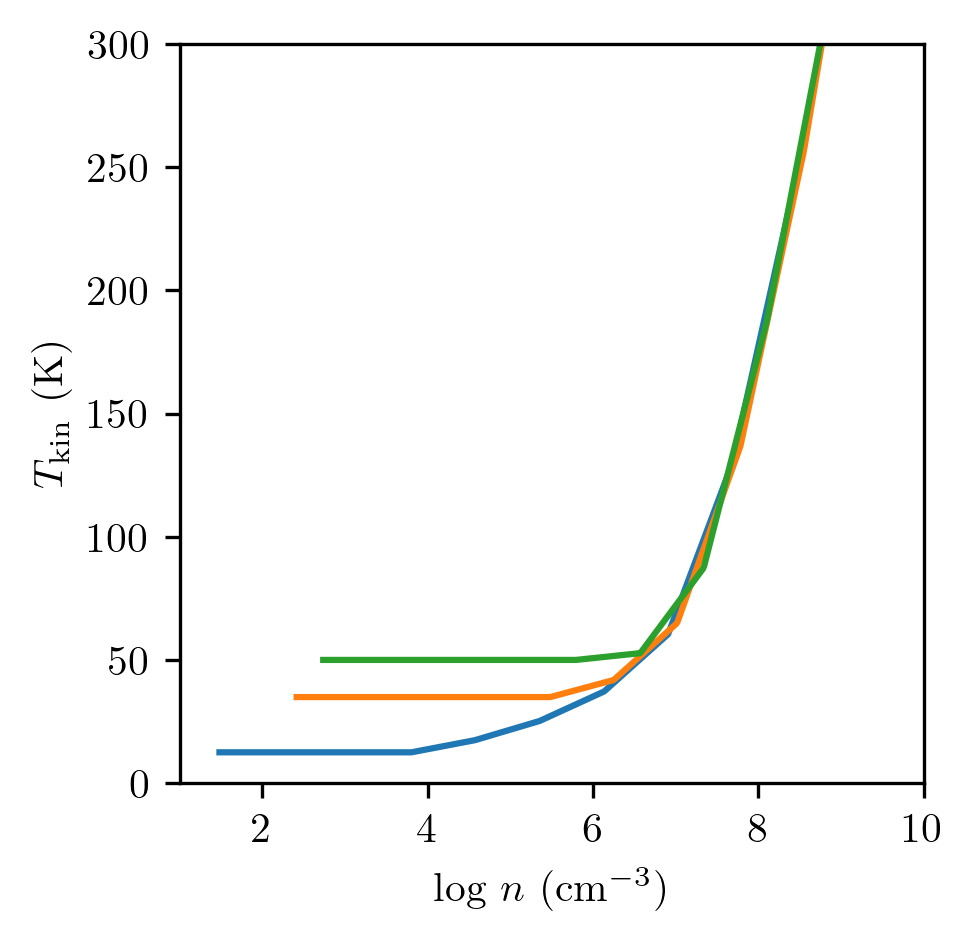}
    \includegraphics[width=0.32\linewidth]{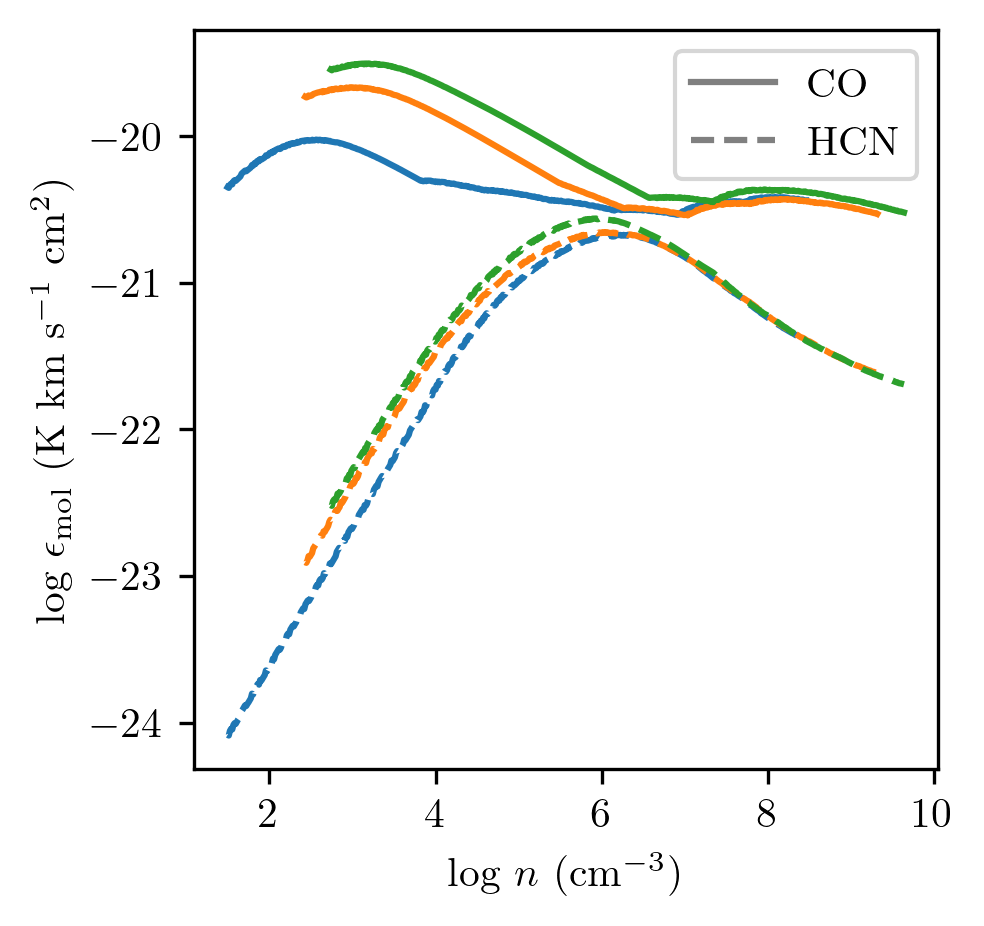}
    \caption{Three models that are representative of clouds in (1) the PHANGS sample (NGC 2903), (2) NGC 4038/9, and (3) NGC 3256. \textit{Left:} Example $n-\mathrm{PDFs}$ from our model parameter space assuming $\alpha_\mathrm{PL}=3$. \textit{Center:} Temperature profiles of the example models.  \textit{Right:} Emissivity profiles of the example models. CO emissivity is shown as   solid lines, and HCN emissivity is shown as   dashed lines. The mass-weighted emissivity, $\left<\epsilon_\mathrm{mol}\right>$, is given by Eq. \ref{eq:emiss_integral}. The range of densities over which radiative transfer is applied is slightly different between models, and depends on the  average gas surface density of the model (see Sect. \ref{sec:param_space}). We plot $p_n$ over a wider range of volume densities (left plot) than those used when performing radiative transfer (center and right plots).}
    \label{fig:ex_pdfs}
\end{figure*}

Each cloud model is comprised of a piecewise lognormal and power-law gas volume density PDF ($n-\mathrm{PDF}$). The lognormal component of the $n-\mathrm{PDF}$ has a characteristic {gas density} variance, $\sigma^2_{n/n_0}$ and mean density, $n_0$. We note that the logarithmic gas density PDF (Eq. 16 in \citetalias{Bemis:2023}) can be converted to its linear form via $p_s = n\,p_{n}$. Likewise, the logarithmic variance (Eq. 17 in \citetalias{Bemis:2023}), $\sigma^2_s$, can be converted to its linear form using $\sigma^2_{n/n_0} = \mathrm{exp}\left(\sigma_s^2 \right)-1$ \citep[cf.][]{Federrath:2008}. The power-law component of the $n-\mathrm{PDF}$ is primarily characterized by its slope, $\alpha_\mathrm{PL}$, and the power-law slope of $p_n$ is related to the slope of $p_s$ via $\alpha_\mathrm{PL}(n)=\alpha_\mathrm{PL}(s) - 1$. The two components of the $n-\mathrm{PDF}$ are analytically connected such that they are smoothly varying \citep{Burkhart:2018}. Similar to \citetalias{Bemis:2023}, we fix $b=0.4$, which represents stochastic mixing between the two turbulent forcing modes \citep{Federrath:2010}, and we neglect magnetic fields and take $\beta\rightarrow\infty$. We illustrate example $n-\mathrm{PDFs}$ in Fig. \ref{fig:ex_pdfs} that are sampled from our model parameter space, described in Sect. \ref{sec:param_space}. 

\subsection{The model parameter space}
\label{sec:param_space}

We constructed our model parameter space to capture observed trends in cloud properties associated with star-forming molecular gas clouds. As described in Sect. \ref{sec:cloud_models}, each unique model is described by its variance, $\sigma_{n/n_0}$, mean density, $n_0$, and power-law slope, $\alpha_\mathrm{PL}$. Within $\sigma_{n/n_0}$ is a dependence on the turbulent gas velocity dispersion, $\sigma_\mathrm{v}$, and gas kinetic temperature, $T_\mathrm{kin}$, via the sonic mach number, $\mathcal{M}=\sqrt{3}\sigma_\mathrm{v}/c_\mathrm{s}$ (assuming isotropy), where $\sigma_\mathrm{v}$ is the 1D turbulent velocity dispersion,  $c_\mathrm{s}=\sqrt{k T_\mathrm{kin}/\mu m_\mathrm{H}}$ is the gas sound speed, $k$ is the Boltzmann constant, $\mu$ is the mean molecular weight of the gas, and $m_{\rm{H}}$ is the mass of a Hydrogen atom. We assume a mean  molecular weight of $\mu=2.33$ \citep[e.g.,][]{Kauffmann:2008}. Each individual model is therefore defined by a unique set of $n_0$, $\sigma_\mathrm{v}$, $T_\mathrm{kin}$, and $\alpha_\mathrm{PL}$. We discuss how we set each of these parameters below.

\subsubsection{Turbulent gas velocity dispersion, $\sigma_\mathrm{v}$:}

In \citetalias{Bemis:2023}, we selected a model parameter space such that the median $\Sigma_\mathrm{mol}-\sigma_\mathrm{v}$ trend generally follows the $\Sigma_\mathrm{mol} -\sigma_\mathrm{v}$ fit to PHANGS \citep[Physics at High Angular resolution in Nearby GalaxieS,][]{Leroy:2021ApJS..257...43L} data in \citet{Sun:2018} and Milky Way cloud-scale studies \citep{Heyer:2009,Field:2011}. In this paper, we used cloud-scale ($R=40 - 45\, \mathrm{pc}$) measurements of cloud properties derived from observations of the CO $J=2-1$ line from the PHANGS sample \citep{Sun:2020}, NGC 3256 \citep{Brunetti:2021} and the Antennae \citep{Brunetti:2024} as the basis of our model parameter space. 
We randomly sampled measurements from each of these cloud-scale studies, that include pairs of molecular gas surface density and velocity dispersion, $\Sigma_\mathrm{mol}$ and $\sigma_\mathrm{v}$. Constructing our model parameter space this way ensures that our models include representatives of cloud-scale observations of normal galaxies (from PHANGS), as well as more extreme systems that are representative of merging galaxies in our study (i.e., NGC 3256 and the Antennae). The Antennae and NGC 3256 are both in our lower-resolution sample from \citetalias{Bemis:2023} and this paper. We plot the corresponding cloud coefficients ($\mathrm{\sigma_v}^2/R$ vs. $\Sigma_\mathrm{mol}$, where $R$ is the size of the molecular component of the cloud, \citealt{Heyer:2015,Field:2011}) of our model parameter space in comparison to those from the PHANGS sample \citep{Sun:2020} and those from our lower-resolution study ($\sim50-900\ \mathrm{pc}$, see Table 1 in \citetalias{Bemis:2023}) in Fig. \ref{fig:model_param_space}. 
\par
Setting our model parameter space this way relies on the assumption that the CO velocity dispersion tracks the turbulent velocity dispersion at these scales. In Appendix \ref{ap:large_scale_motions}, we discuss in detail the uncertainties and evidence for use of observational estimates of gas velocity dispersion from molecular line emission, and summarize the main points here. Using simulations, \citet{Szucs:2016MNRAS.460...82S} find the measured $^{12}$CO velocity dispersion is within $\sim30-40\%$ of the turbulent 1D velocity dispersion in their cloud simulations, on average, and argue that the CO velocity dispersion should trace the turbulent velocity dispersion. This is smaller than the expected uncertainty on the mass conversion factor \citep{Szucs:2016MNRAS.460...82S,Bolatto:2013}, which is up to a factor of two. Additionally, a weak correlation is observed between mach number estimated from various molecular line transitions and gas density variance in resolved clouds in the Milky Way \citep[e.g.,][]{Padoan:1997, Brunt:2010A&A...513A..67B, Ginsburg:2013,KainulainenTan:2013, Federrath:2016ApJ...832..143F, Menon:2021, Marchal:2021ApJ...908..186M, Sharda:2022}, although there is significant scatter potentially due to uncertainties in $b$ \citep{Kainulainen:2017}.
\par
We cannot exclude the possibility of large-scale motions impacting the measured velocity dispersions of studies at $45-50$ pc. {For example, \citet{Federrath:2016ApJ...832..143F} used HNCO to estimate turbulent velocity dispersion in G0.253+0.016 (The Brick) and subtracted a large-scale gradient that appears to contribute $\sim40-50\%$ of the measured velocity dispersion, possibly from shear due to its location in the CMZ of the Milky Way. Using the velocity profiles derived from \citet{Lang:2020}, we conclude that a small fraction of clouds in the PHANGS sample will be impacted by shear motions towards the centres of these disk galaxies (with contributions $>50\%$ to the measured velocity dispersion). It is more difficult to quantify the large-scale motions of gas within the mergers NGC 3256 and NGC 4038/9. Gas streaming motions or shear may bias measured velocity dispersions towards larger values \citep{Sun:2020, Henshaw:2016MNRAS.457.2675H,Federrath:2016ApJ...832..143F}. In general, the measured velocity dispersions are larger in the mergers; however, the gas density PDF is also expected to be wider in mergers (with more dense gas) due to enhanced compressive turbulence \citep[cf.][]{Renaud:2014}. Thus, we conclude that the velocity dispersion measurements from CO likely track the turbulent velocity dispersion, and quote a typical uncertainty of 50\%.}
\par
{Finally, we note that clouds in the PHANGS sample and in NGC 3256 are marginally resolved at $40-45$ pc scales \citep[cf.][]{Rosolowsky:2021, Brunetti:2022}, and we expect the Antennae to have cloud sizes intermediate between those found in the PHANGS galaxies and NGC 3256. For comparison, a typical size of clouds in the Milky Way is 30 pc, with a large range from $<1\, \mathrm{pc}$ to $100\, \mathrm{pc}$ \citep{Miville-Desch:2017}.  Thus, there will be some variation in how resolved each cloud is in the three studies we take measurements from \citep{Sun:2020,Brunetti:2021,Brunetti:2024}. However, we do not expect molecular velocity dispersions in the galaxies to be significantly impacted by observational effects such as beam smearing, since the systems studied are relatively face on \citep{Sun:2020,Brunetti:2021,Brunetti:2024}. Additionally, the gas density variance -- mach number relation (cf. Eq. \ref{eq:pdf_nstd}) will hold on scales smaller than the cloud size, since turbulence is expected to produce self-similar structure \citep[e.g.,][]{Elmegreen:2004ARA&A..42..211E,Dib:2008ApJ...678L.105D,Burkhart:2013}.}

\subsubsection{Mean gas density, $n_0$:} We derived a mean gas density, $n_0$, based on the molecular gas surface density estimates, $\Sigma_\mathrm{mol}$. We estimated a minimum mean gas density by converting $\Sigma_\mathrm{mol}$ to a volume density assuming a spherical geometry ($n(R)$, where $R$ is the assumed size of the molecular cloud that is set by the pixel size), such that $n_0=\Sigma / R$.  We note that we also considered different prescriptions for calculating mean gas density based on the assumption of energy equipartition (i.e., fixed virial parameter) and dynamical equilibrium in a gas disk \citep{Wilson:2019}. We find similar results regardless of the prescription we choose for $n_0$ and therefore only present the results assuming $n_0=\Sigma / R$.\footnote{On larger scales ($\sim 1\,\rm{kpc}$), \citet{2019A&A...622A..64B} find $n_0 \propto \Sigma^{0.6}$ in nearby spiral galaxies; however, on these scales the contribution from H\small{I} becomes significant.} We also acknowledge that the $\Sigma_\mathrm{mol}$ measurements from these cloud scale studies are still prone to uncertainties in the CO conversion factor. However, we confirm that our modeled CO $J=1-0$ intensities with $I_\mathrm{CO}$ are consistent with the intensities measured in our lower resolution sample (see Fig. \ref{fig:inten_compared_to_obs}), with only small offsets.

\subsubsection{Power-law slope of the $n-\mathrm{PDF}$, $\alpha_\mathrm{PL}$:}

We aim to capture observed star formation scaling relations in our study, in addition to capturing observed cloud properties. Following \citetalias{Bemis:2023}, we use the gravoturbulent models of star formation from \citet{Burkhart:2018} and \citet{Burkhart:2019},  which assume the $n-\mathrm{PDF}$ is a combination of a lognormal turbulence-dominated component and gravity-dominated power-law tail at high densities. We use these models to estimate $\epsilon_\mathrm{ff}=t_\mathrm{dep}/t_\mathrm{ff}$ (star formation efficiency per free-fall time), $t_\mathrm{dep}$, and $\Sigma_\mathrm{SFR}$ (the star formation rate surface density). Table 2 in \citetalias{Bemis:2023} describes how each of these quantities are derived. In this framework, the choice of $\alpha_\mathrm{PL}$ (the slope of the power-law component of the $n-\mathrm{PDF}$ in the gravoturbulent star formation models of  \citealt{Burkhart:2018,Burkhart:2019}) has an impact on the resulting $\epsilon_\mathrm{ff}$ such that steeper $\alpha_\mathrm{PL}$ values result in higher $\epsilon_\mathrm{ff}$ and vice versa. We choose $\alpha_\mathrm{PL}=3$ as our fiducial value (which corresponds to $k=1.5$, see Eq. \ref{eq:k}).\footnote{For comparison to the power-law slopes of $p_s$ in \citetalias{Bemis:2023}, subtract one from $\alpha_\mathrm{PL}$.} Similar to \citetalias{Bemis:2023}, we must also assume a local efficiency of $\epsilon_0$ so that values of $t_\mathrm{dep}$ and $\Sigma_\mathrm{SFR}$ derived from our models are consistent with observations.  $\epsilon_0$ accounts for a reduction in star formation efficiency from stellar feedback processes. For $\alpha_\mathrm{PL}=3$ we take $\epsilon_0=0.01$, which returns $\epsilon_\mathrm{ff}\approx 0.01-0.1$, and is consistent with expectations from observations and simulations \citep[e.g.,][]{2015ApJ...806L..36S,Utomo:2018,2018MNRAS.477.4380S,2022MNRAS.511.1431H}.

\subsubsection{Gas kinetic temperature, $T_\mathrm{kin}$:}

We estimated $T_\mathrm{kin}$ following the prescription in \citet{Sharda:2022}, which assumes thermal equilibrium balance of heating and cooling processes in the presence of protostellar radiation feedback:
\begin{equation}
   \Gamma_\mathrm{c} + \Gamma_\mathrm{CR} + \Gamma_\mathrm{H_2} + \Psi_\mathrm{gd} + \Lambda_\mathrm{M} + \Lambda_\mathrm{H_2} + \Lambda_\mathrm{HD} = 0.
\end{equation}
This equation includes compressional heating ($\Gamma_\mathrm{c}$), cosmic ray heating ($\Gamma_\mathrm{CR}$), $\mathrm{H_2}$ formation heating ($\Gamma_\mathrm{H_2}$), metal line cooling ($\Lambda_\mathrm{M}$), $\mathrm{H_2}$ cooling ($\Lambda_\mathrm{H_2}$), and hydrogen deuteride cooling ($\Lambda_\mathrm{HD}$), as well as the dust-gas energy exchange ($\Psi_\mathrm{gd}$), which can serve as either a cooling or heating process. The \citet{Sharda:2022} prescription includes feedback from active star formation in a semi-analytical framework. In addition to aforementioned cooling and heating mechanisms,  \citet{Sharda:2022} consider the impact of radiation feedback from existing protostars in the cloud via the dust-gas energy exchange term, where the dust temperature is set by radiation feedback from active star formation. This prescription is adopted from \citet{Chakrabarti:2005} where the authors developed a framework to treat dust temperatures in the presence of a central radiating source (see also, \citealt{krumholz2011}).

\par
We adopted the prescription for cosmic ray heating from  \citet{Krumholz:2023} that is based on the gas depletion time. \citet{Krumholz:2023} estimate the average cosmic ray ionization rate, $\zeta_\mathrm{CR}$, to be
\begin{equation}
    \zeta_\mathrm{CR}= 1\times10^{-16} \left( \frac{t_\mathrm{dep}}{\mathrm{Gyr}}\right)^{-1}\ \mathrm{s}^{-1}.
\end{equation}
The comic ray heating rate is then
\begin{equation}
    \Gamma_\mathrm{CR} = q_\mathrm{CR} \zeta_\mathrm{CR}
,\end{equation}
where we have taken the average energy per ionization to be $q_\mathrm{CR}=12.25\ \mathrm{eV}$, which is appropriate for molecular gas \citep{Wolfire:2010}. We use $t_\mathrm{dep}$ estimates that are consistent with the molecular gas star formation laws found by \citet{Bigiel:2008} and \citet{Wilson:2019}. The original prescription used by \citet{Sharda:2022} from \citet{Crocker:2021} overestimates the gas temperature for molecular clouds. 
\par 
In addition to these processes included in \citet{Sharda:2022}, we also incorporated mechanical heating,
\begin{equation}
    \Gamma_\mathrm{turb} = 3.3\times10^{-27} \frac{n\ \sigma_\mathrm{v}^3}{R} \ \mathrm{erg\ cm}^{-3}\ \mathrm{s}^{-1}
,\end{equation}
which is potentially important for more turbulent clouds \citep{Pan:2009,Ao:2013}, such as those in mergers or at the centers of barred galaxies. We find that on average over the cloud model, turbulent heating dominates in the models using NGC 4038/9 and NGC 3256 gas surface density and velocity dispersion measurements, while cosmic ray heating dominates in the models using gas surface density and velocity dispersion measurements from PHANGS galaxies. We show example temperature profiles for average PHANGS, NGC 4038/9, and NGC 3256 models in Fig. \ref{fig:ex_pdfs}. For low density regions near the exterior of the model clouds, the heating/cooling model sometimes produces temperature increases, which are likely unphysical. At these low densities we fix the temperature to the minimum value over the model cloud (Fig.~\ref{fig:ex_pdfs}). We note that we assume a solar metallicity composition of the gas for all our cloud models for simplicity, ignoring the metallicity dependence of $T_{\rm{kin}}$. 

\begin{figure}
    \centering
    \includegraphics[width=\columnwidth]{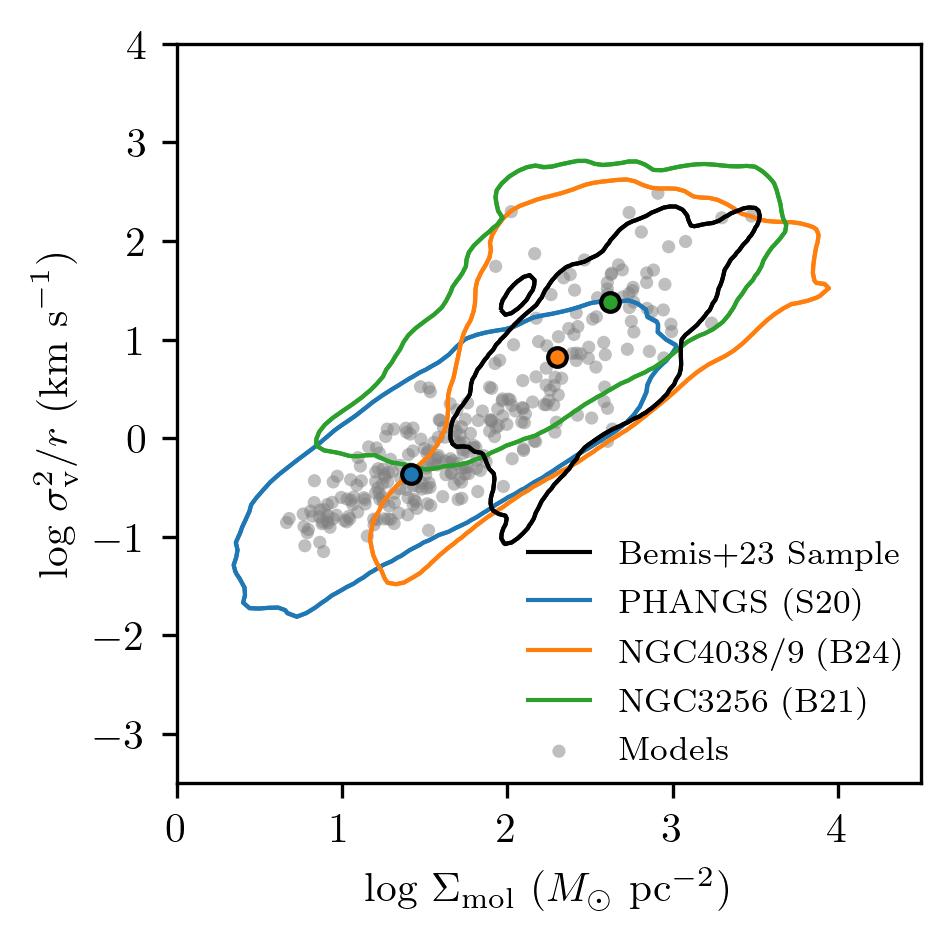}
    \caption{The model parameter space showing the range of gas velocity dispersions and gas surface densities considered in our analysis. The model points are plotted as gray points in the  $\sigma_\mathrm{v}^2/R-\Sigma_\mathrm{mol}$ parameter space, and include a mix of cloud measurements from the PHANGS sample \citep{Sun:2020}, NGC 4038/9 \citep{Brunetti:2024}, and NGC 3256 \citep{Brunetti:2021}. We also outline $\sigma_\mathrm{v}^2/R-\Sigma_\mathrm{mol}$ measurements of the \citet{Sun:2020} PHANGS galaxies at 90 pc resolution ($R=45\ \mathrm{pc}$, blue contours), NGC 4038/9 at 80 pc resolution ($R=40\ \mathrm{pc}$, orange contours), and NGC 3256 at 80 pc resolution ($R=40\ \mathrm{pc}$, green contours). The lower resolution data from \citetalias{Bemis:2023} are outlined by the black solid line. Example models from Fig. \ref{fig:ex_pdfs} are indicated in this plot as the blue (1), orange (2), and green (3) points.}
    \label{fig:model_param_space}
\end{figure}

\subsection{Applying radiative transfer} \label{sec:radiative_transfer}

We used RADEX \citep{vanderTak:2007} to perform radiative transfer and calculate emissivities of our cloud models. Each cloud model is comprised of an $n-$PDF distribution with 500 resolution elements across the PDF in volume density. We run RADEX at each resolution element across the $n-$PDF, using the escape probability formulation and adopting its default uniform sphere geometry. For each resolution element in our cloud model, RADEX computes a line flux, optical depth, and excitation temperature that we later use to calculate PDF-averaged properties of each cloud model (see Sect. \ref{sec:pdf_weighting}). To perform its radiative transfer calculation, RADEX requires the input of $\mathrm{H}_2$ volume density, molecular column density, gas kinetic temperature, and molecular line width at each resolution element. In Sect. \ref{sec:param_space} we describe how we set fiducial values of mean gas density, $n_0$, velocity dispersion $\sigma_\mathrm{v}$, and kinetic temperature, $T_\mathrm{kin}$ for each individual model across our model parameter space. We describe how these physical inputs translate to the range of volume and column densities required for each model in the paragraphs below.
\par

To provide RADEX with a molecular column density for each volume density in the model, we assumed a power-law density distribution for the radial $\mathrm{H}_2$ volume ($n$) and column ($N$) density distributions. One zone models bypass this requirement by assuming a fixed optical depth \citep[e.g.,][]{Leroy:2017a}{}{}, which indirectly determines the $n-N$ relationship, but can underestimate molecular abundances at high densities, and overestimate them at low densities. We therefore adopt power-law radial density distributions that are more realistic for molecular clouds. Spatial density gradients are observed in real molecular clouds, and the slopes of spatial density gradients are potentially connected to the shape of their $n-\mathrm{PDF}$ \citep[cf.][]{Federrath:2013}. Furthermore, these slopes are likely connected to the star formation properties of molecular clouds \citep{Tan:2006,Elmegreen:2011,Parmentier:2014,Kainulainen:2014,Parmentier:2019}. The \citet{Burkhart:2018} and \citet{Burkhart:2019} models predict a connection between the power-law slope of the $n-\mathrm{PDF}$ and $\epsilon_\mathrm{ff}$, and this behaviour has also been observed in Milky Way  clouds \citep{Federrath:2013}. We therefore used the power-law slope of the $n-\mathrm{PDF}$ of our models, $\alpha_\mathrm{PL}$, to determine the gas density gradient of our models.
\par
\citet{Federrath:2013} show that the slope of the gradient in a radially symmetric density distribution will be related to the slope of the corresponding $n-$PDF if they both follow power-law scalings. Using this scaling for spherical geometries in \citet{Federrath:2013}, we connect the slope of the high-density power-law tail of the $n-$PDF ($\alpha_\mathrm{PL}$) to the radial slope of the clump density profile, $k$, via \citep[cf.][]{Federrath:2013}:
\begin{equation} \label{eq:k}
    k = \frac{3}{\alpha_\mathrm{PL}-1}.
\end{equation}
For comparison, a power-law with $k=2$ is consistent with the expectation for isothermal cores \citep{Shu:1977}, and results in an $n-$PDF slope of $\alpha_\mathrm{PL}=2.5$. Shallower $n-$PDF slopes then correspond to steeper spatial density gradients and vice versa. 
\par
The radially symmetric approximation assumed above is only analytically exact for the gravitationally bound gas in the power-law tail of the $n-$PDF. The gas outside of the power-law tail is primarily governed by turbulence, which produces fractal, self-similar structure \citep[e.g.,][]{ElmegreenFalgarone:1996,Schneider:2011}. Self-similarity implies there is no characteristic scale of the gas, but this is not inconsistent with the existence of density gradients in turbulent gas.  In the interest of simplicity we also adopt the same power-law density gradient for the gas that we attribute to the lognormal component of the $n-$PDF. We implement this by adopting a power law for the radial volume and column density distributions, normalised to surface values (see Eqs. \ref{eq:radial_density_dist} and \ref{eq:radial_column_dens_dist}, respectively.) The radial volume density distribution is then given by 
\begin{equation}\label{eq:radial_density_dist}
    n(r) = n(R)\left( \frac{r}{R}\right)^{-k}
,\end{equation}
where $r$ is the radial coordinate, $R$ is the size of the molecular component of the cloud, and $n(R)$ is the volume density of the molecular cloud. We assumes that, in general, the radial profile of the column density will track the radial profile of the volume density. This general trend is consistent with the assumption of either a stiff equation of state (temperature increases with density) or   an isothermal equation of state \citep{Federrath:2015}. The gas-temperature relationship in our models is, on average, consistent with a stiff equation of state. Since the exact trend varies from model to model, we simply adopt
\begin{equation}\label{eq:radial_column_dens_dist}
    N(r) = N(R)\left( \frac{r}{R}\right)^{-(k-1)}
,\end{equation}
where $N(R)$ is the column density at the surface of the molecular cloud and is consistent with an isothermal cloud following and ideal gas equation of state. We then derived molecular column density distributions by multiplying the radial $\mathrm{H}_2$ column density distribution (Eq. \ref{eq:radial_column_dens_dist}) by the appropriate absolute molecular abundance. Although abundance variations are possible within our sources, we present the results of our models assuming fixed molecular abundances $x_\mathrm{HCN}=10^{-8}$ and $x_\mathrm{CO}=1.4\times10^{-4}$ relative to H$_2$ when converting $N$ to a molecular column density (i.e., $N_\mathrm{HCN}$ or $N_\mathrm{CO}$) \citep{DraineBook:2011}. We show example emissivity profiles for several models in Fig. \ref{fig:ex_pdfs}.  We note that we assumed fixed abundances so that the results of our modeling focus on the impact of variations in turbulent velocity dispersion on HCN and CO emissivities. We run additional models to assess the impact of varying the absolute abundance of HCN and CO on model output, and we present these results in Appendix \ref{ap:mol_abundance}. In summary, we find that variations in molecular abundance have a moderate impact on the optical depths of HCN and CO emission, but that these variations do not significantly impact the various correlations between HCN, CO, and molecular cloud properties considered in this work. 

\subsection{Deriving emissivity and intensity from molecular cloud models}
\label{sec:pdf_weighting}

Using the framework described in Sects. \ref{sec:emiss} -- \ref{sec:radiative_transfer}, we numerically solved for CO and HCN $J=1-0$ emissivities. Similar to \citet{Leroy:2017a}, we weighted the unintegrated emissivities (Eq. \ref{eq:emiss}) by the mass distribution of model clouds, $\mathrm{p}_n$, and integrate to determine the mass-weighted emissivity,
\begin{equation}\label{eq:emiss_integral}
    \left< \epsilon_\mathrm{mol} \right> = \frac{\int \ \epsilon_\mathrm{mol}(n)\  n\ \mathrm{p}_n\ \mathrm{d}n}{\int \ n\ \mathrm{p}_n\ \mathrm{d}n}
,\end{equation}
where we have re-written Eq. \ref{eq:emiss} in terms of $n$ and ``$\mathrm{mol}$'' denotes HCN or CO. When calculating $\left< \epsilon_\mathrm{mol} \right>$, we numerically integrated the PDF over densities that are relevant to molecular gas, roughly $\sim10-10^{8}$ cm$^{-3}$. This produces a mass-averaged molecular line emissivity with units of $\mathrm{K\ km\ s^{-1}\ cm^2}$. As \citet{Leroy:2017a} point out, $1/\left< \epsilon_\mathrm{mol} \right>$ can be recast in units of $M_\odot / \mathrm{pc^{2}}\ [\mathrm{K\ km\ s^{-1}]}^{-1}$, similar to a molecular line luminosity-to-mass conversion factor. We note that in this work we primarily model quantities that are surface densities (i.e., molecular intensity and column density). However, we are still able to compare the relative mass conversion factors of HCN and CO using properties of the $n-\mathrm{PDF}$, and we explore the difference between emissivity and conversion factors more in Sect. \ref{sec:ratio_gas_fraction}.
\par 
Similar to Eq. \ref{eq:emiss}, the modeled emissivity can be parameterized by an average intensity, $\left<I_\mathrm{mol}\right>$, and an average $\mathrm{H_2}$ column density, $\left< N_\mathrm{H_2, mol} \right>$, over which the molecular transition is sensitive to: $\left< \epsilon_\mathrm{mol} \right> = \left<I_\mathrm{mol}\right>/\left<N_\mathrm{H_2, mol}\right>$. Thus, if we know $\left<N_\mathrm{H_2, mol}\right>$, we can derive intensities that are analogous to what are measured in observations of individual molecular clouds. We estimated the average column of mass that a transition is sensitive to, $\left<N_\mathrm{H_2, mol}\right>$, from our models using
\begin{equation} \label{eq:expec_val_cd}
    \left< N_\mathrm{H_2, mol} \right> = \frac{\int \, N(n)\, \epsilon_\mathrm{mol}(n)\,  n\, \mathrm{p}_n\ \mathrm{d}n}{\int \epsilon_\mathrm{mol}(n)\, n\, \mathrm{p}_n\, \mathrm{d}n}
,\end{equation}
where $N(n)$ is the $\mathrm{H_2}$ column density corresponding to $\mathrm{H_2}$ volume density $n$, and the two quantities are related via Eqs. \ref{eq:radial_density_dist} and \ref{eq:radial_column_dens_dist} in our models. Using $\left< N_\mathrm{H_2, mol} \right>$, we derived intensities from our emissivities that can be compared with those measured in our sample from \citetalias{Bemis:2023} and the EMPIRE sample \citep{Jimenez_Donaire:2019}.
\par
RADEX also returns optical depth and excitation temperature for a given molecular transition at each $n$ across the $n-\mathrm{PDF}$. We determined a fiducial optical depth, $\left< \tau_\mathrm{mol} \right>$, and excitation temperature, $\left< T_\mathrm{ex,mol} \right>$, for a given molecular transition of each cloud model by calculating the expectation values of these quantities weighted by emissivity via
\begin{equation} \label{eq:expec_val_tau}
    \left< \tau_\mathrm{mol} \right> = \frac{\int \, \tau_\mathrm{mol}(n)\, \epsilon(n)\,  n\, \mathrm{p}_n\ \mathrm{d}n}{\int \epsilon(n)\, n\, \mathrm{p}_n\, \mathrm{d}n},\ \mathrm{and}
\end{equation}
\begin{equation} \label{eq:expec_val_tex}
    \left< T_\mathrm{ex,mol} \right> = \frac{\int \, T_\mathrm{ex,mol}(n)\, \epsilon(n)\,  n\, \mathrm{p}_n\ \mathrm{d}n}{\int \epsilon(n)\, n\, \mathrm{p}_n\, \mathrm{d}n}.
\end{equation}
These estimates of $\left< \tau_\mathrm{mol} \right>$ and $\left< T_\mathrm{ex,mol} \right>$ are useful for comparison to observations.
\par 

\section{Model results}
\label{sec:model_results}

We present the model results in the following sections in Figs. \ref{fig:inten_compared_to_obs} to \ref{fig:eff_Pturb_emiss}. In Sect. \ref{sec:excitation_optical_depth} we discuss the impact of excitation and optical depth on the modeled CO and HCN intensities and illustrate these results using the first set of plots (Figs. \ref{fig:inten_compared_to_obs}, \ref{fig:inten_tex_tau}, \ref{fig:pair_phys}). We constrain the characteristic gas densities that the HCN/CO ratio is sensitive to in Sect. \ref{sec:ratio_gas_fraction} (cf. Fig. \ref{fig:ratio_gas_frac}). We explore  trends in the CO and HCN emissivity ($\left<\epsilon_\mathrm{CO}\right>$ and $\left<\epsilon_\mathrm{HCN}\right>$) and luminosity-to-mass conversion factors ($\alpha_\mathrm{CO}$ and $\alpha_\mathrm{HCN}$, cf. Fig. \ref{fig:co_conversion_factor}) in Sect. \ref{sec:co_hcn_emiss}. In Sects. \ref{sec:ratio_fgrav} and \ref{sec:sf_relations}, we explore if the $I_\mathrm{HCN}/I_\mathrm{CO}$ ratio traces the fraction of gravitationally bound gas (Sect. \ref{sec:ratio_fgrav}), as well as how variations in CO and HCN emissivity impact our interpretation of star formation scaling relations (Sect. \ref{sec:sf_relations}). We note that we sometimes differentiate between models that represent clouds from different star-forming regimes (i.e., PHANGS-type vs. NGC 3256-type and NGC 4038/9-type) and color the results presented in some figures accordingly.
\par
In Sects. \ref{sec:ratio_gas_fraction}, \ref{sec:ratio_fgrav}, and \ref{sec:sf_relations}, we combine the predictions of the LN+PL analytical models of star formation \citep{Burkhart:2018} with the results of our radiative transfer modeling. For convenience, we produce a summary of the most relevant equations in the bottom of Table 2 in \citetalias{Bemis:2023} describing how various quantities are calculated. In these sections we explore how variations in CO and HCN emissivity, as well as variations in the CO and HCN luminosity-to-mass conversion factors, may impact observed star formation scaling relations. We mimic the results of observational studies by applying common conversion factors to our modeled molecular line intensities to derive gas surface densities (method one), and we compare these results with the true model predictions (method two). For method one, the modeled molecular intensities are  multiplied by constant conversion factors, as we have done with our sample from \citetalias{Bemis:2023} and the EMPIRE sample. We choose a value that is intermediate between the Milky Way and starburst values for $\alpha_\mathrm{CO}$: $\alpha_\mathrm{CO}=3\,[\mathrm{M}_\odot\,(\mathrm{K}\,\mathrm{km}\,\mathrm{s}^{-1}\,\mathrm{pc}^{2})^{-1}]$ and $\alpha_\mathrm{HCN}=3.2\,\alpha_\mathrm{CO}$ to produce estimates of gas mass surface densities, which are the same values used in \citetalias{Bemis:2023}.

\subsection{Excitation and optical depth}
\label{sec:excitation_optical_depth}

\begin{figure}
    \includegraphics[width=\columnwidth]{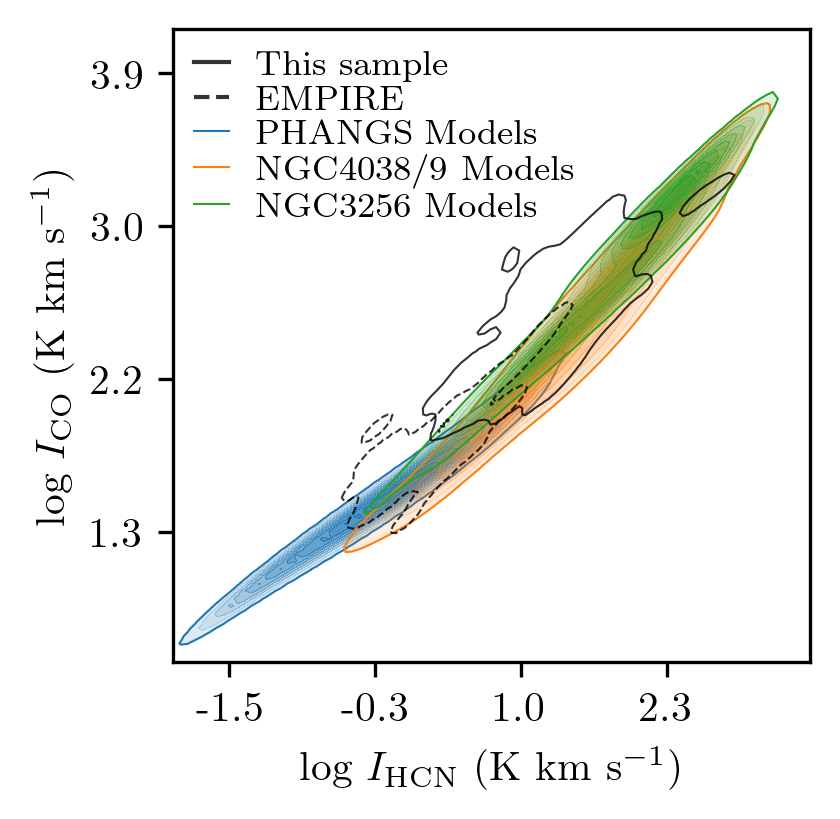}
    \caption{Modeled $I_\mathrm{CO}$ and $I_\mathrm{HCN}$ compared with the measured intensities of the \citetalias{Bemis:2023} sample (solid contours) and the EMPIRE sample \citep[dashed contours][]{Jimenez_Donaire:2019}. The blue filled contours are models whose $\Sigma$ and $\sigma_\mathrm{v}$ are taken from the PHANGS sample \citep{Sun:2020}, the orange filled contours are those taken from NGC 4038/9 \citep{Brunetti:2024}, and the green filled contours are those taken from NGC 3256 \citep{Brunetti:2021}. Ten contours are drawn in even steps from the $16\mathrm{th}$ to $100\mathrm{th}$ percentile. }
    \label{fig:inten_compared_to_obs}
\end{figure}

We show the modeled CO and HCN intensities compared to the intensities measured in our sample from \citetalias{Bemis:2023} and the EMPIRE sample from \citep{Jimenez_Donaire:2019} in Fig. \ref{fig:inten_compared_to_obs}. The ranges of HCN and CO $J=1-0$ intensities produced by our models encompass those we measure in the disk galaxies of the EMPIRE sample \citep[$I_\mathrm{HCN}=0.4-20.5\ \mathrm{K\ km\ s^{-1}}$ and $I_\mathrm{CO}=21.6-331.5\ \mathrm{K\ km\ s^{-1}}$,][]{Jimenez_Donaire:2019} and our more extreme sample of galaxies including U/LIRGs and galaxy centers \citep[$I_\mathrm{HCN}=0.8-814.5\ \mathrm{K\ km\ s^{-1}}$ and $I_\mathrm{CO}=53.8-2397.4\ \mathrm{K\ km\ s^{-1}}$,][cf. Fig. \ref{fig:inten_compared_to_obs}]{Bemis:2023}. The scatter is less well-matched to observations, which may be due to uncertainties in the relative filling fractions of HCN and CO. We calculate the median absolute deviations (MAD) of our measured and modeled HCN and CO intensities and multiply by 1.4826 to get an estimate of the scatter (standard deviation) that is less sensitive to outliers. We find scatters of $\sigma_\mathrm{HCN}=1.9\ \mathrm{K\ km\ s^{-1}}$ and $\sigma_\mathrm{CO}=27.2\ \mathrm{K\ km\ s^{-1}}$ for the EMPIRE sample, $\sigma_\mathrm{HCN}=13.2\ \mathrm{K\ km\ s^{-1}}$ and $\sigma_\mathrm{CO}=176.2\ \mathrm{K\ km\ s^{-1}}$ for our sample, and $\sigma_\mathrm{HCN}=3.2\ \mathrm{K\ km\ s^{-1}}$ and $\sigma_\mathrm{CO}=51.2\ \mathrm{K\ km\ s^{-1}}$ for all models. We find scatters of HCN and CO intensity for just the PHANGS-type models to be $\sigma_\mathrm{HCN}=1.2\ \mathrm{K\ km\ s^{-1}}$ and $\sigma_\mathrm{CO}=25.0\ \mathrm{K\ km\ s^{-1}}$, which is well-matched to the observations of the EMPIRE sample. In contrast to this, we find $\sigma_\mathrm{HCN}=64.0\ \mathrm{K\ km\ s^{-1}}$ and $\sigma_\mathrm{CO}=500.0\ \mathrm{K\ km\ s^{-1}}$ for the NGC 4038/9 and NGC 3256 models combined. This scatter is less well-matched to our data, although roughly on the same order of magnitude as what is measured in our sample. We discuss the impact of emissivity on the scatter of observations in Sect. \ref{sec:sf_relations}.
\par

\begin{figure*}[tb!]
    \centering
    \includegraphics[width=0.49\textwidth]{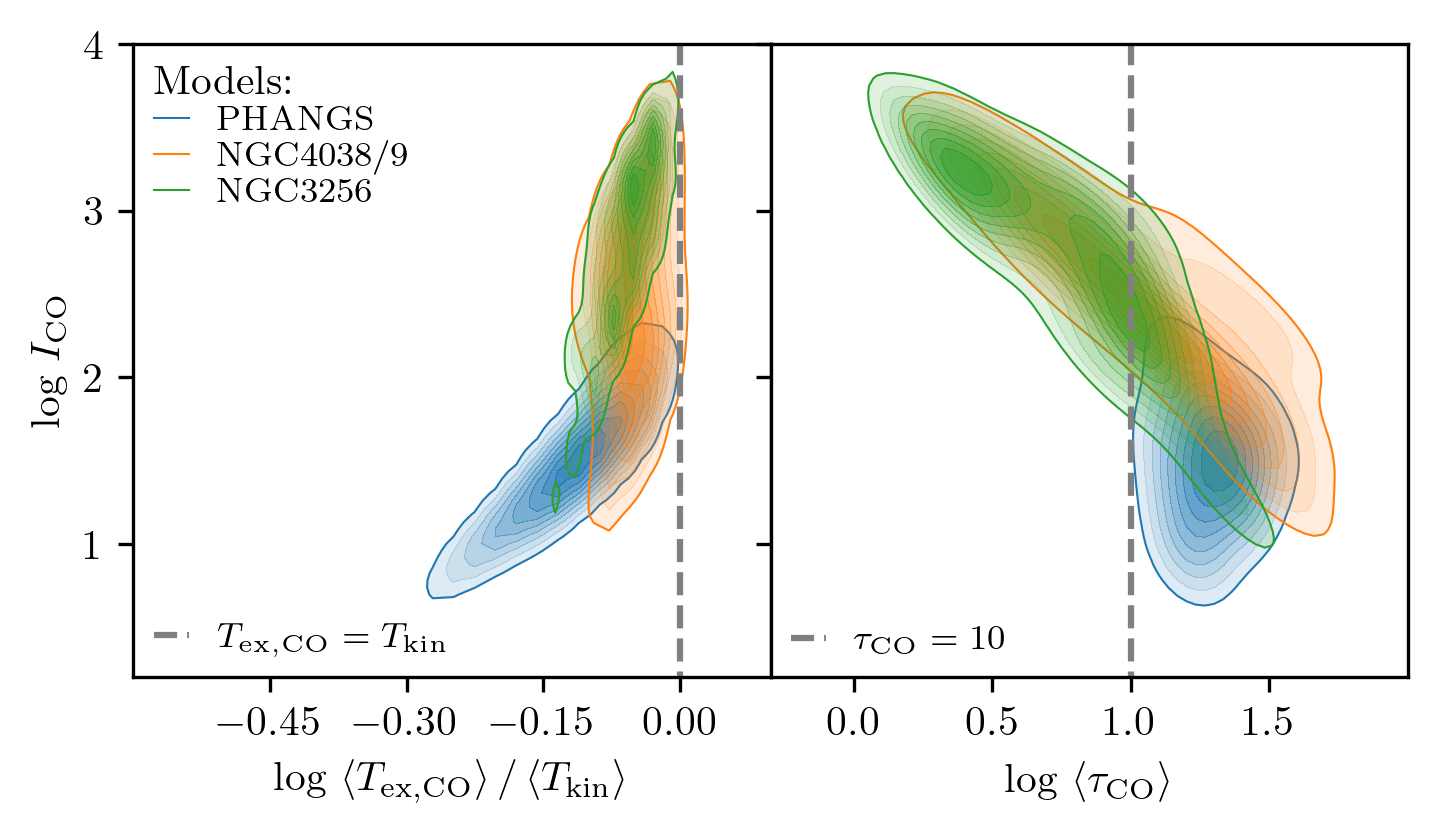}
    \includegraphics[width=0.49\textwidth]{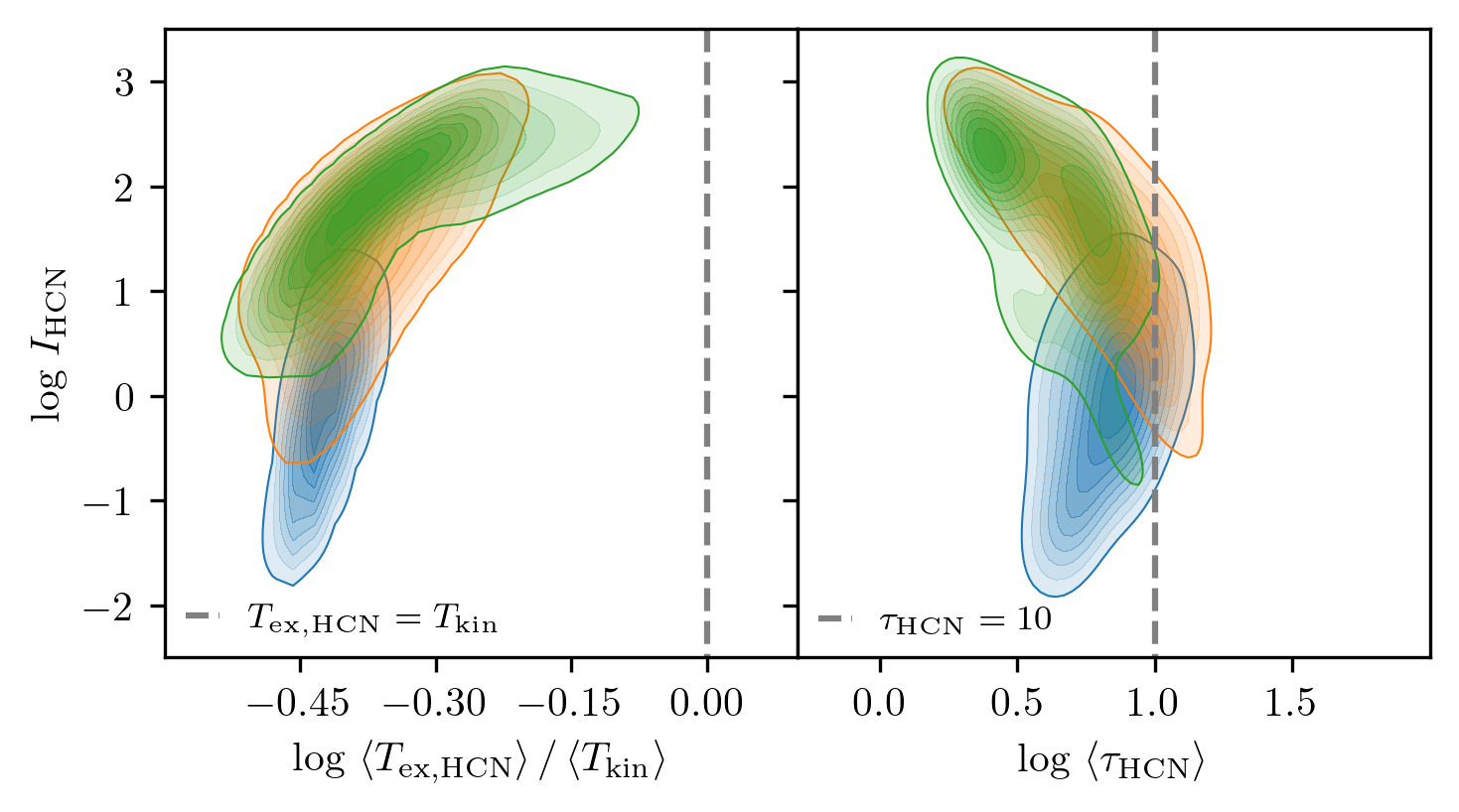}
    \caption{Correlations between modeled $I_\mathrm{CO}$ (left two plots) and $I_\mathrm{HCN}$ (right two plots) and their respective excitation temperatures and optical depths determined using Eqs. \ref{eq:expec_val_tau} and \ref{eq:expec_val_tex}. The formatting is the same as in Fig. \ref{fig:inten_compared_to_obs}. We find that CO optical depth decreases with increasing $I_\mathrm{CO}$ and CO excitation, in general agreement with the findings of previous studies \citep[e.g.,][]{Bolatto:2013, Narayanan:2012, Narayanan:2014}. In our models, HCN appears subthermally excited and moderately optically thick, also in agreement with the findings of previous studies \citep[e.g.,][]{Dame:2023,Jimenez_Donaire:2017}.}
    \label{fig:inten_tex_tau}
\end{figure*}

Figure \ref{fig:inten_tex_tau} presents the modeled CO and HCN $J=1-0$ intensities as a function of the excitation temperature and optical depth. We find that the CO $J=1-0$ transition is close to LTE for the majority of our models when compared to our estimates of $\left< T_\mathrm{kin} \right>$. A subset of PHANGS models show slightly subthermal emission, which is due to the average density of these models being below the critical density for CO $J=1-0$ (i.e.,\ $\sim\!10^{2-3}\,\mathrm{cm^{-3}}$), the density at which the majority of CO emission becomes thermalised. The CO $J=1-0$ transition is, on average, optically thick for the PHANGS-type models. Towards higher $I_\mathrm{CO}$ where the models are dominated by NGC 4038/9- and NGC 3256-type clouds, the CO optical depth drops and approaches $\tau\sim1$ towards the models with the brightest CO emission. This behavior is similar to the results of previous studies of CO excitation \citep[e.g.,][]{Narayanan:2014}, where the optical depth of the $J=1-0$ transition appears to drop towards gas where CO is more excited (and $\Sigma_\mathrm{SFR}$ is high). This is due to the fact that the optical depth of CO $J=1-0$ is inversely proportional to velocity dispersion, and that velocity dispersion tends to increase with $\Sigma_\mathrm{SFR}$.
\par
The HCN $J=1-0$ transition appears subthermally excited, which agrees with a number of studies that assess the excitation of HCN in the Milky Way and nearby galaxies \citep[e.g.,][]{Dame:2023,GarciaRodriguez:2023,Tafalla:2023}. The HCN optical depth is found to be only moderately optically thick for the PHANGS-type clouds in our models, and is in agreement with previous studies towards the centers of disk galaxies \citep{Jimenez_Donaire:2017}. These results suggest that variations in $I_\mathrm{CO}$ may be more strongly impacted by variations in $\tau_\mathrm{CO}$ relative to the impact of $\tau_\mathrm{HCN}$ on $I_\mathrm{HCN}$. We note that the drop in the CO optical depth for the extreme systems coincides with a transition in the dominant heating mechanism from cosmic ray heating to turbulent heating, and is a reflection of an increase in the typical gas velocity dispersion in NGC 4038/9 and NGC 3256-type clouds relative to PHANGS-type clouds.

\begin{figure*}[tb!]
    \raggedright
    \includegraphics[width=0.99\textwidth]{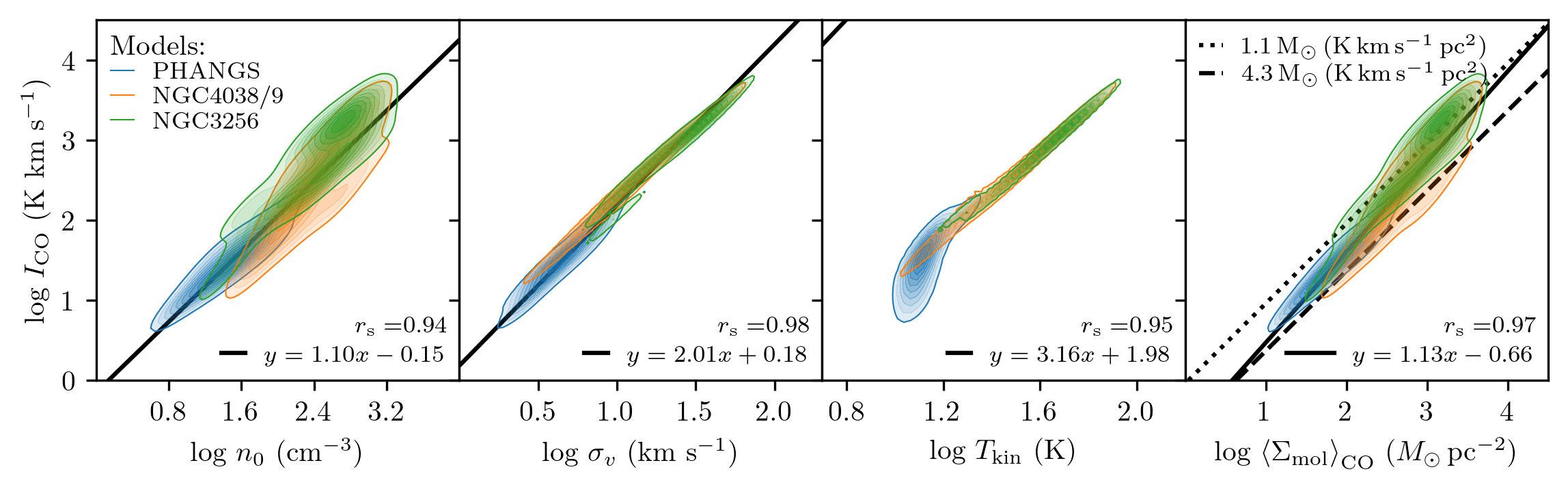}
    \includegraphics[width=0.99\textwidth]{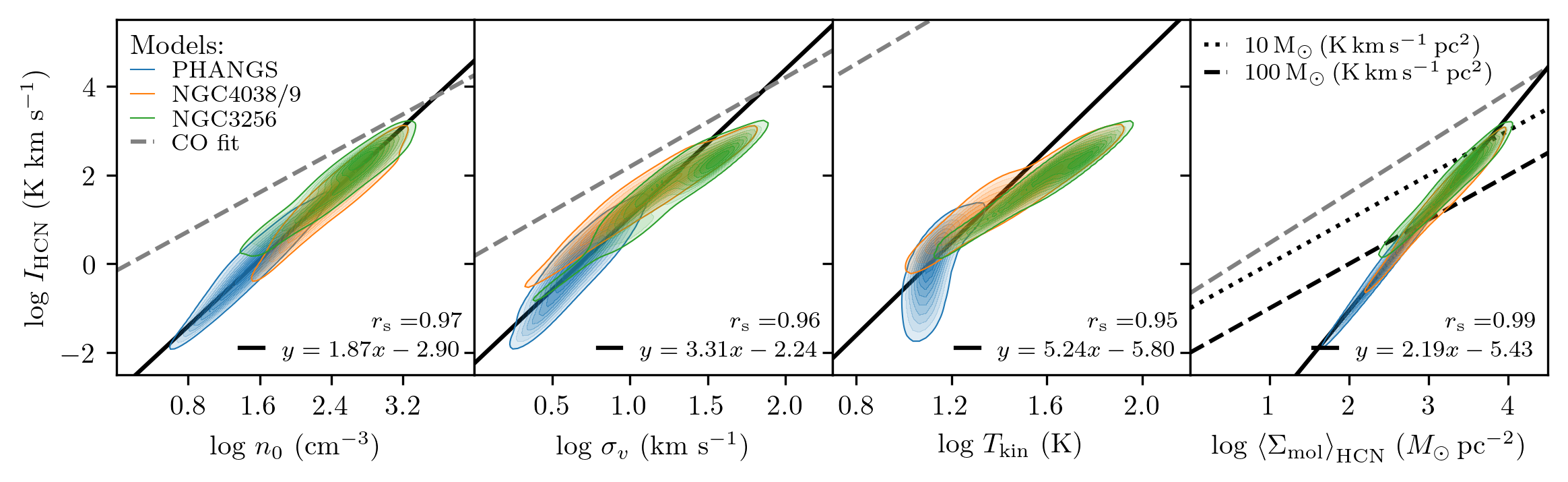}
    \includegraphics[width=0.99\textwidth]{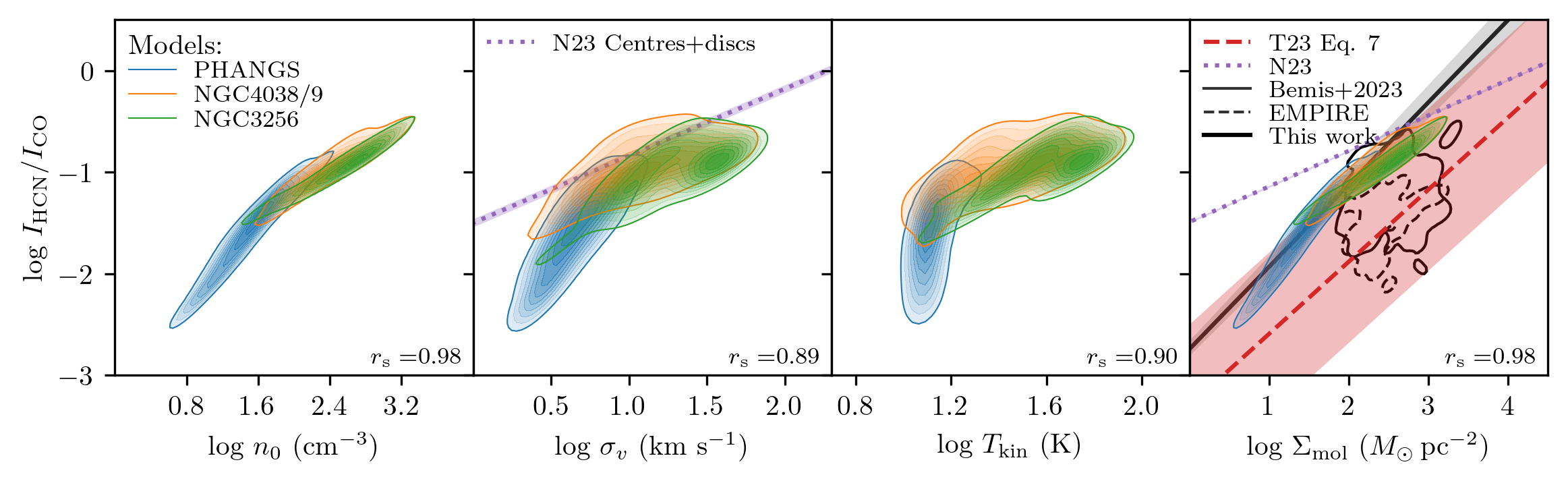}
    \caption{CO intensity (top row), the HCN intensity (center row), and  HCN/CO ratio (bottom row) as a function of mean gas density, velocity dispersion, kinetic temperature, and gas column density. The column densities shown are from Eq. \ref{eq:expec_val_cd} for $I_\mathrm{CO}$ and $I_\mathrm{HCN}$ (top and center rows) and are the fiducial model column densities for the HCN/CO ratio (bottom row). The reference conversion factor values are shown in the intensity vs. column density plots as the dotted and dashed lines. We show the CO fits (top row) in the HCN plots (center row) as the gray dashed lines. Spearman rank coefficients are shown in the lower right corner. The formatting is the same as in Fig.\ref{fig:inten_compared_to_obs}. We plot fits from the results of the ALMOND survey \citep[purple dotted line][]{Neumann:2023} and \citet{Tafalla:2023} (red dashed line). Uncertainties on their respective fits are shown as the shaded areas. For comparison, we plot the results of \citetalias{Bemis:2023} sample (solid contours) and the EMPIRE sample \citep[dashed contours;][]{Jimenez_Donaire:2019} in the HCN/CO ratio vs. gas surface density plot.  Our models show strong positive correlations between the modeled line intensities and mean density, velocity dispersion, mean kinetic temperature, and gas column densities. $I_\mathrm{HCN}$ appears to increase more rapidly with each of these parameters compared to $I_\mathrm{CO}$. The Spearman rank coefficients are shown in the lower right corner of each plot.}
    \label{fig:pair_phys}
\end{figure*}

We also explore how physical quantities impact CO and HCN $J=1-0$ intensities in our models. In Fig. \ref{fig:pair_phys}, we present the modeled CO and HCN $J=1-0$ intensities as a function of mean density, velocity dispersion, kinetic temperature and gas surface density. For simplicity, we use $I_\mathrm{HCN}$ and $I_\mathrm{CO}$ in place of $\left<I_\mathrm{HCN}\right>$ and $\left<I_\mathrm{CO}\right>$ when referring to our modeled intensities (cf. Sect. \ref{sec:pdf_weighting}). We perform fits using orthogonal distance regression. We also calculate Spearman rank coefficients and show these in the lower right corner of each plot. For comparison, we have included the relationships between $I_\mathrm{HCN}/I_\mathrm{CO}$ and $\sigma_\mathrm{v}$ and $\Sigma_\mathrm{mol}$ found in nearby galaxies from the ALMOND survey \citep{Neumann:2023}, as well as the relationship found by \citet{Tafalla:2023} between $I_\mathrm{HCN}/I_\mathrm{CO}$ and gas surface density as determined through extinction measurements in Milky Way clouds. Both $I_\mathrm{HCN}$ and $I_\mathrm{CO}$ are strongly correlated with $n_0$, $\sigma_\mathrm{v}$, and $\left< T_\mathrm{kin} \right>$ in our models. We fit each trend to assess how rapidly $I_\mathrm{HCN}$ and $I_\mathrm{CO}$ change with each parameter (cf Fig. \ref{fig:pair_phys}). We find that $I_\mathrm{HCN}$ increases more steeply than $I_\mathrm{CO}$ with each parameter. Individually, $I_\mathrm{CO}$ and $I_\mathrm{HCN}$ are most strongly correlated with the velocity dispersion, with the ratio $I_\mathrm{HCN}/I_\mathrm{CO}$ instead appearing most strongly correlated with the mean density. The trend in $I_\mathrm{HCN}/I_\mathrm{CO}$ vs. $\sigma_\mathrm{v}$ is more shallow for the ALMOND galaxies than what is found by our models. The trend from ALMOND galaxies intersects with the NGC 3256- and NGC 4038/8-type models relative to the PHANGS-type models. In general, there appears to be slight differences between the PHANGS-type clouds to NGC 4038/9- and NGC 3256-type clouds  in how $I_\mathrm{HCN}$ and $I_\mathrm{CO}$ vary with each physical parameter. This is most obvious when looking at the ratio of $I_\mathrm{HCN}/I_\mathrm{CO}$ relative to each quantity. Most notably, the trends in $I_\mathrm{HCN}/I_\mathrm{CO}$ with $n_0$, $\sigma_\mathrm{v}$, and $\left< T_\mathrm{kin} \right>$ appear to flatten towards the NGC 4038/9- and NGC 3256-type models (relative to the PHANGS-type models).
\par
We find a relationship between $I_\mathrm{HCN}/I_\mathrm{CO}$ and gas surface density in our models (cf. \ref{fig:pair_phys}), which has been found in studies of gas clouds in the Milky Way \citep[e.g][]{Tafalla:2023} as well as nearby galaxies \citep[e.g.,][]{Gallagher:2018b,Neumann:2023}. We perform a fit between $\mathrm{log}\left(I_\mathrm{HCN}/I_\mathrm{CO}\right)$ and log of the mean gas surface density and find a sublinear relationship similar to that found by \citet[eq. 6]{Tafalla:2023}: 
\begin{equation}
    \mathrm{log}\left(\frac{I_\mathrm{HCN}}{I_\mathrm{CO}}\right) = (0.81\pm0.03)\, \mathrm{log}\ \left(\frac{\Sigma_\mathrm{mol}}{\mathrm{M_\odot\,pc^{-2}}}\right) -2.73^{+0.07}_{-0.08}.
\end{equation}
Uncertainties on the fit are determined using bootstrapping. \citealt{Tafalla:2023} find a slope of $0.71$. As \citet{Tafalla:2023} show, other extragalactic studies find sublinear slopes, as well (0.81 in \citealt{Gallagher:2018} and 0.5 in \citealt{Jimenez_Donaire:2019}). Interestingly, the recent results of the ALMOND survey find a much shallower slope of $\sim0.33$ \citep{Neumann:2023}. \citet{Neumann:2023} compare observations of HCN and CO at 2.1 kpc scales with cloud-scale measurements of velocity dispersion and gas surface density from PHANGS galaxies, which may explain this discrepancy. We compare our fit with the results of \citet{Tafalla:2023} and \citet{Neumann:2023} in Fig. \ref{fig:pair_phys}. Our fit is slightly offset from \citet{Tafalla:2023}, which is consistent with the offset we see in our model intensities in Fig. \ref{fig:inten_compared_to_obs}. This relationship is in part due to the gas volume density and gas surface density scaling with each other in our models (cf. Eqs. \ref{eq:radial_density_dist} and \ref{eq:radial_column_dens_dist}), and the overall dense gas fraction increasing with gas volume density.\footnote{We note that simulations find that gas volume density tracks column densities in molecular clouds \citep[cf.][]{Priestly:2023}.} In general, our models are able to reproduce the sublinear relationship observed between the HCN/CO intensity ratio and gas surface density observed in both Milky Way clouds at $\sim\mathrm{parsec}$ scales and nearby galaxies at $\sim\mathrm{kiloparsec}$ scales.

\par 
In summary, our models are able to reproduce the range of HCN and CO $J=1-0$ intensities measured in the disk galaxies of the EMPIRE sample \citep{Jimenez_Donaire:2019} and our more extreme sample of galaxies including U/LIRGs and galaxy centers, presented in \citetalias{Bemis:2023} (cf. Fig. \ref{fig:inten_compared_to_obs}). Furthermore, we show that our models reproduce the expectations of CO excitation and optical depth \citep[cf.][]{Bolatto:2013,Narayanan:2012, Narayanan:2014}. Although HCN is less well-studied than CO, we find that our models agree with results of the current  works. In particular, HCN appears subthermally excited, as has been found via studies of high$-J$ lines of HCN emission in nearby galaxies \citep{GarciaRodriguez:2023}, and inferred from studies in Milky Way clouds \citep{Dame:2023}. Additionally, HCN appears only moderately optically thick ($\tau < 10$), as was found by \citep{Jimenez_Donaire:2017} when comparing HCN and H$^{13}$CN emission towards the centers of nearby disk galaxies. Since gas volume density tracks column density in our models, we find $I_\mathrm{HCN}/I_\mathrm{CO}$ also scales with gas surface density.

\subsection{The fraction of gas traced by the HCN/CO ratio}\label{sec:ratio_gas_fraction}

\begin{figure*}
    \includegraphics[width=\textwidth]{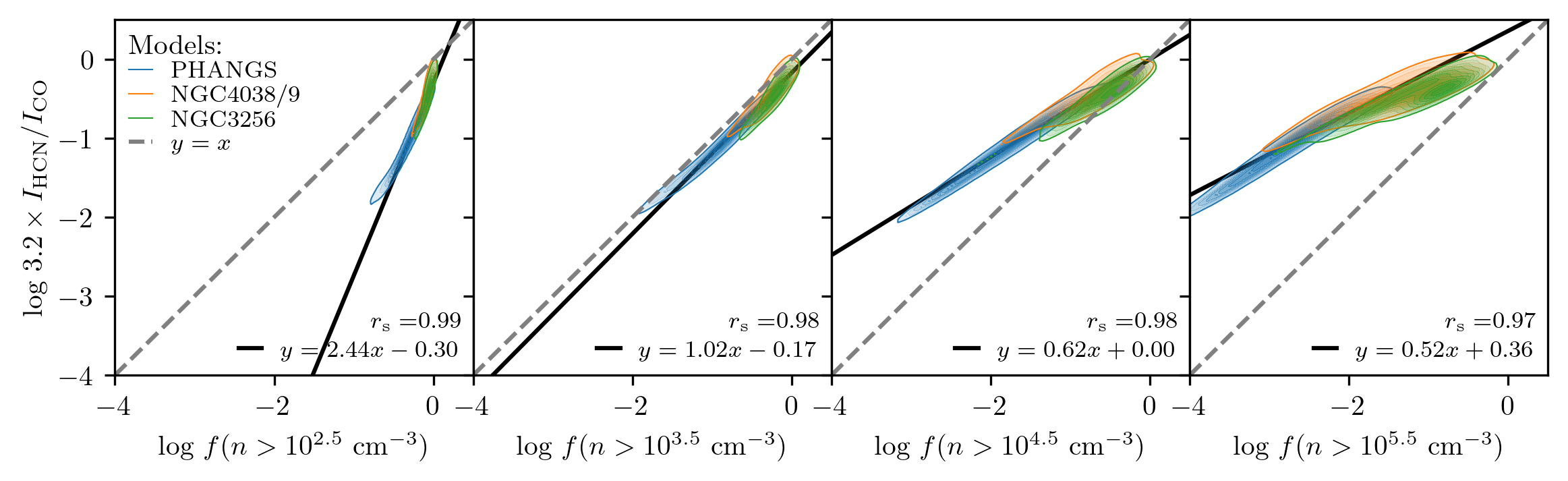}
    \caption{$3.2 \times I_\mathrm{HCN} / I_\mathrm{CO}$ as a function of the fraction of gas above (from left to right) $n\sim 10^{2.5},\,10^{3.5},\,10^{4.5}$, and $10^{5.5}$ cm$^{-3}$.   The formatting is the same as in Fig. \ref{fig:inten_compared_to_obs}. The fits are shown as the solid black line (see  legend). The modeled $I_\mathrm{HCN}/I_\mathrm{CO}$ scales most directly (has a slope closest to unity) with $f(n>10^{3.5}\ \mathrm{cm}^{-3})$, supporting previous findings that this line ratio is sensitive to gas above moderate densities. Although not shown here, the same is found when comparing the emissivity ratio, $\left<\epsilon_\mathrm{HCN}\right>/\left<\epsilon_\mathrm{CO}\right>$, with the same gas fractions. The Spearman rank coefficients are shown in the lower right corner of each plot.}
    \label{fig:ratio_gas_frac}
\end{figure*}

We consider what fraction of gas the $I_\mathrm{HCN}/I_\mathrm{CO}$ ratio is sensitive to in molecular clouds, and whether this changes in more extreme environments, such as those found in galaxy centers. This is motivated by previous studies which have found an increase in the dense gas depletion time towards the centers of some disk galaxies \citep{Gallagher:2018,Querejeta:2019,Jimenez_Donaire:2019,Beslic:2021} and even in the nuclei of the Antennae \citep{Bemis:2019}, despite these regions also having higher $I_\mathrm{HCN}/I_\mathrm{CO}$. Under the assumption that $I_\mathrm{HCN}/I_\mathrm{CO}$ tracks the fraction of dense ($n>10^{4.5}\ \mathrm{cm}$), star-forming gas in molecular clouds, those results appeared in conflict with fixed threshold models of star formation that predict star formation should turn on above a relatively fixed density, and that the star formation rate should increase in the presence of higher $f_\mathrm{dense}$ (see the works by \citealt{Usero:2015,2019MNRAS.488.1407K}). In agreement with previous studies \citep[e.g.,][]{Burkhart:2019}, \citetalias{Bemis:2023} shows that gravoturbulent models of star formation are able to reproduce this increase in dense gas depletion time towards regions with higher fractions of dense gas. This result agrees with the findings of \citet{Gallagher:2018,Querejeta:2019,Jimenez_Donaire:2019,Beslic:2021} and \citet{Bemis:2019}. The major caveats of this conclusion are: 1. that $I_\mathrm{HCN}/I_\mathrm{CO}$ is tracing the fraction of gas above a relatively fixed density (i.e., $n>10^{4.5}\ \mathrm{cm}$), and 2. that the turbulent gas velocity dispersion is also increasing with $I_\mathrm{HCN}/I_\mathrm{CO}$. We have already shown that $I_\mathrm{HCN}/I_\mathrm{CO}$ increases with $\sigma_\mathrm{v}$ in Fig. \ref{fig:pair_phys}. We now consider if  $I_\mathrm{HCN}/I_\mathrm{CO}$ is tracing the fraction of gas above a relatively fixed density.
\par
In \citetalias{Bemis:2023}, we focus on comparing the HCN/CO ratio with the fraction of gas above $n>10^{4.5}$ cm$^{-3}$, which is the assumed threshold density for some clouds in the Milky Way disk. Other studies have shown that HCN is tracing gas primarily at moderate densities, $n\sim10^3$ cm$^{-3}$ (e.g., \citealt{Kauffmann:2017hcn_moderate,Pety:2017,Shimajiri:2017,Barnes:2020,Tafalla:2021,Tafalla:2023,Santa-Maria:2023}, Ngoc Le et al. in prep.), such that it may be more sensitive to mass fractions including densities below $n\sim10^{4.5}$. We compare the modeled HCN/CO ratio to several gas fractions derived from the model $n-$PDFs in Fig. \ref{fig:ratio_gas_frac}. To determine the fraction of gas above an arbitrary threshold density, we integrate the $n-$PDF above that threshold ($n_\mathrm{thresh}$):
\begin{equation}
    f(n>n_\mathrm{thresh}) = \frac{\int_{n>n_\mathrm{thresh}} \, n\, \mathrm{p}_n\ \mathrm{d}n}{\int n\, \mathrm{p}_n\, \mathrm{d}n}.
\end{equation}
We calculate gas fractions using the $n-$PDF above densities $\mathrm{log}\,(n)=2.5,\,3.5,\,4.5,\,5.5$ cm$^{-3}$, denoted by $f_{2.5}$, $f_{3.5}$, $f_{4.5}$, and $f_{5.5}$, respectively. We numerically integrate over a wide range in densities when calculating these fractions to ensure the $n-\mathrm{PDF}$ is fully sampled. We note that we multiply the modeled $I_\mathrm{HCN}/I_\mathrm{CO}$ ratio by a fixed factor, $\alpha_\mathrm{HCN}/\alpha_\mathrm{CO}=3.2$, which is the ratio of the \citet{Gao:2004a,Gao:2004b} HCN-to-$\mathrm{dense\ H_2}$ mass and Milky Way CO-to-$\mathrm{total\ H_2}$ mass conversion factors, and is the same factor we have we have applied to the HCN/CO ratio measured in the sources of our sample to estimate dense gas fractions in \citetalias{Bemis:2023}.
\par
We calculate Spearman rank coefficients ($r_\mathrm{s}$) between $I_\mathrm{HCN}/I_\mathrm{CO}$ and the gas fractions shown in Fig. \ref{fig:ratio_gas_frac} (i.e., $f_{2.5}$, $f_{3.5}$, $f_{4.5}$, and $f_{5.5}$). The $I_\mathrm{HCN}/I_\mathrm{CO}$ from our models is strongly ($| r_\mathrm{s}|>0.7$) correlated with all of the gas fractions we consider, with little difference between their Spearman rank coefficients. We fit the correlations between $3.2 \times I_\mathrm{HCN}/I_\mathrm{CO}$ and each of the fractions shown in Fig. \ref{fig:ratio_gas_frac} to assess the directness of each relationship. With a slope close to unity, the $I_\mathrm{HCN}/I_\mathrm{CO}$ ratio appears to have the most direct relationship with the fraction of gas above $n\sim 10^{3.5}$ cm$^{-3}$. The relationship between $3.2 \times I_\mathrm{HCN}/I_\mathrm{CO}$ and $f(n> 10^{4.5}\ \mathrm{cm}^{-3})$ appears sublinear in our models, such that $3.2\times I_\mathrm{HCN}/I_\mathrm{CO}$ overestimates $f(n> 10^{4.5}\ \mathrm{cm}^{-3})$ for the PHANGS-type clouds. These results suggest that $I_\mathrm{HCN}/I_\mathrm{CO}$ is tracing gas above a relatively constant fraction of gas, but that this includes gas at moderate densities below $n < 10^{4.5}\ \mathrm{cm}^{-3}$. 
\par
In summary, our models predict that, on average,  $I_\mathrm{HCN}/I_\mathrm{CO}$ does appear to track gas above a relatively fixed density, but that this fraction includes more moderately dense gas (i.e., $n>10^{3.5}\ \mathrm{cm}$) as opposed to strictly dense gas above $n>10^{4.5}\ \mathrm{cm}$. This result appears in agreement with more recent studies of HCN emission in Milky Way clouds that find HCN emission includes more moderate gas densities, for example  $n\sim800\ \mathrm{cm}^{-3}$ \citep[e.g.,][]{Kauffmann:2017hcn_moderate}. Our models include a range of cloud properties, including those found in normal galaxies (i.e.,\ PHANGS models) as well as more extreme cloud models based on cloud properties from the Antennae and NGC 3256. We find evidence that $I_\mathrm{HCN}/I_\mathrm{CO}$ tracks a gas fraction including more moderate gas densities even in the more extreme environments. 

\subsection{Using estimates of CO and HCN Emissivity to derive gas masses} \label{sec:co_hcn_emiss}

We explore if the dense gas fraction can be consistently recovered from observations of HCN and CO using luminosity-to-mass conversion factors, which are commonly used to estimate molecular gas masses from molecular line observations. We recall from Sect. \ref{sec:pdf_weighting} that emissivity can be recast in units of luminosity-to-mass conversion factors, such that $\alpha_\mathrm{mol} \propto 1/\left< \epsilon_\mathrm{mol} \right>$, with the caveat that emissivities derived in this work are relative to true cloud surface densities, rather than integrated quantities such as mass and molecular line luminosity. By construction, the ratio of our modeled intensities will be proportional to the ratio of molecular line luminosities analogous to those measured in resolved or unresolved observations, or the ratio of line intensities of resolved observations. We note that for the remainder of this work, we use $\left<\alpha_\mathrm{mol}\right>$ when we are referring to the inverse of modeled emissivity of a molecular transition, and $\alpha_\mathrm{mol}$ when referring to an estimate of the idealised mass conversion factor of a molecular transition (which may also include additional factors, such as the filling fraction).
\par
\begin{figure*}
    \centering
    \includegraphics[width=0.49\textwidth]{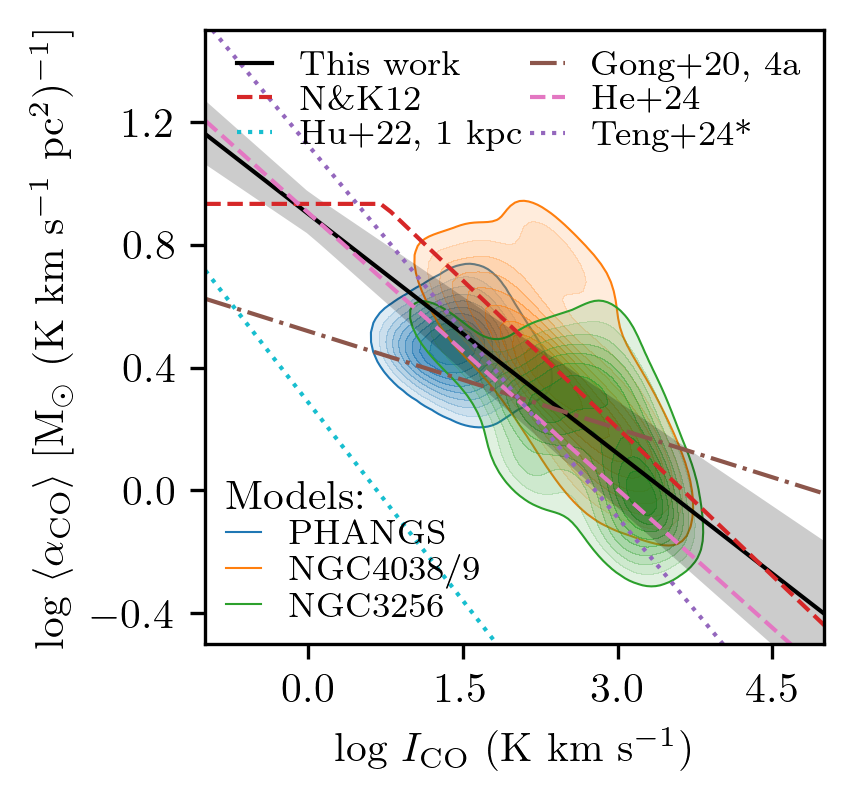}
    \includegraphics[width=0.47\textwidth]{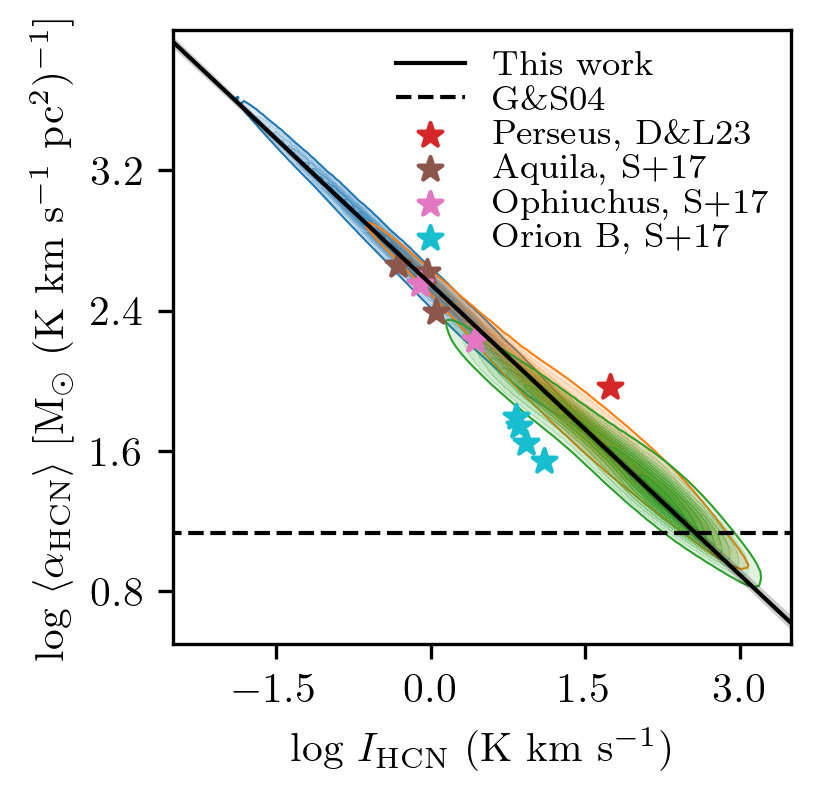} 
    \caption{The modeled emissivities of CO and HCN in units of $M_\odot\, \mathrm{(K\, km\, s^{-1}\, pc^2)^{-1}}$ as a function of CO and HCN intensity, respectively. \textit{Left:} Inverse of the CO emissivity (in units of the CO conversion factor) as a function of CO intensity, $\left<\alpha_\mathrm{CO}\right>=\left<\epsilon_\mathrm{CO}\right>^{-1}$. We include the fit to our models (Eq. \ref{eq:alpha_co}, solid line) and the 1$\sigma$ uncertainty on the fit (gray shaded area). We also include the results of several numerical studies \citep[][ red dashed line, cyan dotted line,  brown dash-dotted line, respectively]{Narayanan:2012,Hu:2022,Gong:2020ApJ...903..142G} as well as the results of observational studies at $\sim150\,\mathrm{pc}$ scales in the Antennae \citep[pink dashed line, ][]{He:2024} and PHANGS galaxies \citep[purple dotted line, ][]{Teng:2024}. We note that we have recast the $\alpha_\mathrm{CO} -\sigma_\mathrm{v}$ fit from \citet[][]{Teng:2024} in terms of $I_\mathrm{CO}$ using the fit between $I_\mathrm{CO}$ and $\sigma_\mathrm{v}$ from our models. \textit{Right:}  Inverse of the HCN emissivity (in units of the HCN conversion factor) as a function of HCN intensity, $\left<\alpha_\mathrm{HCN}\right>=\left<\epsilon_\mathrm{HCN}\right>^{-1}$. We include a fit to our models (Eq. \ref{eq:alpha_hcn}, solid line) and the 1$\sigma$  uncertainty on the fit (gray shaded area). For comparison, we include several published values of $\alpha_\mathrm{HCN}$ from observations of Milky Way clouds \citep{Dame:2023,Shimajiri:2017}, and we show the \citet{Gao:2004a,Gao:2004b} value as the black dashed line. This figure demonstrates how well our models are able to reproduce previous numerical prescriptions of $\alpha_\mathrm{CO}$, as well as observationally constrained values of $\alpha_\mathrm{CO}$ and $\alpha_\mathrm{HCN}$. The formatting of the model output is the same as in Fig. \ref{fig:inten_compared_to_obs}.} 
    \label{fig:co_conversion_factor}
\end{figure*}

In Fig. \ref{fig:co_conversion_factor}, we present the CO and HCN emissivities from our models, and contrast these against idealised luminosity-to-mass conversion factors. We fit the relationship between $\left<\alpha_\mathrm{CO}\right>=1/\left<\epsilon_\mathrm{CO}\right>$ and $I_\mathrm{CO}$ using orthogonal distance regression and find the following:
\begin{multline}\label{eq:alpha_co}
    \mathrm{log}\,\left(\frac{\alpha_\mathrm{CO}}{\mathrm{M_\odot\, (K\,km\,s^{-1}\,pc^2)^{-1}}}\right) = \\
    (-0.26^{0.03}_{-0.04}) \mathrm{log}\, \left(\frac{I_\mathrm{CO}}{\mathrm{K\,km\,s^{-1}}}\right) + 0.90\pm0.07.
\end{multline}
Uncertainties are determined using bootstrapping. We show in Fig. \ref{fig:co_conversion_factor} that our CO emissivities agree well with the \citet{Narayanan:2012} prescription for the CO-to-$\mathrm{H_2}$ conversion factor. We find a similar slope, -0.26, compared to -0.32 in \citep{Narayanan:2012}. We also compare with the numerical works of \citet{Hu:2022} and \citet{Gong:2020ApJ...903..142G}. The prescription taken from \citet{Hu:2022}, in particular, is for 1 kpc scales, which might explain the offset between their prescription and ours, but has roughly a similar slope ($-0.43$). In their work they also include modeling of $\alpha_\mathrm{CO}$ at higher resolution and do find higher values more consistent with our modeling. The \citet{Gong:2020ApJ...903..142G} relationship has a shallower slope than our trend, which appears inconsistent with some of the most recent studies of $\alpha_\mathrm{CO}$ in nearby galaxies \citep[e.g.,][]{He:2024,Teng:2024}.  We also compare with observationally derived estimates of $\alpha_\mathrm{CO}$ at $\sim150\,\mathrm{pc}$ scales in the Antennae \citep{He:2024} and PHANGS galaxies \citep{Teng:2024}. We find good agreement with these studies. We note that we have recast the $\alpha_\mathrm{CO} -\sigma_\mathrm{v}$ fit from \citet[][]{Teng:2024} in terms of $I_\mathrm{CO}$ using the fit between $I_\mathrm{CO}$ and $\sigma_\mathrm{v}$ from our models. Additionally, we find that there is little difference between $1/\left<\epsilon_\mathrm{CO}\right>$ and our model estimates of $\alpha_\mathrm{CO}$, where we divide the model column density by $I_\mathrm{CO}$ directly. This agreement is a reflection of how well the column of mass traced by CO $J=1-0$ tracks the mean $\mathrm{H_2}$ column density of our models, and further reinforces the utility of CO $J=1-0$ as a tracer of the total molecular gas content in molecular clouds in nearby galaxies. On average, $\alpha_\mathrm{CO}$ decreases with increasing $I_\mathrm{CO}$ which is a reflection of increasing CO excitation as well as variations in CO optical depth. Overall, our model estimates of $\left<\alpha_\mathrm{CO}\right>=1/\left<\epsilon_\mathrm{CO}\right>$ appear to agree well with prescriptions from numerical work \citep[e.g.,][]{Narayanan:2012} as well as recent, high-resolution studies of molecular gas in galaxies we have used as our model templates\citep[e.g.,][]{He:2024,Teng:2024}. 
\par
We see a similar decrease in $1/\left<\epsilon_\mathrm{HCN}\right>$ with increasing $I_\mathrm{HCN}$, but find that values of $1/\left<\epsilon_\mathrm{HCN}\right>$ span over $\sim2.5$ dex, while values of $1/\left<\epsilon_\mathrm{CO}\right>$ span $\sim1$ dex in our models. We fit the relationship between $\left<\alpha_\mathrm{HCN}\right>=1/\left<\epsilon_\mathrm{HCN}\right>$ and $I_\mathrm{HCN}$ using orthogonal distance regression and find
\begin{multline}\label{eq:alpha_hcn}
    \mathrm{log}\,\left(\frac{\alpha_\mathrm{HCN}}{\mathrm{M_\odot\, (K\,km\,s^{-1}\,pc^2)^{-1}}}\right) = \\ \left(-0.55\pm0.01\right) \mathrm{log}\, \left(\frac{I_\mathrm{HCN}}{\mathrm{K\,km\,s^{-1}}}\right) + 2.55\pm0.01.
\end{multline}
Again, uncertainties are determined using bootstrapping.  We also see in Fig. \ref{fig:co_conversion_factor} that the \citet{Gao:2004a,Gao:2004b} value for $\alpha_\mathrm{HCN}$ is only consistent with the brightest $I_\mathrm{HCN}$ in our models. Several recent studies of  Milky Way clouds find evidence of larger values of $\alpha_\mathrm{HCN}$ relative to the original estimate by \citet{Gao:2004a,Gao:2004b}. An estimate of $\alpha_\mathrm{HCN}=92\ \mathrm{(M_\odot\ [K\ km\ s^{-1} pc^2])^{-1}}$ in the Perseus Molecular Cloud from \citet{Dame:2023} falls within the range of $1/\left<\epsilon_\mathrm{HCN}\right>$ found in our models. They derive $\alpha_\mathrm{HCN}$ by comparing observations of HCN $J=1-0$  luminosity with gas mass estimates derived from extinction measurements of dust. \citet{Dame:2023} also note that HCN brightness has a significant effect on the value of $\alpha_\mathrm{HCN}$, and when the original \citet{Gao:2004a,Gao:2004b} value is scaled by an HCN brightness temperature more appropriate for Galactic GMCs they derive a value more consistent with their measurement from Perseus. We also compare with the results of \citet{Shima:2017} in Fig. \ref{fig:co_conversion_factor}, and find good agreement with the values they derive for Aquila, Ophiuchus, and Orion B. \citet{Tafalla:2023} also derive estimates of $\alpha_\mathrm{HCN}$ in Milky Way clouds using extinction estimates. They find $\alpha_\mathrm{HCN}=23,\ 46,
 \mathrm{and}\ 73\ M_\odot\ (\mathrm{K\ km\ s^{-1}\ pc^2)^{-1}}$ for the California, Orion A, and Perseus molecular clouds, respectively. Additionally, \citet{Forbrich:2023MNRAS.525.5565F} find evidence of deviations in $\alpha_\mathrm{HCN}$ from the original estimate of \citet{Gao:2004a,Gao:2004b}. They find $\alpha_\mathrm{HCN}\approx1\ \mathrm{(M_\odot\ [K\ km\ s^{-1} pc^2])^{-1}}$ in six GMCs in Andromeda by comparing estimates of dust with HCN emission. When assuming the Milky way dust-to-gas mass ratio, they find a much larger value of $\alpha_\mathrm{HCN}\approx109\ \mathrm{(M_\odot\ [K\ km\ s^{-1} pc^2])^{-1}}$, similar to that of \citet{Dame:2023}. We note that the \citet{Tafalla:2023} estimate for Perseus is slightly lower than that quoted by \citet{Dame:2023}, which they argue is potentially from extended HCN emission not included in the mapping area of the \citet{Dame:2023} study. However, we find the opposite effect on $\alpha_\mathrm{HCN}$ ($1/\left<\epsilon_\mathrm{HCN}\right>$) when we exclude HCN emission from lower column densities in our models (cf. Appendix \ref{ap:sens_limits}) and conclude that this discrepancy could, in part, be due to the sensitivity limit of the \citet{Tafalla:2023} study.

\par
\begin{figure}
    \centering
    \includegraphics[width=0.9\columnwidth]{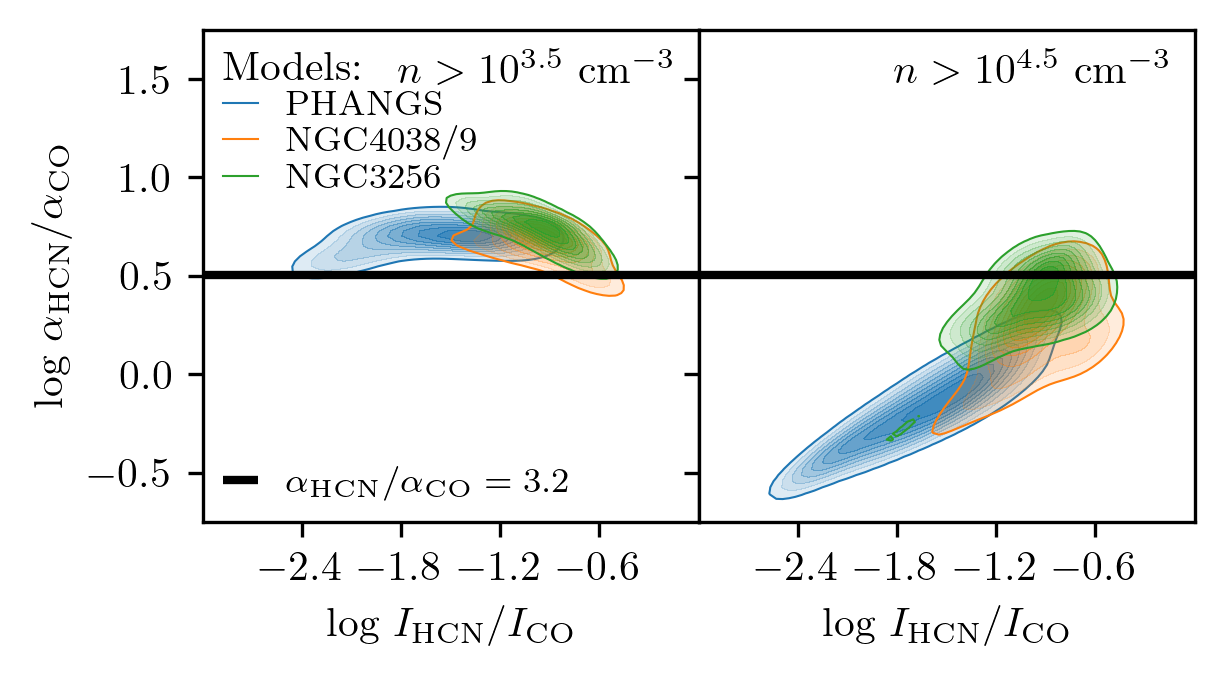}
    \caption{Ratio of the  HCN and CO conversion factors (given in Eq. \ref{eq:alpha_hcn_alpha_co}) as a function of the ratio of the HCN and CO intensities, where we consider $\alpha_\mathrm{HCN}$ as a conversion factor for the total mass above $n>10^{3.5}\ \mathrm{cm}^{-3}$ (left) and  $n>10^{4.5}\ \mathrm{cm}^{-3}$ (right). For comparison, we plot $\alpha_\mathrm{HCN}/\alpha_\mathrm{CO}=3.2$ (solid black line), which is the ratio of the \citet{Gao:2004a,Gao:2004b} conversion factor and Milky Way CO conversion factor. The formatting is the same as in Fig. \ref{fig:inten_compared_to_obs}. This figure demonstrates how emissivity is sensitive to the intensity per mass traced by a particular transition, whereas luminosity-to-mass conversion factors account for additional factors that allow us to estimate specific masses (e.g., total gas mass and dense gas mass) that may not be fully reflected in the molecular emissivity. We find that due to the subthermal excitation of HCN $J=1-0$, this transition is a poor tracer of the of the gas mass above $n>10^{4.5}\ \mathrm{cm}^{-3}$ and is a better tracer of moderate gas densities ($n>10^{3.5}\ \mathrm{cm}^{-3}$), as found in previous observational studies.} 
    \label{fig:alpha_hcn_alpha_co_frac}
\end{figure}

 
Despite the broader range in HCN emissivity relative to CO emissivity, we find that $I_\mathrm{HCN}/I_\mathrm{CO}$ closely tracks the fraction of gas above $n>10^{3.5}\ \mathrm{cm}^{-3}$, which implies a nearly constant HCN and CO luminosity-to-mass ratio, $\alpha_\mathrm{HCN}/\alpha_\mathrm{CO}$, can be used to estimate $f(n>10^{3.5}\ \mathrm{cm}^{-3})$ from observations (cf. Fig. \ref{fig:alpha_hcn_alpha_co_frac}).  Regardless of the absolute value of $\alpha_\mathrm{HCN}/\alpha_\mathrm{CO}$, the results of our modeling suggest that the fraction of gas above $n>10^{3.5}\ \mathrm{cm}^{-3}$ can be roughly estimated by applying a fixed $\alpha_\mathrm{HCN}/\alpha_\mathrm{CO}$ ratio to $I_\mathrm{HCN}/I_\mathrm{CO}$, although this ratio appears to be larger than our initially assumed value of $\alpha_\mathrm{HCN}/\alpha_\mathrm{CO}=3.2$. These results suggest that (1) the HCN intensity scales with the fraction of mass above moderate gas densities, and (2) a constant ratio between $\alpha_\mathrm{HCN}/\alpha_\mathrm{CO}$ can be assumed to derive this fraction of gas using $I_\mathrm{HCN}/I_\mathrm{CO}$. Furthermore, our models predict that $\alpha_\mathrm{HCN}$ does not scale directly with the emissivity of HCN. This difference in behaviour between $1/\left<\epsilon_\mathrm{HCN}\right>$ and $\alpha_\mathrm{HCN}$ in our models is a reflection of HCN $J=1-0$ being primarily subthermally excited. 
\par 
We reframe the results above in terms of the ratio of the HCN and CO luminosity-to-mass conversion factors, $\alpha_\mathrm{HCN}/\alpha_\mathrm{CO}$ by multiplying the ratio of $I_\mathrm{HCN}/I_\mathrm{CO}$ by the fraction of mass with densities above $n_\mathrm{thresh}>10^{3.5}\ \mathrm{cm}^{-3}$ and $n_\mathrm{thresh}>10^{4.5}\ \mathrm{cm}^{-3}$, for example 
\begin{equation}
    f(n>n_\mathrm{thresh}) = \frac{\alpha_\mathrm{HCN}} {\alpha_\mathrm{CO}} \times \frac{I_\mathrm{HCN}}{I_\mathrm{CO}}.
    \label{eq:alpha_hcn_alpha_co}
\end{equation}Thus, when the ratio of  $\alpha_\mathrm{HCN}/\alpha_\mathrm{CO}$ is multiplied with $I_\mathrm{HCN}/I_\mathrm{CO}$, we get an estimate of said gas mass fraction:
\begin{equation}
    f(n>n_\mathrm{thresh}) = \frac{M_\mathrm{H_2}(n>n_\mathrm{thresh})}{M_\mathrm{H_2,tot}}.
\end{equation}
We show these results in Fig. \ref{fig:alpha_hcn_alpha_co_frac} as a function of $I_\mathrm{HCN}/I_\mathrm{CO}$. We find that  $\alpha_\mathrm{HCN}/\alpha_\mathrm{CO}$ is relatively constant when assuming $I_\mathrm{HCN}$ tracks the mass above $n>10^{3.5}\ \mathrm{cm}^{-3}$. To derive the fraction of gas above $n>10^{3.5}\ \mathrm{cm}^{-3}$, our models predict that one can apply $\alpha_\mathrm{HCN}/\alpha_\mathrm{CO}\approx5$ to $I_\mathrm{HCN}/I_\mathrm{CO}$. Contrary to this, $\alpha_\mathrm{HCN}/\alpha_\mathrm{CO}$ must increase with $I_\mathrm{HCN}/I_\mathrm{CO}$ when assuming $I_\mathrm{HCN}$ tracks the mass above $n>10^{4.5}\ \mathrm{cm}^{-3}$. Although not shown in Fig. \ref{fig:alpha_hcn_alpha_co_frac}, this relationship is even steeper when considering $f(n>10^{5.5}\ \mathrm{cm}^{-3})$. This analysis is consistent with our findings in Fig. \ref{fig:ratio_gas_frac}, where we see that  $I_\mathrm{HCN}/I_\mathrm{CO}$ scales most directly (linearly) with the fraction of gas above $n>10^{3.5}\ \mathrm{cm}^{-3}$.
\par
These results suggest that, in theory, the fraction of dense gas above $n>10^{4.5}\ \mathrm{cm}^{-3}$ can be derived from $I_\mathrm{HCN}/I_\mathrm{CO}$ if one adopts a prescription for $\alpha_\mathrm{HCN}/\alpha_\mathrm{CO}$ that increases with $I_\mathrm{HCN}/I_\mathrm{CO}$. However, our models show that estimates of dense gas mass using the original estimate of $\alpha_\mathrm{HCN}$ from \citet{Gao:2004a,Gao:2004b} likely overestimate the true dense gas mass, except in the most extreme cases like galaxy mergers and U/LIRGs. This overestimate is more significant for disk galaxies, such as the Milky Way and galaxies in the PHANGS sample. It may be more useful to  observe other molecular line transitions that are exclusively sensitive to higher gas densities, such as $\mathrm{N_2H^+}$ \citep{Kauffmann:2017hcn_moderate,Pety:2017,Priestly:2023} to estimate $f(n>10^{4.5}\ \mathrm{cm}^{-3})$, rather than attempting to calibrate the relationship between $I_\mathrm{HCN}/I_\mathrm{CO}$ and $f(n>10^{4.5}\ \mathrm{cm}^{-3})$.
\par
Despite HCN having a higher critical density than $\mathrm{N_2H^+}$, $\mathrm{N_2H^+}$ appears to more reliably trace cool, dense gas in Milky Way molecular clouds \citep{Kauffmann:2017hcn_moderate,Pety:2017,Tafalla:2021}, whereas HCN emission tends to originate from gas at more moderate temperatures \citep{Pety:2017,Barnes:2020} and more moderate gas densities \citep{Kauffmann:2017hcn_moderate,Pety:2017}. There are several chemical processes that limit $\mathrm{N_2H^+}$ emission to regions of primarily dense, cool gas ($T<20\ \mathrm{K}$). $\mathrm{N_2H^+}$ is destroyed in the presence of CO via ion–neutral interactions \citep{Meier:2005}. Furthermore, the creation of $\mathrm{N_2H^+}$ depends on the availability of $\mathrm{H_3^+}$ to react with $\mathrm{N_2}$, which is a chemical process in competition with the creation of CO. Thus, $\mathrm{N_2H^+}$ is primarily abundant in regions where CO is depleted onto dust grains \citep{Caselli:2012}, unlike HCN which is present also at moderate densities of gas overlapping with CO \citep{Kauffmann:2017hcn_moderate,Pety:2017}. Thus, $\mathrm{N_2H^+}$ may be a better tracer of the cold, dense gas that serves as the direct fuel for star formation.
\par
Interestingly, \citet{Jimenez-Donaire:2023} find that $\mathrm{N_2H^+}$ and HCN have a nearly constant ratio over a large range of spatial scales. They compare observations of $\mathrm{N_2H^+}$ and HCN in NGC 6946 at 1 kpc scales with existing literature values of other galaxies \citep{Mauersberger:1991,Nguyen:1992,Watanabe:2014,Aladro:2015,Nishimura:2016,Takano:2019,Eibensteiner:2022} and Milky Way clouds \citep{Jones:2012,Pety:2017,Barnes:2020,Yun:2021}, and find this ratio is $\mathrm{I_\mathrm{N_2H^+}/I_\mathrm{HCN}=0.15\pm0.02}$ averaged over parsec scales and kiloparsec scales. Due to the segregation of $\mathrm{N_2H^+}$ in CO-depleted regions of molecular clouds, \citet{Jimenez-Donaire:2023} conclude that the linear scaling between HCN and $\mathrm{N_2H^+}$ must be a product of the self-similarity of clouds, and that HCN emission may still be a valuable dense gas tracer to assess the properties of the cooler, denser $\mathrm{N_2H^+}$-emitting gas. However, extragalactic observations of $\mathrm{N_2H^+}$ are so far limited to a handful of nearby galaxies, and have yet to be completed at cloud scales. Thus, it remains an important next step to perform comparable observations of $\mathrm{N_2H^+}$ and HCN over a large sample of cloud environments in nearby galaxies.

\subsection{$I_\mathrm{HCN}/I_\mathrm{CO}$ and the fraction of gravitationally bound gas} \label{sec:ratio_fgrav}

We explore how well the $I_\mathrm{HCN}/I_\mathrm{CO}$ ratio tracks gravitationally bound fraction of gas ($f_\mathrm{grav}$) as predicted by the LN+PL analytical models of star formation. We emphasize here that we are interested in general trends that are predicted by turbulent models of star formation, and the LN+PL analytical models of star formation \citep{Burkhart:2018} agree closely with those of the LN-only models \citep{Krumholz:2005,Padoan:2011}. As such, we only compare against the results of the LN+PL analytical models of star formation.
\par
\begin{figure*}
    \centering
    \includegraphics[width=0.32\textwidth]{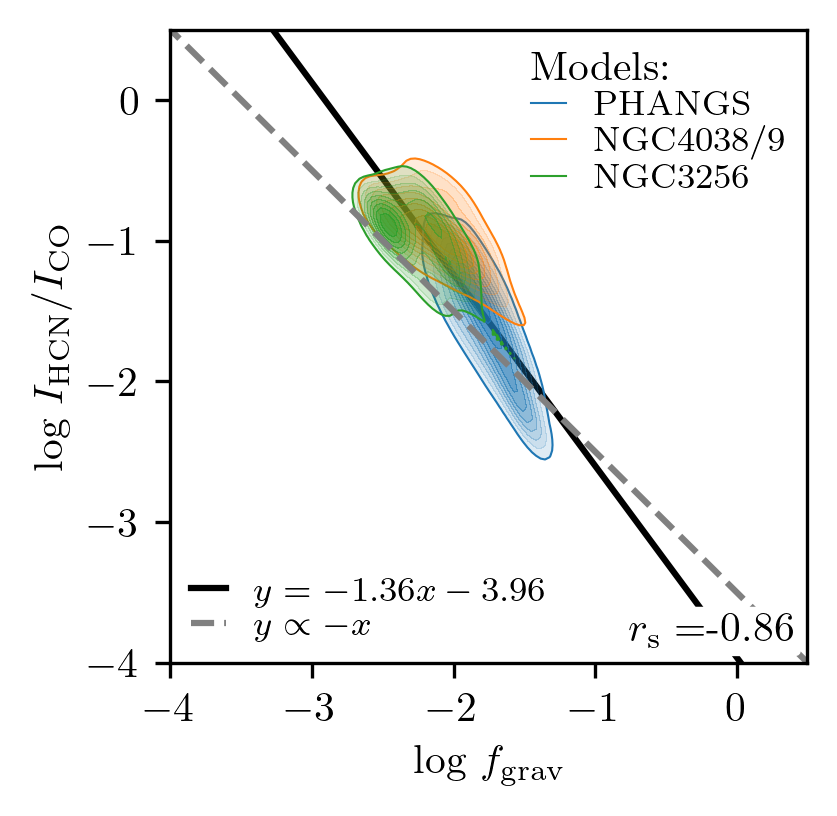}
    \includegraphics[width=0.32\textwidth]{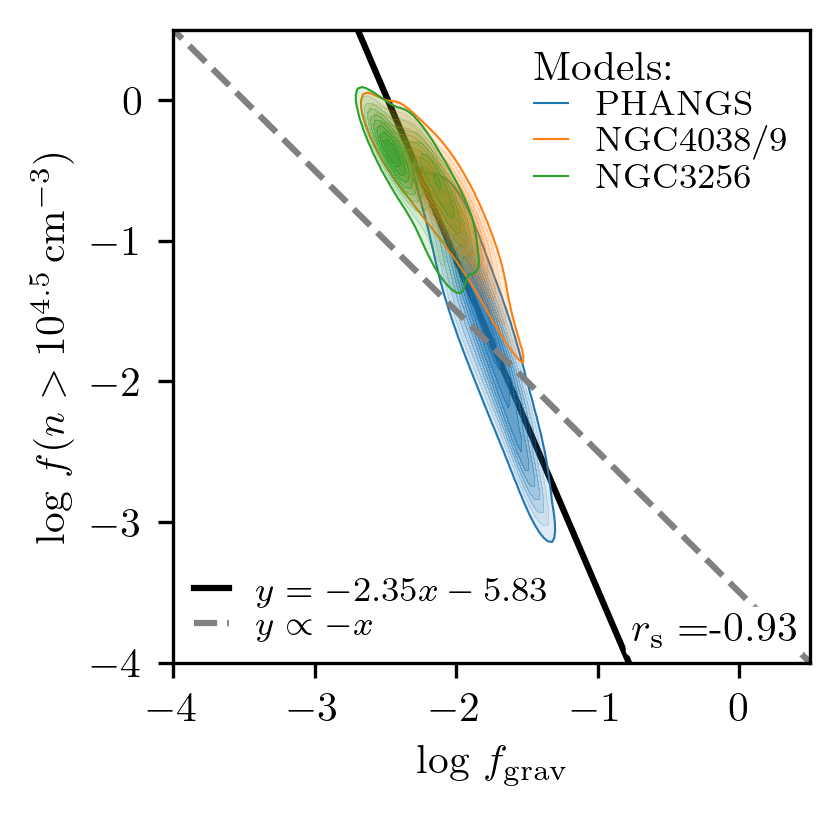}
    \includegraphics[width=0.32\textwidth]{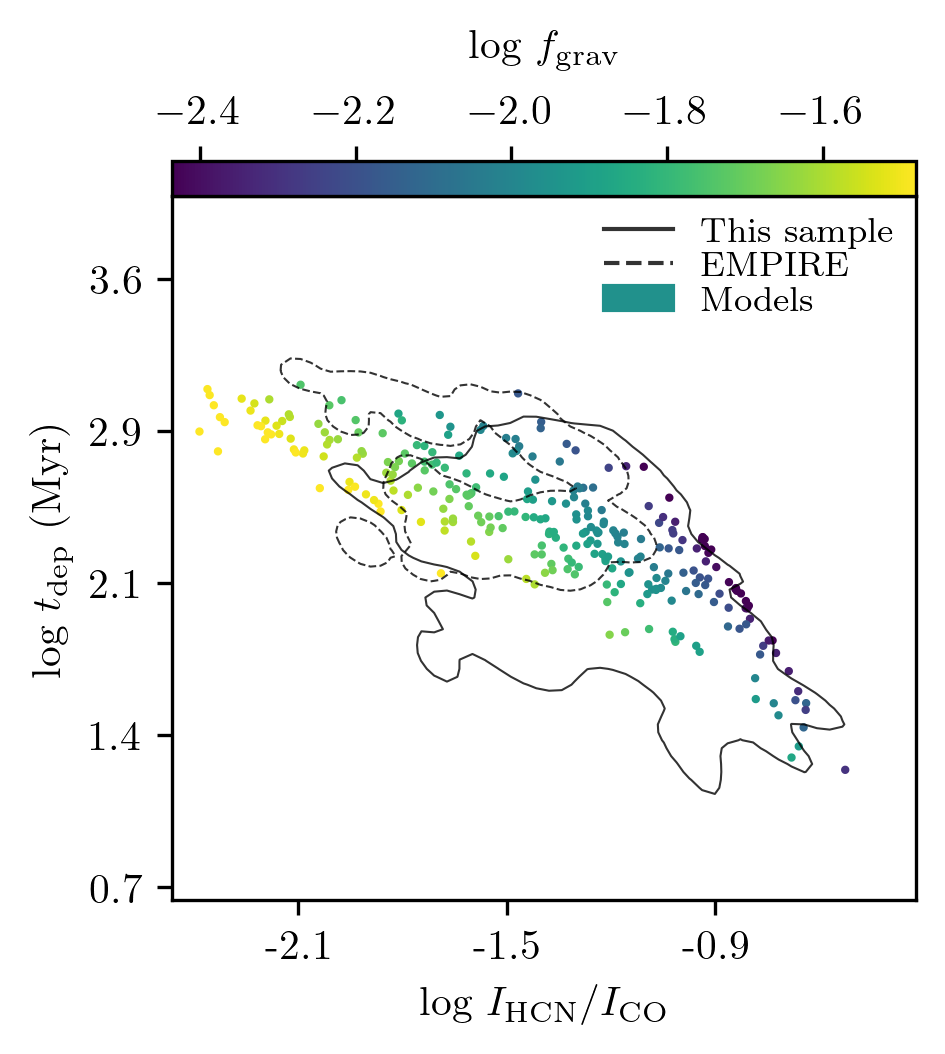}
    \caption{The relationships between modeled $I_\mathrm{HCN}/I_\mathrm{CO}$, $f(n>10^{4.5}\,\mathrm{cm}^{-3})$,  $f_\mathrm{grav}$, and $t_\mathrm{dep}$. \textit{Left:} Modeled $I_\mathrm{HCN}/I_\mathrm{CO}$ as a function of the gravitationally bound fraction of gas ($f_\mathrm{grav}$) predicted by the LN+PL analytical models of star formation. \textit{Center:} Fraction of dense gas above $n>10^{4.5}\,\mathrm{cm}^{-3}$ as a function of the gravitationally bound fraction of gas ($f_\mathrm{grav}$) predicted by the LN+PL analytical models of star formation. The formatting is the same as in Fig. \ref{fig:inten_compared_to_obs}. The Spearman rank coefficients are shown in the lower right corner of the left and center panels.
    \textit{Right:} Total gas depletion time ($t_\mathrm{dep}$) as a function of $I_\mathrm{HCN}/I_\mathrm{CO}$ ratio. The models are shown as   colored points. The measurements of $t_\mathrm{dep}$ and $I_\mathrm{HCN}/I_\mathrm{CO}$ from our sample of galaxies and the EMPIRE sample are shown as the solid black and dashed black contours, respectively. Our models find that the $I_\mathrm{HCN}/I_\mathrm{CO}$ ratio is negatively correlated with $f_\mathrm{grav}$ as predicted by  gravoturbulent models of star formation. Additionally, the fraction of dense gas above $n>10^{4.5}\,\mathrm{cm}^{-3}$ has an even steeper negative correlation with $f_\mathrm{grav}$, as predicted by gravoturbulent models of star formation \citep[e.g.,][]{Burkhart:2018,Burkhart:2019}. Thus, although $I_\mathrm{HCN}/I_\mathrm{CO}$ is sensitive to gas above moderate densities, we conclude that a single molecular line ratio, such as HCN/CO, does not necessarily scale with the fraction of directly star-forming gas in clouds. We also find that models with the lowest estimates of $f_\mathrm{grav}$ and highest $f(n>10^{4.5}\,\mathrm{cm}^{-3})$ have the shortest depletion times, and the corresponding modeled $I_\mathrm{HCN}/I_\mathrm{CO}$ and predicted depletion times are consistent with our data.}
    \label{fig:pairwise_fracs}
\end{figure*}

In Fig. \ref{fig:pairwise_fracs}, we plot the modeled $I_\mathrm{HCN}/I_\mathrm{CO}$ ratio and dense gas fraction, $f(n>10^{4.5}\,\mathrm{cm}^{-3})$ against $f_\mathrm{grav}$. We take $f_\mathrm{grav}$ to be the fraction of gas in the power-law component of the LN+PL model (see Eq. 20 in \citealt{Burkhart:2019}). We find that $I_\mathrm{HCN}/I_\mathrm{CO}$ has a strong, negative correlation with $f_\mathrm{grav}$.  This is consistent with the results of \citetalias{Bemis:2023}, where we made a similar conclusion without including radiative transfer in our analysis. $f(n>10^{4.5}\,\mathrm{cm}^{-3})$ has an even steeper, negative correlation with $f_\mathrm{grav}$. We note that the primary driver of the decrease in $f_\mathrm{grav}$ towards higher $I_\mathrm{HCN}/I_\mathrm{CO}$ and $f(n>10^{4.5}\,\mathrm{cm}^{-3})$ is a reflection of the higher gas velocity dispersion in these models (which correspond to models with higher gas surface density and wider $n-\mathrm{PDFs}$). We also find that models with the lowest estimates of $f_\mathrm{grav}$ and highest $f(n>10^{4.5}\,\mathrm{cm}^{-3})$ have the shortest depletion times, and the corresponding modeled $I_\mathrm{HCN}/I_\mathrm{CO}$ and predicted depletion times are consistent with our data (cf. Fig \ref{fig:pairwise_fracs}). In general, the models of star formation we consider predict that turbulence acts as a supportive process that prevents gravitational collapse of gas. Indeed, we find that the transition density (the density at which gas becomes self-gravitating in our models) increases across our model parameter space from $n=10^{4.5}\,\mathrm{cm}^{-3}$ in the PHANGS-type models to $n=10^{5.9}\,\mathrm{cm}^{-3}$ and $n=10^{6.6}\,\mathrm{cm}^{-3}$ in the NGC 4038/9- and NGC 3256-type models, respectively. We also note that the transition density for the PHANGS-type models agrees well with the estimation for the threshold density for star formation in the Milky Way \citep[e.g., $n\gtrsim10^{4}\,\mathrm{cm}^{-3}$,][]{Lada:2010,Lada:2012}.
\par
We also show in Fig. \ref{fig:pairwise_fracs} that models with higher  $I_\mathrm{HCN}/I_\mathrm{CO}$ and lower $f_\mathrm{grav}$ are, on average, still consistent with observations and have overall shorter total gas depletion times ($t_\mathrm{dep}$), as is also seen in observations and is in agreement with the results of \citetalias{Bemis:2023}. The above results also have important implications for the interpretation of dense gas depletion times. These results support that the longer $t_\mathrm{dep,dense}$ observed towards higher $I_\mathrm{HCN}/I_\mathrm{CO}$ in our data (assuming fixed $\alpha_\mathrm{HCN}/\alpha_\mathrm{CO}$) do not necessarily imply lower star formation efficiencies of the directly star-forming gas, but rather that a lower fraction of the dense gas is unstable to collapse in these systems (see right panel of Fig \ref{fig:pairwise_fracs}). We confirm that $f_\mathrm{grav}$ is predicted to decrease from $1.7\%$ in the PHANGS-type models to $0.8\%$ and $0.6\%$ in the NGC 4038/9- and NGC 3256-type models, respectively. In contrast to this, the fraction of dense gas above $n=10^{4.5}\,\mathrm{cm}^{-3}$ increases from $1.9\%$ in the PHANGS-type models to $22\%$ and $30\%$ in the NGC 4038/9- and NGC 3256-type models, respectively. It is also interesting to note that in the PHANGS-type models $f_\mathrm{grav}$ is well-matched to $f_\mathrm{dense}$.

\subsection{The impact of CO and HCN emissivity on star formation relations} \label{sec:sf_relations}
\par
We consider here how variations in emissivity can impact the scatter as well as the general trends of some star formation relations. One of the differences between the results shown in \citetalias{Bemis:2023} and this work is the origin of the scatter in the various star formation scaling relationships. In \citetalias{Bemis:2023}, the scatter produced in the modeled star formation scaling relationships is partially from changes in $\epsilon_\mathrm{ff}$ due to variations in PL slope for the LN+PL models or variations in  turbulence (quantified by the sonic Mach number) for the LN-only models. In this work, we also show that variations in the emissivity of CO contribute to and may even account for the majority of the scatter in observational star formation scaling relationships. 
\par
\begin{figure*}
    \centering
    \includegraphics[width=0.35\textwidth]{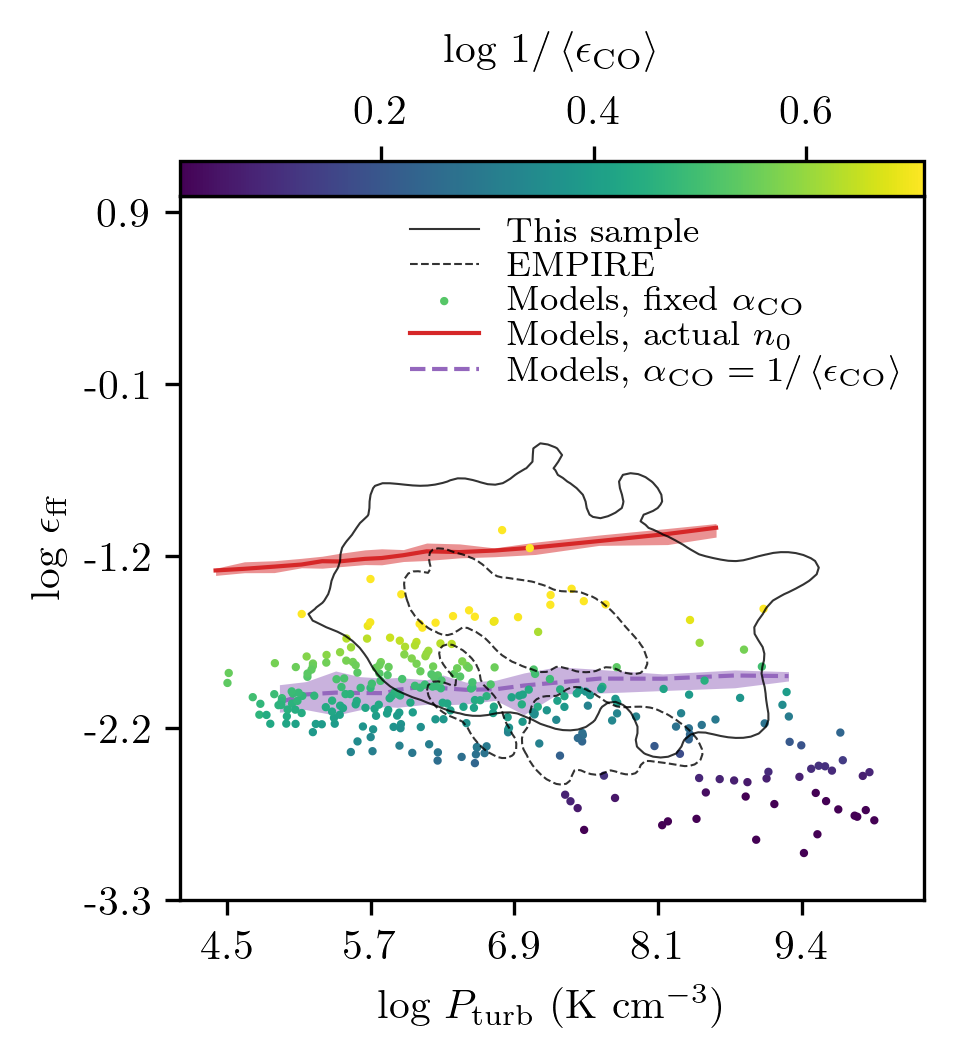}
    \includegraphics[width=0.35\textwidth]{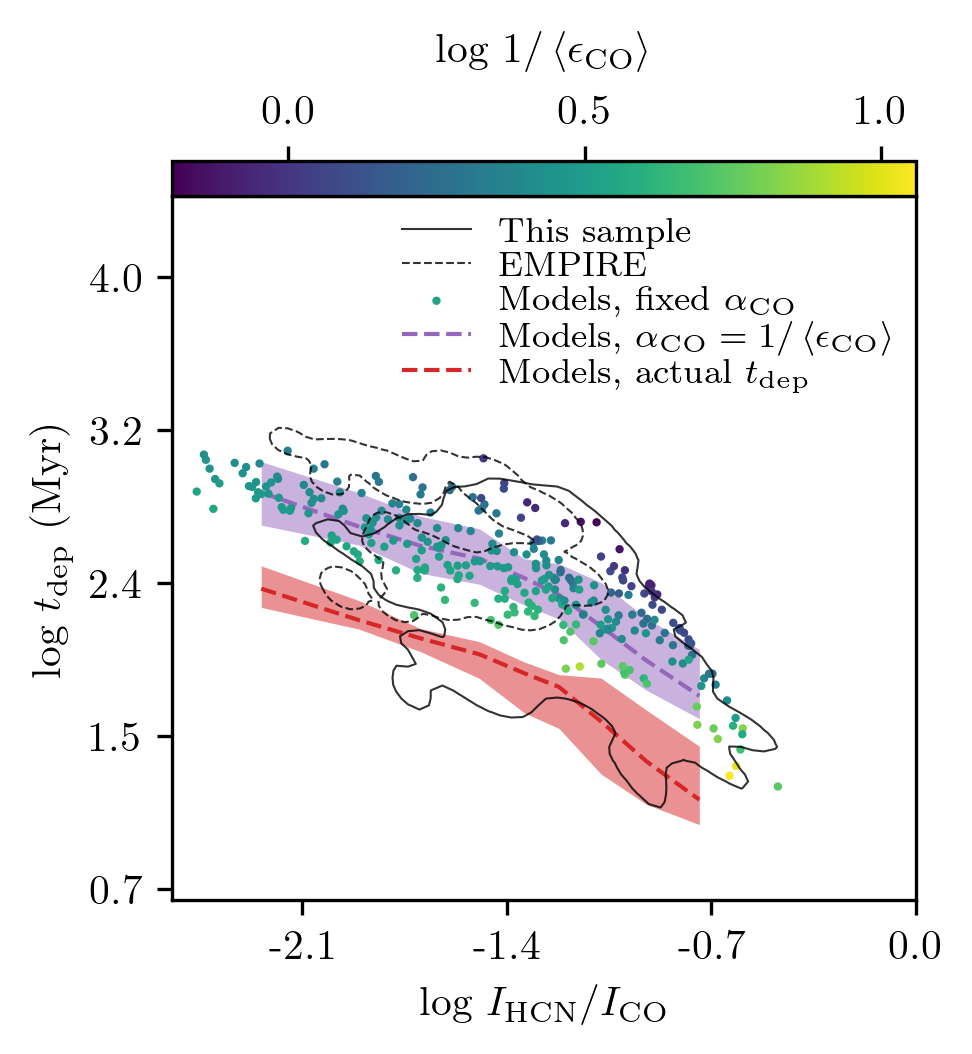}
    \caption{Comparisons between theoretical predictions and our radiative transfer modeling of the relationships between $\epsilon_\mathrm{ff}$, $P_\mathrm{turb}$, $t_\mathrm{dep}$, and $I_\mathrm{HCN}/I_\mathrm{CO}$. \textit{Left:} Model efficiency per free-fall time as a function of turbulent pressure shown three ways: (1) using modeled $I_\mathrm{CO}$ and a fixed CO conversion factor for gas surface density estimates (colored points), (2) using the actual theoretical $\epsilon_\mathrm{ff}$ and $P_\mathrm{turb}$ values (red line), and (3) using modeled $I_\mathrm{CO}$ with $\left<\alpha_\mathrm{CO}\right>=1/\left<\epsilon_\mathrm{CO}\right>$ (purple trend line). The shaded regions represent the full range of model scatter. Estimates of these quantities from our sample of galaxies and the EMPIRE sample are shown as the solid black and dashed black contours, respectively. \textit{Right:} Model depletion time as a function of HCN/CO ratio shown three ways, using the same approach. We find that the scatter produced by variations in CO emissivity can account for a significant portion of the scatter seen in observations. Additionally, the trend in $\epsilon_\mathrm{ff}$ with $P_\mathrm{turb}$ is dependent on the assumed CO conversion factor. Pixel-by-pixel estimates of $\alpha_\mathrm{CO}$ may be necessary for accurate studies of star formation trends. We note that the offset in $\epsilon_\mathrm{ff}$ and $t_\mathrm{dep}$ between methods (2) and (3) is a result of the modeled CO emission missing a fraction of the lower surface density gas in our models,  to CO-dark gas.}
    \label{fig:eff_Pturb_emiss}
\end{figure*}

For example, we show in Fig. \ref{fig:eff_Pturb_emiss} that the modeled trend in $\epsilon_\mathrm{ff}$ with $P_\mathrm{turb}$ agrees well with observations under the assumption of a fixed $\alpha_\mathrm{CO}$ and assuming mean density scales with gas surface density. We plot $\epsilon_\mathrm{ff}$ versus $P_\mathrm{turb}$ using method one described at the beginning of the results section, which is analogous to the method used to derive gas surface densities from observations. For comparison, we also plot $t_\mathrm{dep}$ as a function of the HCN/CO ratio in Fig. \ref{fig:eff_Pturb_emiss}. In these two plots the scatter in our models primarily comes from variations in CO intensity (since we have fixed PL slope). The scatter is also correlated with variations in CO emissivity. This is apparent in the gradient in $\left<\alpha_\mathrm{CO}\right>=1/\left<\epsilon_\mathrm{CO}\right>$ across the colored points in the left two panels of Fig. \ref{fig:eff_Pturb_emiss}. Models with lower CO intensity (which in general have higher $\alpha_\mathrm{CO}$ and lower CO emissivity, see Fig. \ref{fig:co_conversion_factor}) appear to have higher $\epsilon_\mathrm{ff}$ and vice versa (Fig. \ref{fig:eff_Pturb_emiss}). This agrees with the trend we observe in our data in \citet{Bemis:2023} (also shown in Fig. \ref{fig:eff_Pturb_emiss}) where we have adopted a fixed CO conversion factor and assumed mean gas density scales with gas surface density. These results show that variations in CO emissivity can account for a significant amount of scatter observed in star formation scaling relations. When we apply $\left<\alpha_\mathrm{CO}\right>=1/\left<\epsilon_\mathrm{CO}\right>$ to our modeled $I_\mathrm{CO}$ to estimate gas surface density (while still using the assumption that the mean gas volume density scales with with gas surface density), we produce tighter trends in $\epsilon_\mathrm{ff}$ with $P_\mathrm{turb}$ and $t_\mathrm{dep}$ with HCN/CO (purple lines) that are qualitatively more consistent with the actual model predictions (red lines, left two panels of Fig. \ref{fig:eff_Pturb_emiss}). The offset in $\epsilon_\mathrm{ff}$ and $t_\mathrm{dep}$ between the model prediction and what is obtained when we apply $\left<\alpha_\mathrm{CO}\right>=1/\left<\epsilon_\mathrm{CO}\right>$ to our modeled $I_\mathrm{CO}$ in Fig. \ref{fig:eff_Pturb_emiss} is a result of modeled CO emission missing a fraction of the lower surface density gas in our models, analogous to CO-dark gas \citep[cf.][]{Bolatto:2013}. When we scale $\left<\epsilon_\mathrm{CO}\right>^{-1}$ by the ratio between the true model gas surface density and $\left<N_\mathrm{CO}\right>$ we find nearly identical trends.
\par
We quantify the scatter in $\epsilon_\mathrm{ff}$ vs. $P_\mathrm{turb}$ by fitting a line to the relationship and calculating the standard deviation on the $y-$residuals. Assuming constant conversion factors, the scatter in the $\epsilon_\mathrm{ff}$ vs. $P_\mathrm{turb}$ relationship is $\sim0.36$ using method one and becomes $\sim0.05$ when we apply $\left<\alpha_\mathrm{CO}\right>=1/\left<\epsilon_\mathrm{CO}\right>$, which is similar to the scatter in the theoretical prediction ($\sim0.02$). The scatter in $t_\mathrm{dep}$ with HCN/CO is $\sim0.21$ when assuming constant conversion factors and becomes $\sim0.12$ when we apply $\left<\alpha_\mathrm{CO}\right>=1/\left<\epsilon_\mathrm{CO}\right>$. The scatter in the theoretical prediction is $\sim0.18$. We conclude that a significant portion of the scatter in these relationships originates from variations in emissivity in our models.
\par
We calculate Spearman rank coefficients ($r_\mathrm{s}$) between $\left<\alpha_\mathrm{CO}\right>=1/\left<\epsilon_\mathrm{CO}\right>$ and  $\left<\tau_\mathrm{CO}\right>$, $\left<T_\mathrm{ex,CO}\right>$, $ \sigma_\mathrm{v}$, $\Sigma_\mathrm{mol}$, $n_0$, $T_\mathrm{kin}$, and $\left<I_\mathrm{CO}\right>$ to assess the strength of the correlation between these parameters. We find that $\left<\alpha_\mathrm{CO}\right>$ only strongly ($|r_\mathrm{s}| > 0.7$) correlates with $\left<\tau_\mathrm{CO}\right>$ ($r_\mathrm{s}=0.9$) in our models. $\left<\alpha_\mathrm{CO}\right>$ is moderately ($0.5 < |r_\mathrm{s}| < 0.7$) correlated with $\sigma_\mathrm{v}$ ($r_\mathrm{s}=-0.6$) and $\left<I_\mathrm{CO}\right>$ ($r_\mathrm{s}=-0.5$). $\alpha_\mathrm{CO}$ is weakly ($|r_\mathrm{s}| < 0.5$) correlated with the remainder of the parameters ($\left<T_\mathrm{ex,CO}\right>,\, \Sigma,\, n_0,\,\mathrm{and}\, T_\mathrm{kin}$). This suggests that variations in CO emissivity are primarily driven by changes in optical depth in our models. Furthermore, the connection between $\left<\alpha_\mathrm{CO}\right>$ and $\left<\tau_\mathrm{CO}\right>$ likely stems from variations in gas velocity dispersion, since higher gas velocity dispersions are connected to lower optical depths in our models and higher CO intensities, as shown in Fig. \ref{fig:inten_tex_tau}. This also explains the variations we see in $\left<\alpha_\mathrm{CO}\right>$, for example in the scatter of  $\epsilon_\mathrm{ff}$ with $P_\mathrm{turb}$ and $t_\mathrm{dep}$ with HCN/CO, since the scatter of our model parameter space (shown in Fig. \ref{fig:model_param_space}) is set by variations in velocity dispersion. We can also conclude that variations in $\left<\alpha_\mathrm{CO}\right>$ are, to a lesser effect, driven by variations in the mean gas density that the CO emission originates from, but $\left<\alpha_\mathrm{CO}\right>$ is more strongly correlated with $T_\mathrm{kin}$ ($r_\mathrm{s}=-0.4$) relative to $n_0$ ($r_\mathrm{s}=-0.2$) in our models. We note that CO emissivity is also impacted by the width of the $n-\mathrm{PDF}$, which is set by a combination of the turbulent velocity dispersion and gas kinetic temperature in our models. Thus, inconsistencies between observationally derived quantities and model predictions, such as $\epsilon_\mathrm{ff}$, may in part be due to uncertainties in mean gas density, but are also likely driven by a number of other quantities (i.e., gas velocity dispersion and kinetic temperature) that we expect to vary consistently across trends in star formation. 
\par
Recent work on the Antennae \citep{He:2024} shows that $\alpha_\mathrm{CO}$ in this system has a negative correlation with their measurements of velocity dispersion. They make a similar argument that $\alpha_\mathrm{CO}$ may have a connection to the dynamics of the gas. \citet{He:2024} find a strong, positive correlation between $\alpha_\mathrm{CO}$ and $\tau_\mathrm{CO}$. We note that optical depth has a dependence on $\sigma_\mathrm{v}$ and gas surface density ($\tau_\mathrm{CO}\propto \sigma_\mathrm{v}/\Sigma$). Thus, these results suggest some variations in the dynamics of the gas also impact CO optical depth, which is reflected in $\alpha_\mathrm{CO}$. The importance of dynamics has also been discussed in earlier works by \citet{1987ApJ...319..730S}, and \citet{1997ApJ...478..144S,1999ApJ...512L..99G}.
\par
Work on the PHANGS galaxies at 150 pc scales shows there is a negative correlation between $\alpha_\mathrm{CO}$ and velocity dispersion which appears to have lower scatter ($\sim0.1$ dex) relative to previous prescriptions relying on gas or stellar surface density \citep{Teng:2024}. They also argue that this correlation is tied to variations in emissivity of CO. We note that \citet{Teng:2024} find a slightly steeper correlation than \citet{He:2024} (slope $-0.81$ vs. $-0.46$, respectively). When we fit this relationship in our models, we find a slope of $\sim\,-0.5$, more consistent with the Antennae relationship. When we estimate the scatter relative to our fit, we find $\sim0.16$ dex, similar that found by \citet{Teng:2024}, $0.12$ dex. In summary, our models are able to reproduce the negative correlation between $\alpha_\mathrm{CO}$ and $\sigma_\mathrm{v}$ seen in high-resolution studies of PHANGS galaxies and the Antennae and can produce similar scatter. Furthermore, the physical origin of the variations in CO emissivity in the scatter of our models can be interpreted as arising from variations in optical depth tied to the dynamics of the gas, analogous to what is observed across the Antennae and PHANGS galaxies \citep[i.e.,][]{He:2024,Teng:2024}.
\par
Additionally, we find that uncertainties in CO emissivity can lead to different slopes in star formation scaling relations that can have significantly different implications. In \citetalias{Bemis:2023}, we find a discrepancy between the predictions of gravoturbulent models of star formation and observations such that $\epsilon_\mathrm{ff}$ is predicted to increase with $P_\mathrm{turb}$ by these models, but observations instead show a decrease in $\epsilon_\mathrm{ff}$ with $P_\mathrm{turb}$. We also show this in Fig. \ref{fig:eff_Pturb_emiss}, where we plot the model-predicted $\epsilon_\mathrm{ff}$ as a function of $P_\mathrm{turb}$, using the actual mean gas volume density to estimate $t_\mathrm{ff}$ and therefore $\epsilon_\mathrm{ff}$ (as well as $P_\mathrm{turb}$). In \citetalias{Bemis:2023}, we assume this discrepancy between our data and model predictions arises from uncertainties in mean volume density and our assumption that mean volume density scales with gas surface density (see Fig. 7 in \citetalias{Bemis:2023}). In Fig. \ref{fig:eff_Pturb_emiss}, we show that all model estimates of  $t_\mathrm{dep}$ as a function of the HCN/CO ratio have a negative trend, highlighting that this effect may be most important for observational relationships where subtle trends are expected. For example, the theoretical prediction for $\epsilon_\mathrm{ff}$ as a function of $P_\mathrm{turb}$ in our models has a slope of only $\sim0.05$, and while our models (and data) show negative slopes around $\sim-0.27$. This comparison suggests that accurate pixel-by-pixel estimates of CO emissivity are required to derive much more accurate star formation scaling relations from observations. Such estimates can be derived via resolved studies of molecular line transitions with independent mass estimates, such as those derived from dust.
\par
\begin{figure*}
    \centering
\includegraphics[width=0.65\textwidth]{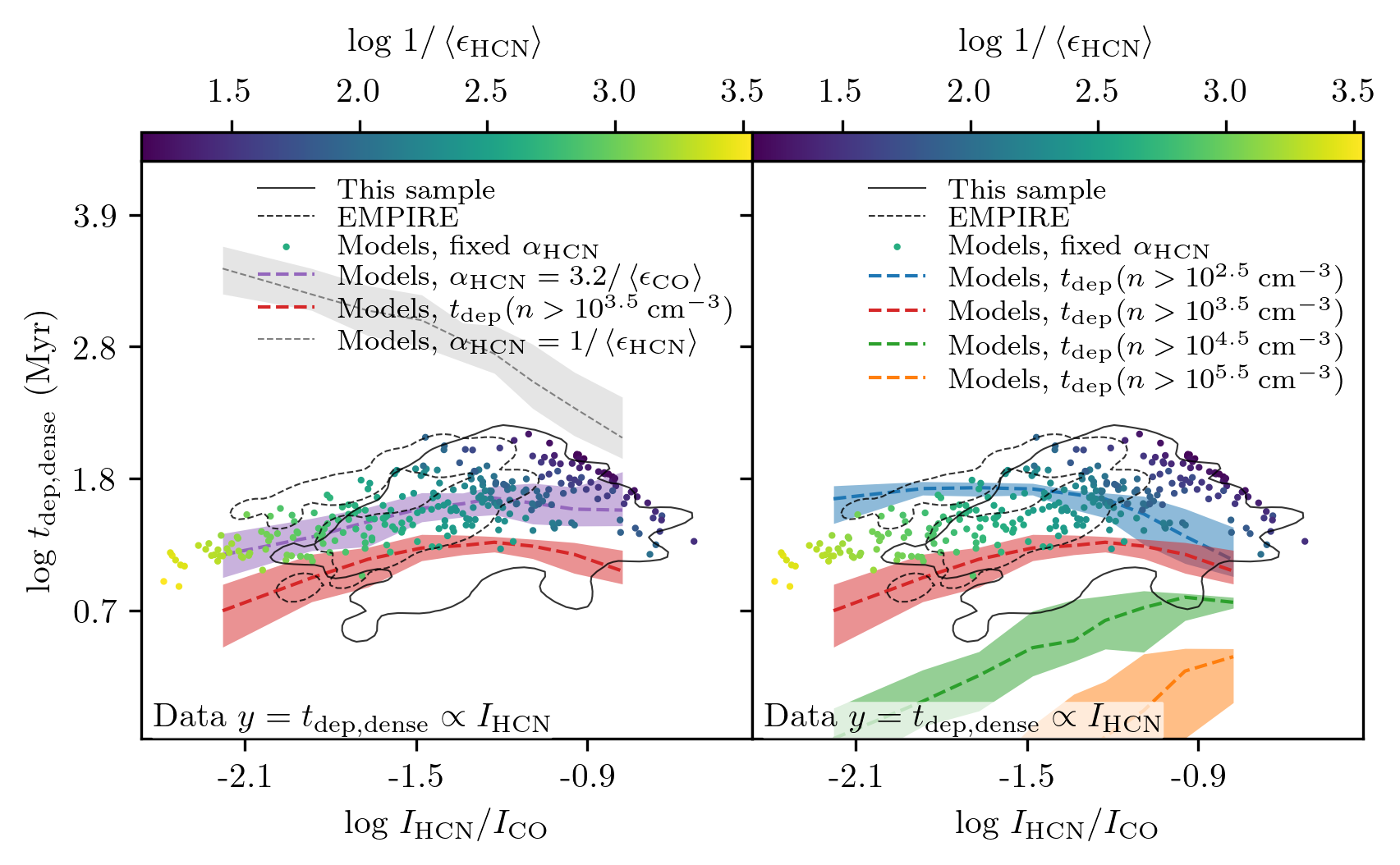}
    \caption{Comparisons between theoretical predictions and our radiative transfer modeling of the relationships between $t_\mathrm{dep, dense}$ and $I_\mathrm{HCN}/I_\mathrm{CO}$. \textit{Left:} Model dense gas depletion time as a function of HCN/CO ratio shown three ways: (1) using modeled $I_\mathrm{HCN}$ and a fixed HCN conversion factor for gas surface density estimates  (colored points), (2) using the theoretical depletion time of the fraction of gas above $n>10^{3.5}\,\mathrm{cm}^{-3}$ (red line), and (3) using modeled $I_\mathrm{HCN}$ with $\left<\alpha_\mathrm{HCN}\right>=3.2\left<\alpha_\mathrm{CO}\right>$ applied to estimate dense gas surface density (purple line). We also show the resulting trend using modeled $I_\mathrm{HCN}$ with $\left<\alpha_\mathrm{HCN}\right>=1/\left<\epsilon_\mathrm{HCN}\right>$ (gray dashed line) for comparison. \textit{Right:} Model dense gas depletion time plotted as a function of HCN/CO ratio shown three ways, this time overlaying depletion times of several fractions of gas:  that above $n>10^{2.5},\,10^{3.5},\,10^{4.5},\,10^{5.5},\,\mathrm{cm}^{-3}$ (blue, red, green, and orange lines, respectively) as a function of the  HCN/CO ratio. The shaded regions represent the full range of model scatter.  We find that the HCN emissivity does not appear to contribute to the scatter of these relationships in the way that CO contributes to those in Fig. \ref{fig:eff_Pturb_emiss}, which is likely due to the stronger dependence of HCN emissivity on excitation relative to optical depth. Additionally, the trend in dense gas depletion time with the HCN/CO ratio depends critically on the assumed HCN conversion factor. This figure also demonstrates that the HCN emissivity does not necessarily track the mass of star-forming gas (gray trend in the left panel), but that applying $\left<\alpha_\mathrm{HCN}\right>=3.2\left<\alpha_\mathrm{CO}\right>$ to $I_\mathrm{HCN}$ when estimating $t_\mathrm{dep,dense}$ does a reasonable job at reproducing the depletion time of gas above moderate gas densities (red lines, both panels; see also Figs. \ref{fig:ratio_gas_frac} and \ref{fig:co_conversion_factor}  and Sect. \ref{sec:ratio_fgrav}).}
    \label{fig:tdepdense_ratio}
\end{figure*}

 Finally, we plot $t_\mathrm{dep,dense}$ as a function of the HCN/CO ratio in Fig. \ref{fig:tdepdense_ratio} to consider how variations in HCN emissivity may impact observations of star formation relations. Interestingly, we do not see the same variation of the HCN emissivity in the scatter in $t_\mathrm{dep,dense}$ as a function of the HCN/CO ratio that we see in CO emissivity in Fig. \ref{fig:eff_Pturb_emiss}. This is likely because variations in HCN emissivity are more strongly driven by HCN excitation, and only weakly driven by optical depth in our models. When we calculate the Spearman rank coefficient between $\left<\alpha_\mathrm{HCN}\right>=1/\left<\epsilon_\mathrm{HCN}\right>$ and the same paramaters that we consider for CO, we find $\alpha_\mathrm{HCN}$ is strongly correlated ($|r|\geq0.95$) with $\left<T_\mathrm{ex,HCN}\right>,\, \sigma_\mathrm{v},\, \Sigma,\, n_0,\, T_\mathrm{kin},\, \left<I_\mathrm{HCN}\right>$ and only weakly correlated with $\left<\tau_\mathrm{HCN}\right>$ ($r_\mathrm{s}=0.21$). This further supports the idea that variations in HCN emissivity will not necessarily closely track variations in CO emissivity in observations.
\par
Following our discussion in Sect. \ref{sec:ratio_gas_fraction} (the HCN/CO ratio tracks the fraction of gas above $n\gtrsim10^{3.5}\,\mathrm{cm}^{-3}$) and \ref{sec:co_hcn_emiss} (application of a constant ratio in $\alpha_\mathrm{HCN}/\alpha_\mathrm{CO}$ may still roughly yield $f(n\gtrsim10^{3.5}\,\mathrm{cm}^{-3})$), we consider how applying $\left<\alpha_\mathrm{HCN}\right>=3.2/\left<\epsilon_\mathrm{CO}\right>$ to $I_\mathrm{HCN}$ impacts the observed trend in $t_\mathrm{dep,dense}$ as a function of the HCN/CO ratio and how that compares to  $t_\mathrm{dep}(n>10^{3.5}\,\mathrm{cm}^{-3})$. Assuming constant $\alpha_\mathrm{HCN}$, the scatter in $t_\mathrm{dep,dense}$ as a function of the HCN/CO is $\sim0.21$ and becomes $\sim0.12$ when we apply $\left<\alpha_\mathrm{HCN}\right>=3.2/\left<\epsilon_\mathrm{CO}\right>$. This scatter in the relationship between the predicted $t_\mathrm{dep}(n>10^{3.5}\,\mathrm{cm}^{-3})$ and modeled HCN/CO ratio is $\sim0.18$. When applying $\left<\alpha_\mathrm{HCN}\right>=3.2/\left<\epsilon_\mathrm{CO}\right>$, the trend in $t_\mathrm{dep,dense}$ as a function of the HCN/CO agrees well with the predicted $t_\mathrm{dep}(n>10^{3.5}\,\mathrm{cm}^{-3})$ and modeled HCN/CO ratio. To highlight this agreement, we also show a plot of $t_\mathrm{dep,dense}$ as a function of the HCN/CO ratio in Fig. \ref{fig:eff_Pturb_emiss} compared against depletion times of different fractions of gas (i.e., $n>10^{2.5},\,10^{3.5},\,10^{4.5},\,10^{5.5}\,\mathrm{cm}^{-3}$). Lower cuts in gas density produce shallow or negative relationships, while higher cuts produce steeper relationships. We also emphasize that the HCN/CO intensity ratio still appears to track a fairly constant fraction of gas, which in our models is at moderate gas densities ($n>10^{3.5}\,\mathrm{cm}^{-3}$). Thus, assuming a constant ratio in the conversion factors of HCN and CO (e.g., $\left<\alpha_\mathrm{HCN}\right>=3.2/\left<\epsilon_\mathrm{CO}\right>$) may still be useful for determining the fraction of gas above this density.
\par
We conclude that variations in HCN emissivity do not contribute significantly to the scatter of the considered star formation relationships. The scatter (e.g., in $\epsilon_\mathrm{ff}$ and $t_\mathrm{dep}$ as a function of $P_\mathrm{turb}$ and $t_\mathrm{dep,dense}$ as a function of HCN/CO ratio) primarily originates from variations in gas velocity dispersion in our models, which has a stronger effect on CO emissivity relative to HCN emissivity. This does not exclude the possibility of variations in HCN emissivity occuring in the scatter of real observations. We do expect variations in HCN emissivity in the case where variations in the physical conditions of the gas (i.e., mean gas density and kinetic temperature) impact the excitation of HCN. Due to the strong dependence of HCN emissivity on excitation, it is necessary to perform multi-line studies of HCN to assess variations of this quantity. It still remains a challenge to determine a method for assessing the fraction of star-forming gas in molecular clouds in nearby galaxies, which may ultimately require highly resolved studies of star-forming gas clouds. 

\section{Discussion and conclusions} \label{sec:conc}

In this work we explored the properties of HCN and CO emission across a range of cloud models with realistic gas density distributions, and we combined the results of this analysis with the predictions of gravoturbulent models of star formation. Our models use measurements of cloud properties based on observations of CO emission in nearby galaxies and incorporate a range of heating and cooling mechanisms to produce realistic gas temperatures \citep{Sharda:2022}. This prescription also includes the impact of radiation feedback from active star formation  via the dust-gas energy exchange, which is important for star-forming clouds \citep[cf.][]{Sharda:2022}. Our models span cloud properties found in  Milky Way-type clouds (e.g., some of the PHANGS-type models of our study) through to more extreme cloud models, based on cloud properties measured in the Antennae and NGC 3256. We also incorporated radiative transfer (RADEX, \citealt{vanderTak:2007}) in order to calculate emissivities corresponding to these cloud models. This analysis allowed us to constrain the impact of various physical properties (e.g., excitation, optical depth, mean density, velocity dispersion, temperature) on observed emission from CO and HCN across a broad range of galactic environments.  Furthermore, we evaluated the sensitivity of the HCN-to-CO ratio to different gas densities, and to the fraction of gravitationally bound star-forming gas, as predicted by analytic models of star formation (e.g., \citealt{Burkhart:2018,Burkhart:2019}, which we used for this work, and also see, e.g., \citealt{Krumholz:2005,Federrath:2012,Hennebelle:2011,Burkhart:2018}, for which the results are still broadly applicable). Below we provide an itemized summary as well as a brief discussion of the primary scientific results from this work:

\begin{enumerate}
    \item Simple models of clouds that combine realistic gas volume density distributions with radiative transfer are successful at reproducing observed $I_\mathrm{HCN}/I_\mathrm{CO}$ ratios \citep[cf. Sect. \ref{sec:model_framework} and Figs.  \ref{fig:ex_pdfs} and \ref{fig:inten_compared_to_obs}; see also][]{Leroy:2017a,Shirley:2015}. Furthermore, they are successful at reproducing CO and HCN emissivities, optical depths, and trends in excitation that have been constrained from numerical work \citep[cf. Sect. \ref{sec:excitation_optical_depth} and Figs. \ref{fig:inten_tex_tau} and \ref{fig:pair_phys}; ][]{Narayanan:2012,Narayanan:2014,Gong:2020ApJ...903..142G,Hu:2022}. Additionally, we find agreement between the trends in our model CO and HCN emissivities as a function of CO and HCN intensity and observationally derived values of the CO and HCN conversion factors in nearby galaxies and Milky Way clouds \citep[cf. Sect. \ref{sec:co_hcn_emiss} and Fig \ref{fig:co_conversion_factor}; ][]{Teng:2024,He:2024,Dame:2023,Shimajiri:2017,Tafalla:2023}.
    \item $I_\mathrm{HCN}/I_\mathrm{CO}$ is linearly correlated with the fraction of gas above moderate gas densities (e.g., $n\sim10^{3.5}$ cm$^{-3}$), and the relationship between $I_\mathrm{HCN}/I_\mathrm{CO}$ and the fraction of dense gas above $n\sim10^{4.5}$ cm$^{-3}$ is sublinear in our models (cf. Sect. \ref{sec:ratio_gas_fraction} and Fig. \ref{fig:ratio_gas_frac}). Thus, our models predict that  $I_\mathrm{HCN}/I_\mathrm{CO}$   traces the fraction of gas above a roughly constant, moderate gas density, in agreement with the results of previous studies \citep[e.g.,][]{Kauffmann:2017hcn_moderate,Pety:2017}, and this ratio is still useful in the determination of the fraction of gas above moderate densities (cf. Fig. \ref{fig:tdepdense_ratio}). One can still apply a roughly constant ratio in the HCN and CO conversion factors to $I_\mathrm{HCN}/I_\mathrm{CO}$ to estimate  $f(n\sim10^{3.5}$ cm$^{-3})$, for example. This is roughly $\alpha_\mathrm{HCN}/\alpha_\mathrm{CO}\approx5$ in our models.
    \item The modeled $I_\mathrm{HCN}/I_\mathrm{CO}$ and HCN/CO emissivity ratios are negatively correlated with $f_\mathrm{grav}$, as predicted by gravoturbulent models of star formation with varying star formation thresholds (cf. Sect. \ref{sec:ratio_fgrav} and Fig. \ref{fig:pairwise_fracs}). Thus, models with the lowest estimates of $f_\mathrm{grav}$ appear to have the highest dense gas fractions (i.e., $f(n>10^{4.5}\,\mathrm{cm}^{-3}))$ and highest $I_\mathrm{HCN}/I_\mathrm{CO}$. We find that $f_\mathrm{grav}$ is predicted to decrease from $1.7\%$ in the PHANGS-type models to $0.8\%$ and $0.6\%$ in the NGC 4038/9-type and NGC 3256-type models, respectively. In contrast to this, the fraction of dense gas above $n=10^{4.5}\,\mathrm{cm}^{-3}$ increases from $1.9\%$ in the PHANGS-type models to $22\%$ and $30\%$ in the NGC 4038/9- and NGC 3256-type models, respectively. The transition density (the density at which gas becomes self-gravitating in our models) increases across our model parameter space from $n=10^{4.5}\,\mathrm{cm}^{-3}$ in the PHANGS-type models to $n=10^{5.9}\,\mathrm{cm}^{-3}$ and $n=10^{6.6}\,\mathrm{cm}^{-3}$ in the NGC 4038/9- and NGC 3256-type models, respectively. Thus, in the PHANGS-type models $f_\mathrm{grav}$ is well matched to $f_\mathrm{dense}$, and the transition density for the PHANGS-type models agrees well with the estimation for the threshold density for star formation in the Milky Way \citep[e.g., $n\gtrsim10^{4}\,\mathrm{cm}^{-3}$, cf. Sect. \ref{sec:ratio_fgrav}, ][]{Lada:2010,Lada:2012}.
    \item Models with the lowest estimates of $f_\mathrm{grav}$ (highest  $f(n>10^{4.5}\,\mathrm{cm}^{-3})$ and highest $I_\mathrm{HCN}/I_\mathrm{CO}$)  appear to have the shortest gas depletion times (i.e., the NGC 4038/9- and NGC 3256-type models). Thus, lower $f_\mathrm{grav}$ does not necessarily mean longer depletion times in the case where sufficient mass is available to star formation. We find that the trend in the modeled gas  depletion times and $I_\mathrm{HCN}/I_\mathrm{CO}$ are consistent with the trend observed in our data (cf. Sect. \ref{sec:ratio_fgrav} and Fig. \ref{fig:pairwise_fracs}).
    
    \item The scatter observed in star formation trends, such as $\epsilon_\mathrm{ff}$ and $t_\mathrm{dep}$ as a function of $P_\mathrm{turb}$ and HCN/CO ratio, can largely be attributed to variations in CO emissivity. We find that the scatter in these relationships is reduced by a factor of $\sim2-3$ when we apply modeled CO emissivity to $I_\mathrm{CO}$ to estimate gas surface density (relative to the assumption of a fixed CO conversion factor). We find variations in CO emissivity are primarily driven by variations in the optical depth of CO due to the dynamics of the gas (cf. Sect. \ref{sec:sf_relations} and Fig. \ref{fig:eff_Pturb_emiss}). We do not see the same variations in HCN emissivity in the scatter of $t_\mathrm{dep,dense}$ as a function of HCN/CO ratio, and find that HCN emissivity is more strongly correlated with excitation. Thus, variations in HCN and CO emissivity have different physical origins according to our models (cf. Figs. \ref{fig:pairwise_fracs} and \ref{fig:tdepdense_ratio}).
    \item The assumption of constant conversion factors can alter the slope of star formation trends, which is particularly important for trends with subtle slopes. For example, we find the assumption of a constant CO conversion factor can produce a negative trend in $\epsilon_\mathrm{ff}$ with turbulent pressure (slope $\sim-0.27$) in our models that also agrees with the negative trend we find in our sample in \citetalias{Bemis:2023}. Our models predict that the actual trend in $\epsilon_\mathrm{ff}$ with turbulent pressure is marginally positive ($\sim0.05$, cf. Fig. \ref{fig:eff_Pturb_emiss}).
    
\end{enumerate}

A key prediction of our models is that, on average,  $I_\mathrm{HCN}/I_\mathrm{CO}$ does appear to track gas above a relatively fixed density. However, this fraction includes more moderately dense gas (i.e., $n>10^{3.5}\ \mathrm{cm}$) as opposed to strictly dense gas above $n>10^{4.5}\ \mathrm{cm}$. This conclusion generally agrees with more recent studies of HCN emission in Milky Way clouds that find HCN emission includes more moderate gas densities \citep[e.g.,][]{Kauffmann:2017hcn_moderate,Pety:2017}.  We find evidence that $I_\mathrm{HCN}/I_\mathrm{CO}$ tracks a gas fraction including more moderate gas densities even in the more extreme environments. This analysis implies that previous estimates of dense gas fractions likely overestimate the true fraction of gas above $n>10^{4.5}$.
\par
Furthermore, we show that the fraction of gravitationally bound gas, as predicted by turbulent models of star formation (i.e., \citealt{Burkhart:2018,Burkhart:2019}, which we use in this work, and also see, e.g., \citealt{Krumholz:2005,Federrath:2012,Hennebelle:2011,Burkhart:2018}), decreases with $I_\mathrm{HCN}/I_\mathrm{CO}$. This result agrees with \citetalias{Bemis:2023}, and combined with the subthermal excitation of HCN, suggests that it may not be appropriate to interpret $I_\mathrm{HCN}/I_\mathrm{CO}$ as a straightforward tracer of the dense gas associated with ongoing star formation in galaxies.  While $I_\mathrm{HCN}/I_\mathrm{CO}$ does scale with the fraction of moderate to high density gas, this is not necessarily equivalent to the fraction of gas contributing to star formation, especially in more extreme environments.
\par
A critical observational uncertainty in the study of gas traced by HCN in extragalactic systems is the lack of observational constraints on the HCN conversion factor. Most observational prescriptions assume that HCN is optically thick, but as we show with our modeling, HCN appears only moderately thick, and these findings are consistent with the study of HCN and H$^{13}$CN in the centers of nearby galaxies \citep{Jimenez_Donaire:2017}. Additionally, we also show that HCN is primarily subthermally excited, which also agrees with the recent findings of HCN emission in Perseus by \citet{Dame:2023}. Despite these complications, it may still be possible to use HCN as a tracer of dense gas. Recent work shows that on galactic scales, the ratio of HCN to N$_2$H+ is nearly constant \citep{Jimenez-Donaire:2023}. Since N$_2$H$^+$ has been shown to be a tracer of even denser gas than that traced by HCN \citep{Kauffmann:2017hcn_moderate,Pety:2017,Priestly:2023}, it may indicate that it is still possible to calibrate a conversion between HCN luminosity and total dense gas mass in molecular clouds.
\par
One potential limitation of the use of an HCN conversion factor is if the fraction of dense gas mass does not increase linearly with the total mass of molecular clouds. Our models show that the ratio of the CO to HCN conversion factors, $\alpha_\mathrm{HCN}/\alpha_\mathrm{CO}$, would need to increase with $I_\mathrm{HCN}/I_\mathrm{CO}$ for $I_\mathrm{HCN}$ to accurately trace the fraction of gas $n>10^{4.5}\ \mathrm{cm}^{-3}$ predicted by an $n-\mathrm{PDF}$ with both a lognormal and  power-law component. This result needs to be confirmed through more resolved studies of molecular clouds, particularly in the Milky Way. It is crucial to map the density structure down to small scales in clouds and directly compare this with mappings of multiple molecular line transitions over a range of cloud types \citep[e.g.,][]{Dame:2023, Tafalla:2023, Shimajiri:2017}. Including an analysis of the distribution of gas densities can also shed light on the physics of molecular clouds, and how much dense star-forming gas there is in relation to various molecular line emissivities. Additionally, multi-line studies targeting higher-J transitions are necessary to constrain the mean volume density and gas temperature traced by a particular molecular species, which are also important for determining total gas masses.

\begin{acknowledgements}
We thank the anonymous referee for their comments and feedback on the manuscript which improved this work. Part of this work was supported by an Ontario Trillium Scholarship. The research of
CDW is supported by grants from the Natural Sciences and
Engineering Research Council of Canada and the Canada
Research Chairs program. PS is supported by the Leiden University Oort Fellowship and the International Astronomical Union -- Gruber Foundation Fellowship. IDR is supported by the Banting Fellowship program.  This paper makes use of the following
ALMA data: ADS/JAO.ALMA\#2011.0.00467.S, ADS/JAO.ALMA\#2011.0.00525.S, ADS/JAO.ALMA\#2011.0.00772.S,
ADS/JAO.ALMA\#2012.1.00165.S, ADS/JAO.ALMA\#2012.1.00185.S, ADS/JAO.ALMA\#2012.1.01004.S, ADS/JAO.ALMA\#2013.1.00218.S, ADS/JAO.ALMA\#2013.1.00247.S,
ADS/JAO.ALMA\#2013.1.00634.S, ADS/JAO.ALMA\#2013.1.00885.S, ADS/JAO.ALMA\#2013.1.00911.S, ADS/JAO.ALMA\#2013.1.01057.S, ADS/JAO.ALMA\#2015.1.00993.S,
ADS/JAO.ALMA\#2015.1.01177.S, ADS/JAO.ALMA\#2015.1.01286.S, ADS/JAO.ALMA\#2015.1.01538.S. ALMA is a partnership of ESO (representing its member states), NSF
(USA) and NINS (Japan), together with NRC (Canada), MOST
and ASIAA (Taiwan), and KASI (Republic of Korea), in
cooperation with the Republic of Chile. The Joint ALMA
Observatory is operated by ESO, AUI/NRAO and NAOJ. This work made use of the following software: \textsc{RADEX} \citep{vanderTak:2007}, \textsc{ASTROPY} \citep{astropy:2013,Astropy:2018,astropy:2022}, \textsc{PANDAS} \citep{mckinney-proc-scipy-2010, reback2020pandas}, \textsc{MATPLOTLIB} \citep{Hunter:2007}, \textsc{NUMPY} \citep{harris:2020}, and \textsc{SCIPY} \citep{virtanen:2020}.
\end{acknowledgements}

\bibliographystyle{aa}
\bibliography{refs}

\begin{appendix}

\section{Velocity dispersion estimates from molecular lines as a measure of mach number}
\label{ap:large_scale_motions}

In this section we discuss how velocity dispersion estimates from measured line emission may be connected to mach number in gas clouds, as this is an assumption of our models.

\subsection{Observational evidence for the $\sigma_{n/n_0}^2$ -- $\mathcal{M}$ relation}
{Comparisons between independent measurements of $\sigma_{n/n_0}^2$ and $\mathcal{M}$ in resolved studies of clouds provide crucial tests to these theories.  Studies focusing on clouds in the Solar neighborhood only find weak observational evidence of a scaling between the 2D gas density variance and velocity dispersion derived from molecular transitions \citep[e.g.,][]{Kainulainen:2009,KainulainenTan:2013}, which may in part be due the uncertainty in confounding factors, such as $b$ \citep[i.e.,][]{Kainulainen:2017}. Alternatively, this may be due to lack of dynamic range in $\mathcal{M}$ in Solar neighborhood clouds; a stronger correlation is seen when including a range of galactic environments (e.g., of HI clouds) in the Milky Way \citep{Gerrard:2024MNRAS.530.4317G} and SMC \citep{Gerrard:2023MNRAS.526..982G} in addition to those of molecular clouds in the Milky Way and LMC \citep[e.g.,][]{Padoan:1997, Brunt:2010A&A...513A..67B, Ginsburg:2013, Federrath:2016ApJ...832..143F, Menon:2021, Marchal:2021ApJ...908..186M, Sharda:2022}, although this is still limited to small number statistics. Additionally, many of studies assessing the properties of gas density PDFs use different methodologies and observational tracers \citep[cf.][and references therein]{Schneider:2022A&A...666A.165S}, as well as different atomic or molecular transitions to assess the kinematics of gas. Although there is still clearly much to understand about these relationships, there is strong theoretical support of a connection between gas density variance and mach number, and emerging observational support for this relationship from observations. Furthermore, there is numerical evidence that the CO molecular linewidth tracks the turbulent velocity dispersion \citep[e.g.,][]{Szucs:2016MNRAS.460...82S}, thus providing support for the use of molecular transitions as probes of the initial velocity field of a molecular cloud.}

\subsection{Estimates of turbulent gas velocity dispersion}

{There are multiple possible contributions to velocity dispersion measured from line emission in molecular clouds. We summarize the various contributions as follows:
\begin{equation}
    \sigma^2_\mathrm{v,obs} \approx \sigma^2_\mathrm{v,inst} + \sigma^2_\mathrm{v,T} + \sigma^2_\mathrm{v,NT} + \sigma^2_\mathrm{v,ls} 
\end{equation}
where $\sigma_{v,\mathrm{obs}}$ is the total measured dispersion, $\sigma_\mathrm{v,T} = c_\mathrm{s}$ is the thermal contribution to the velocity dispersion, $\sigma_\mathrm{v,ls}$ is the background contribution (from  large-scale motions due to shear, streaming, or rotation), $\sigma_\mathrm{v,inst}$ is the instrumental contribution from limited observational velocity resolution, and $\sigma_\mathrm{v,NT}$ is the nonthermal contribution.}
\subsubsection{$\sigma^2_\mathrm{v,inst}$}
{The instrumental contribution is easily accounted for, and is subtracted in quadrature from the measured velocity dipsersion, $\sigma^2_\mathrm{{v,corr}} = \sigma^2_\mathrm{v,obs} - \sigma^2_\mathrm{v,inst}$, where  $\sigma_\mathrm{{v,corr}}$ is the corrected velocity dispersion, $\sigma_\mathrm{v,inst} = \Delta v / 2\pi$, and $\Delta v$ is the  velocity channel width of the data \citep[cf.][]{Rosolowsky:2006}. Velocity dispersion measurements from the previous studies we refer to have been corrected for the instrumental contribution \citep{Sun:2020,Brunetti:2021,Brunetti:2024}. }
\subsubsection{$\sigma^2_\mathrm{v,T}$}
{Typical molecular gas kinetic temperatures of clouds in the Milky Way disk range from $T_\mathrm{kin}\approx10-20\,\mathrm{K}$ \citep[cf.][]{Heyer:2015}, resulting in thermal sound speeds of $c_\mathrm{s} \approx 0.2-0.3\,\mathrm{km\,s}^{-1}$. Higher kinetic temperatures are estimated for some clouds in the CMZ, possibly due to the enhanced turbulent heating \citep[e.g., $T_\mathrm{kin}\approx55-125\,\mathrm{K}$ from $\mathrm{H_2CO}$,][]{Ao:2013}, giving rise to higher sounds speeds of $c_\mathrm{s} \approx 0.4-0.7\,\mathrm{km\,s}^{-1}$. Additionally, \citet{Friesen:2017ApJ...843...63F} find that gas kinetic temperature derived from ammonia ($\mathrm{NH_3}$) increases with star formation activity in Milky Way clouds. Thus, molecular gas temperature in clouds depends on both galaxy environment and star formation evolutionary stage.}
\par
{Typical temperatures of clouds in mergers can range from those typical of disk galaxies (e.g., $T_\mathrm{kin}\approx10-20\,\mathrm{K}$ in Arp 55, an intermediate stage merger) to temperatures similar to those in the CMZ \citep[e.g., $T_\mathrm{kin}\approx110\,\mathrm{K}$ in NGC 2623, a merger remnant,][]{Sliwa:2017ApJ...840....8S}. We can therefore also expect a range of average molecular gas kinetic temperatures across galaxy types. In our models, we our estimates of $\left<T_\mathrm{kin}\right>$ (described in Sect. \ref{sec:param_space}) range from $\left<T_\mathrm{kin}\right>=10\,\mathrm{K}$ to $\left<T_\mathrm{kin}\right>=65\,\mathrm{K}$, and find sound speeds ranging from a typical $c_\mathrm{s}\approx0.2\, \mathrm{km/s}$ in the PHANGS-type galaxy cloud models to $c_\mathrm{s}\approx0.3\, \mathrm{km/s}$ and $c_\mathrm{s}\approx0.4\, \mathrm{km/s}$ in the NGC 4038/9- and NGC 3256-type galaxy cloud models. We note that thermal contributions are not subtracted from the velocity dispersion measurements we use \citep{Sun:2020,Brunetti:2021,Brunetti:2024}, but this contribution will be small relative to the nonthermal contributions as we discuss below. } 
\subsubsection{$\sigma^2_\mathrm{v,NT}$}
{The relative thermal and nonthermal contributions to velocity dispersion are still uncertain in molecular clouds.  Constraints on the ratio between $\sigma^2_\mathrm{v,T}$ and  $\sigma^2_\mathrm{v,NT}$ in the literature arise largely from Milky Way studies of ammonia, $\mathrm{NH_3}$, \citep[e.g.,][]{Myers:1991ApJ...372L..95M, 
Pineda:2010ApJ...712L.116P,Chen:2019ApJ...886..119C,Choudhury:2021A&A...648A.114C,Friesen:2024ApJ...969...70F}, a known tracer of molecular gas kinetic temperature \citep{Ho:1983ARA&A..21..239H}, with some studies including other molecular line transitions (e.g., $\mathrm{C_3H_2}$ \citealt{Myers:1991ApJ...372L..95M}, $\mathrm{CCS}$ \citealt{Foster:2009ApJ...696..298F}, $\mathrm{N_2H^+}$ \citealt{Sokolov:2019ApJ...872...30S, Yue:2021RAA....21...24Y}). These molecular lines are primarily sensitive to gas denser than $n>10^3\,\mathrm{cm}^{-3}$. Thus, these Milky Way studies tend to focus on dense clumps and cores on $\sim$subparsec scales, as opposed to the bulk of molecular gas in clouds ($n\gtrsim10\,\mathrm{cm}^{-3}$) existing on tens of parsecs. In a study of cores in Perseus ($\sim0.1$ pc) in ammonia and $\mathrm{CCS}$, \citet{Foster:2009ApJ...696..298F} find that a typical ratio of nonthermal to thermal linewidths of $\sim0.6$ in protostellar and starless cores, with a range from $\sim0.2$ to $\sim1$. In a study of ammonia and $\mathrm{N_2H^+}$ in the Orion Molecular Cloud 2 and 3, \citet{Yue:2021RAA....21...24Y} find that there is a transition from supersonic to transonic turbulence at $\sim$0.05 pc, and a transition from transonic to subsonic turbulence between $\sim$0.05 pc and $\sim$0.006 pc. Thus, at small scales we expect the thermal linewidth to be comparable to the nonthermal linewidth. In this work, we are primarily concerned with the largest scales relevant to molecular clouds.}
\par
{Power-law relationships between the size, measured velocity dispersion, and gas surface density (i.e., Larson's relations) of whole molecular clouds are well-established in the Milky Way and nearby galaxies \citep[cf.][]{Larson:1981,Heyer:2015,Miville-Desch:2017,Rosolowsky:2021,Schinnerer:2024ARA&A..62..369S}. Studies of of the velocity field within clouds also find a power-law scaling, with larger, supersonic velocity dispersions associated with larger ($\gtrsim$parsec) scales \citep[e.g.,][]{Choudhury:2021A&A...648A.114C, Yue:2021RAA....21...24Y}. At the largest scales of molecular clouds (corresponding to densities $n\sim10\, \mathrm{cm}^{-3}$), gas temperature only weakly varies with gas density \citep[cf.][]{Glover:2007ApJS..169..239G,Glover:2007ApJ...659.1317G}, suggesting that molecular gas temperatures will not change significantly at the scales associated with larger velocity dispersions measured by  CO, for example. Thus, at small scales we expect the thermal contribution to the velocity dispersion to be comparable to nonthermal component, and for the relative contribution of the nonthermal component to increase towards larger scales in the turbulent envelopes of clouds. We note that the interpretation of the nature of large linewidths are still debated \citep{Ballesteros-Paredes:2011a}, but it is clear that the nonthermal component of molecular linewidth increases towards larger scales in clouds.} 
\subsubsection{$\sigma^2_\mathrm{v,ls}$}
We also consider how large-scale motions of gas, such as galactic shear or cloud rotation, might contribute to velocity dispersion measurements from CO. The impact of shear on a molecular cloud in a normal disk galaxy can be estimated using the Oort Constant $A$ \citep{Fleck:1981,Utomo:2015}, and depends on the radius at which the molecular cloud resides ($R_0$), as well as the rotational speed of the galaxy at that radius ($V_0$):
\begin{equation}\label{eq:shear}
    \Omega_\mathrm{shear} = \left| \frac{\Delta v}{\Delta r}\right| = \left| 2A \right| = \left| \frac{V_0}{R_0} - \left( \frac{dV}{dR}\right)_0\right| 
\end{equation}
where $V(R)$ is the rotation curve of the galaxy. Clouds that experience the most significant shear in a disk galaxy are therefore those closest to the galaxy center. We use the analytical fits to the rotation curves of the PHANGS galaxies from \citet{Lang:2020} to estimate shear as a function of galaxy radius. We assume the internal rotational velocity of a cloud is equivalent to the shear it experiences from Eq. \ref{eq:shear} ($\Omega = \Omega_\mathrm{shear}$), and estimate the ratio of cloud rotational energy to turbulent energy in the PHANGS galaxies using \citep{Goodman:1993,Utomo:2015}:
\begin{equation}
    \gamma_\mathrm{rot} = \frac{p\Omega^2R^2}{3\sigma_\mathrm{v}^2}
\end{equation}
where $R=45\, \mathrm{pc}$ and $p=2/5$ for a uniform sphere. Using this estimation, we find less than 0.02\% of the clouds measured in the PHANGS sample (also with analytical velocity curves from \citealt{Lang:2020}) have $\gamma_\mathrm{rot}>1$. Therefore, velocity dispersion measurements of the PHANGS galaxies are  turbulent velocity motions. We also note that PHANGS galaxies are selected to have low inclination, which reduces blending of molecular line emission along our line of sight and thus optimizes estimates of cloud properties such as turbulent velocity dispersion.
\par
Shear estimates are more difficult to obtain in more disturbed galaxies, such as mergers, as gas motions are noncircular and potentially driven more by streaming motions \citep[e.g.,][]{Bournaud:2008MNRAS.389L...8B,Barnes:1996ApJ...471..115B}. \citet{Brunetti:2022} assess whether clouds around the northern nucleus of NGC 3256 show evidence of alignment, which may be an indication of cloud shear. However, they find no clear evidence of cloud alignment. Thus, it is possible that shear does not play a significant role in the dynamics of molecular clouds in this merger. An analysis of this kind has not been performed on clouds in NGC 4038/9.
\par 
For comparison, gas within the bars of barred galaxies do experience more shear and shocks as a result of streaming motions \citep[e.g.,][]{Kim:2024ApJ...968...87K}, but will also have lower associated $b$ values. For example, simulations predict $b\approx0.22$ for clouds in the CMZ, which also appear to have higher total and turbulent velocity dispersions relative to the disk \citep{Federrath:2016ApJ...832..143F}. As we mention in Sect. \ref{sec:cloud_models}, due to the overall uncertainty in $b$ we assume $b=0.4$ which represents stochastic mixing of turbulent modes (divergence-free and curl-free, or solenoidal and compressive, respectively). Ultimately, contributions to $\sigma_\mathrm{v}$ from large scale motions as well as the uncertainty in $b$ may impact our estimates of the variance of the $n-\mathrm{PDF}$ (Eq. \ref{eq:pdf_nstd}). It is therefore possible that this will contribute some scatter to the intensities and emissivities of our models, but ultimately these uncertainties will not change the overall trends we explore in this paper.

\section{Variations in molecular abundance}
\label{ap:mol_abundance}

\begin{figure*}[tbh]
    \centering
    \includegraphics[width=0.24\textwidth]{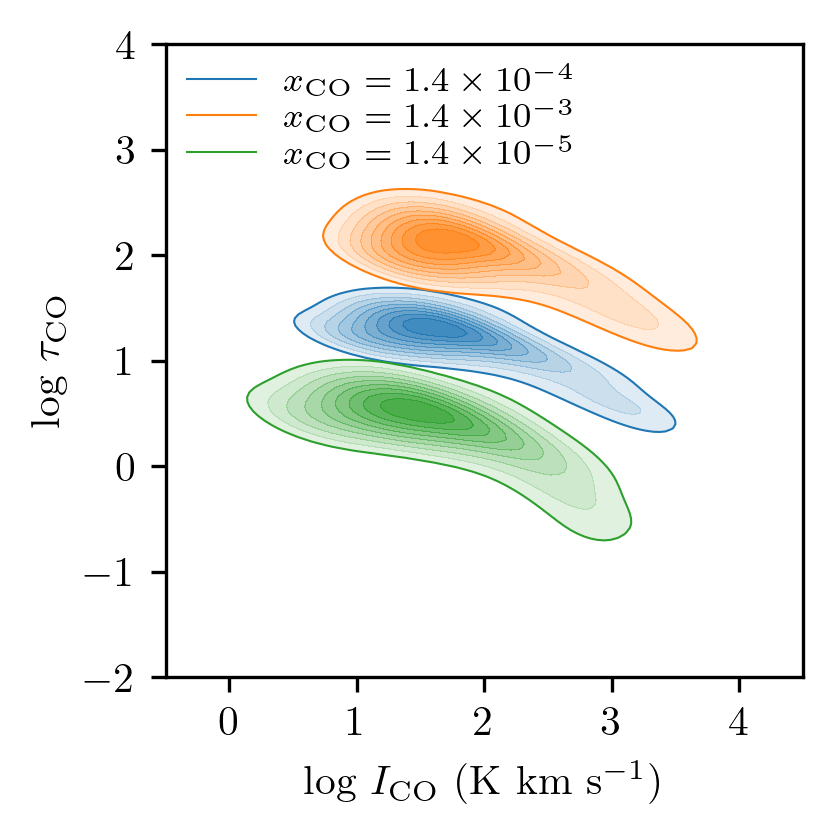}
    \includegraphics[width=0.24\textwidth]{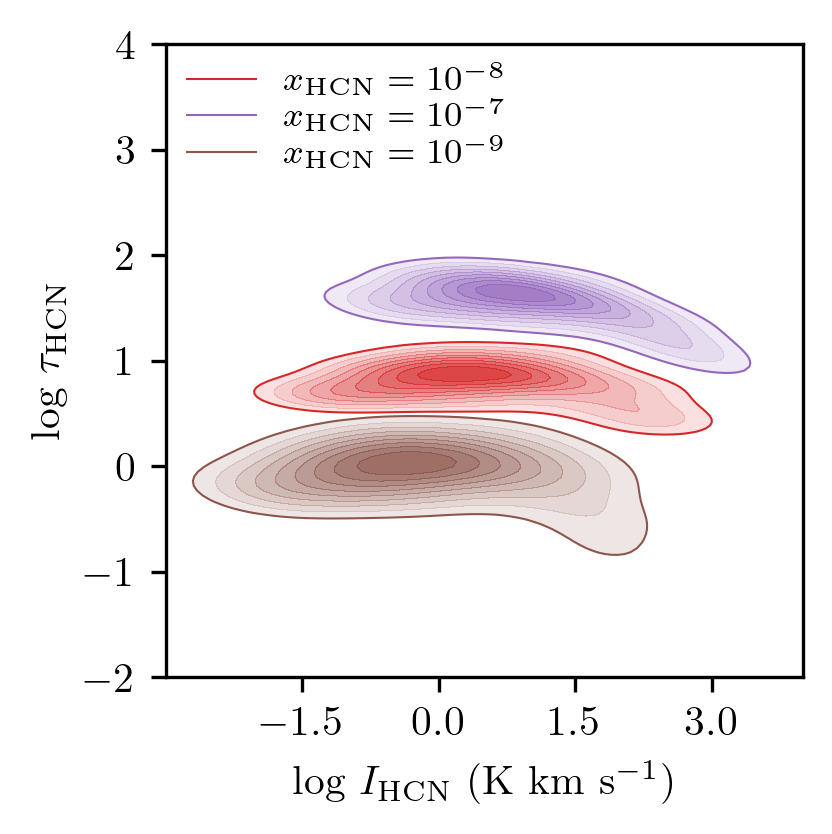}
    \includegraphics[width=0.24\textwidth]{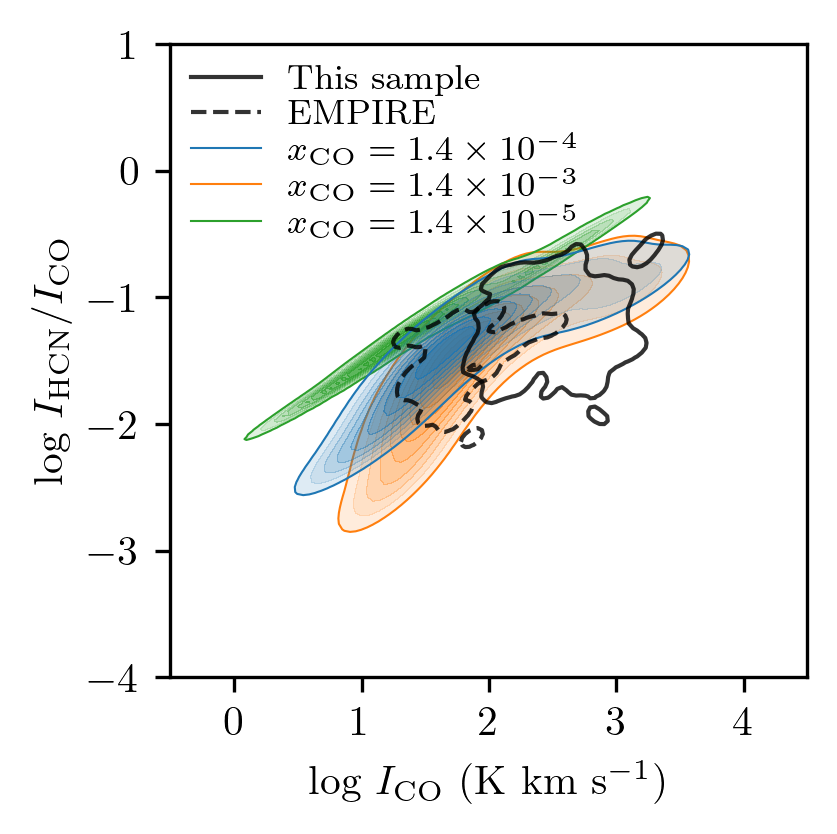}
    \includegraphics[width=0.24\textwidth]{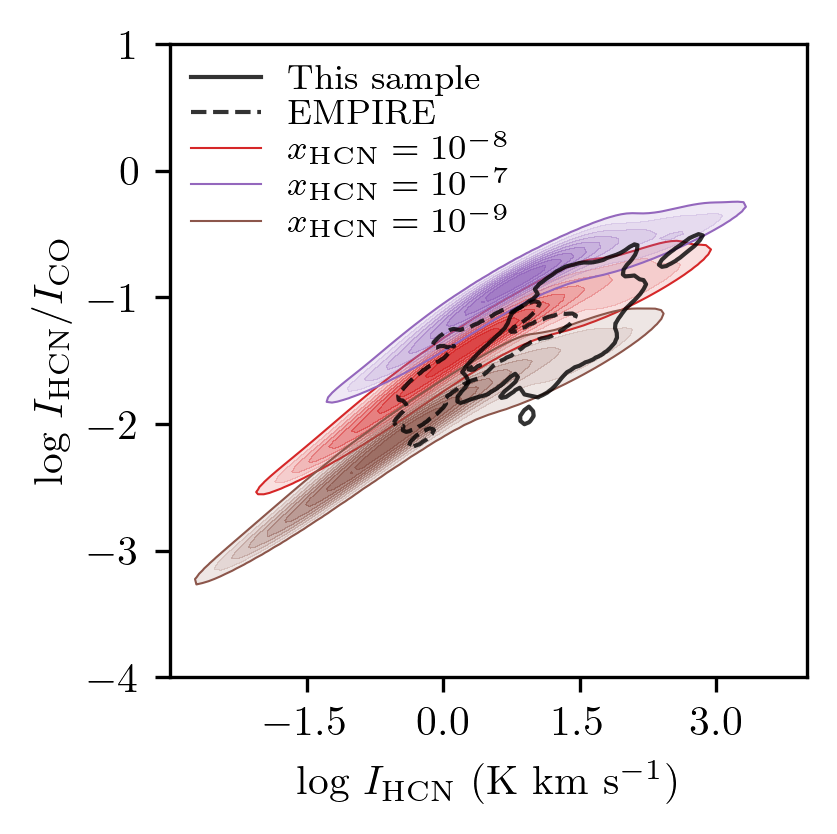}
    \caption{The effect of molecular abundance variations on optical depth and intensity. \textit{Left:} Modeled CO optical depth as a function of CO intensity for $x_\mathrm{CO}=1.4\times10^{-5}$ (green contours) $x_\mathrm{CO}=1.4\times10^{-4}$ (blue contours) $x_\mathrm{CO}= 1.4\times10^{-3}$ (orange contours). \textit{Left center:} Modeled HCN optical depth as a function of HCN intensity for $x_\mathrm{HCN}=10^{-9}$ (brown contours), $x_\mathrm{HCN}=10^{-8}$ (red contours), and $x_\mathrm{HCN}=10^{-7}$ (purple contours). \textit{Right center:} Modeled HCN/CO intensity ratio as a function of CO optical depth for varying CO abundance. The HCN intensities are for $x_\mathrm{HCN}=10^{-8}$. \textit{Right:} Modeled HCN/CO intensity ratio as a function of CO optical depth for varying HCN abundance. The CO intensities are for $x_\mathrm{CO}=1.4\times10^{-4}$. Ten contours are drawn in even steps from the $16\mathrm{th}$ to $100\mathrm{th}$ percentile. Contours are shown in the right two panels for our sample of galaxies and the EMPIRE sample as the solid black and dashed black contours, respectively.}
    \label{fig:vary_abun}
\end{figure*}

We consider the impact of varying HCN and CO abundance on our model results, and we run four additional sets of models using $x_\mathrm{HCN}=10^{-9}$, $x_\mathrm{HCN}=10^{-7}$, $x_\mathrm{CO}=1.4\times10^{-3}$, and $x_\mathrm{CO}=1.4\times10^{-5}$. We note that $x_\mathrm{CO}=1.4\times10^{-3}$ is higher than we expect to find in real molecular clouds \citep[cf.][]{Bolatto:2013}, but we use this value to consider a wide range of CO abundances.  We find that increasing (decreasing) HCN and CO abundance has the effect of increasing (decreasing) the optical depths of the molecular gas tracers, which is a product of  molecular column density scaling with molecular abundance in our models. We find mean CO optical depths of $\tau_\mathrm{CO}\approx2.7,\ 16.9,\ \mathrm{and}\ 107.2$ for $x_\mathrm{CO}=1.4\times10^{-5},\ 1.4\times10^{-4},\ \mathrm{and}\ 1.4\times10^{-3}$, respectively. Mean HCN optical depths are found to be $\tau_\mathrm{HCN}\approx0.88,\ 6.1,\ \mathrm{and}\ 36.5$ for $x_\mathrm{HCN}=10^{-9},\ 10^{-8},\ \mathrm{and}\ 10^{-7}$, respectively. The impact of varying abundance on CO optical depth is therefore more significant relative to HCN and is a result of bright CO emission spanning a broader range of column densities relative to HCN in our models (see Fig. \ref{fig:ex_pdfs}). Optical depths are plotted against molecular intensities for each of these abundances in Fig. \ref{fig:vary_abun}.

We also find that increasing (decreasing) molecular abundance slightly increases (decreases) the median intensity (and, similarly, emissivity, see Eq. \ref{eq:emiss}) in our models. We find $I_\mathrm{CO}\approx63.0,\ 47.9,\ \mathrm{and}\, 37.1\ \mathrm{K\ km\ s}^{-1}$  for $x_\mathrm{CO}=1.4\times10^{-3}, 1.4\times10^{-4},\mathrm{and}\, 1.4\times10^{-5}$, respectively. HCN intensity appears to vary in roughly regular steps with molecular abundance (see Fig. \ref{fig:vary_abun}), with $I_\mathrm{HCN}\approx8.9,\ 2.2,\ \mathrm{and}\ 0.7\ \mathrm{K\ km\ s}^{-1}$ for $x_\mathrm{HCN}=10^{-7},\ 10^{-8},\ \mathrm{and}\ 10^{-9}$, respectively. Furthermore, $x_\mathrm{HCN}=10^{-9}$ appears to produce HCN intensities that are also consistent with measurements from our sample \citepalias[cf.][]{Bemis:2023} (relative to $x_\mathrm{HCN}=10^{-8}$), and some of the higher $I_\mathrm{HCN}$ measurements of the EMPIRE sample \citep{Jimenez_Donaire:2019}. As shown in Fig. \ref{fig:vary_abun}, there is a subsample of measurements with higher $I_\mathrm{CO}$ that have HCN/CO ratios that fall below those produced by our models assuming $x_\mathrm{HCN}=10^{-8}$. It may be possible that these sources (which are mostly comprised of the more extreme systems in our sample) have a lower HCN abundance, on average. A higher HCN abundance ($x_\mathrm{HCN}=10^{-7}$) is likely not realistic for most molecular clouds \citep[cf.][]{Viti:2017}, and also appears to produce higher HCN/CO ratios than what is measured in our sample and the EMPIRE sample.
\par
In summary, the optical depth for $x_\mathrm{CO}=1.4\times10^{-3}$ is $\sim6$ times larger than that for $x_\mathrm{CO}=1.4\times10^{-4}$, but this makes little difference to the main conclusions of this paper as $\tau_\mathrm{CO}>10$ and CO is in LTE for both of these values for the majority of our models, resulting in similar intensities and emissivities, and likewise similar results. For $x_\mathrm{CO}=1.4\times10^{-5}$, the modeled CO ranges from only moderately optically thick to optically thin ($\mathrm{log} \tau < 0$). Since CO emission is likely optically thick in most molecular clouds, these results are not considered for analysis. We therefore focus on the results using the Solar CO abundance, $x_\mathrm{CO}=1.4\times10^{-4}$, in the main text. We find that both $x_\mathrm{HCN}=10^{-8}$ and $x_\mathrm{HCN}=10^{-9}$ produce HCN intensities and HCN/CO ratios consistent with measurements in our sample and the EMPIRE sample. It is possible that the actual HCN abundances in the galaxies considered here vary between $x_\mathrm{HCN}=10^{-8}$ and $x_\mathrm{HCN}=10^{-9}$ \citep[cf.][]{Viti:2017}. It is also possible that more extreme systems trend towards lower HCN abundance, but this is highly uncertain due to a lack of data of optically thin dense gas tracers. We therefore focus on the results of  $x_\mathrm{HCN}=10^{-8}$ in the main text and we note that many of main conclusions would remain similar using a different value of $x_\mathrm{HCN}$, with slight offsets in HCN intensity and emissivity.

\section{The impact of sensitivity limits} \label{ap:sens_limits}

\begin{figure*}[tbh]
    \centering
    \includegraphics[width=\textwidth]{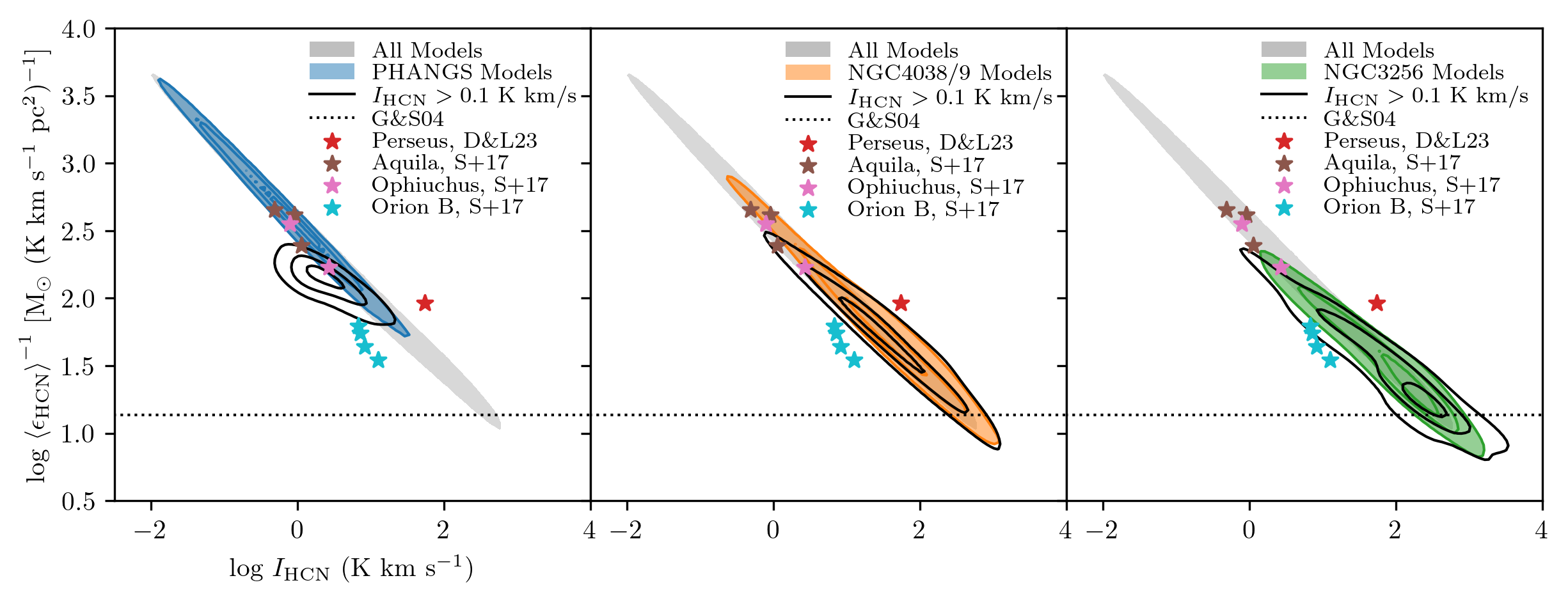}
    \caption{ Inverse of the HCN emissivity (in units of the HCN conversion factor) as a function of HCN intensity.  We plot the PHANGS-, NGC 4038/9-, and NGC 3256-type models from left to right as the blue, orange, and green filled contours. The black contours represent $\left< \epsilon_\mathrm{HCN} \right>^{-1}$ for the models where we make a cut at $0.1\,\mathrm{K\,km\,s^{-1}}$, which is analogous to a sensitivity limit in observations. For comparison, we also show the \citet{Gao:2004a,Gao:2004b} value as the black, dotted horizontal line and several published values of $\alpha_\mathrm{HCN}$ from observations of Milky Way clouds \citep{Dame:2023,Shimajiri:2017}. The gray, filled contour represents all models. We find the HCN emissivities from the PHANGS-type models to be most strongly impacted by a sensitivity cut. These models appear to have artificially lower $\left< \epsilon_\mathrm{HCN} \right>^{-1}$ (higher emissivity) as a result of the cut.}
    \label{fig:alpha_hcn_limit}
\end{figure*}

We re-derive emissivity from our models by excluding the low-density regions of our models that have HCN intensities below $\sim0.1\ \mathrm{K\ km\ s^{-1}}$ (roughly the sensitivity limit in \citealt{Tafalla:2023}) and show our results in Fig. \ref{fig:alpha_hcn_limit}. We find that this has the effect of shifting $\alpha_\mathrm{HCN}$ to smaller values and more strongly impacts $\alpha_\mathrm{HCN}$ from the PHANGS-type models. This is because much of the emission in our PHANGS-type models resides at lower gas column densities. Additionally, the $\alpha_\mathrm{HCN}$ derived by \citet{Tafalla:2023} appear more consistent with models in our sample that have larger gas surface density and larger velocity dispersion (i.e., the models based on measurements from the Antennae and NGC 3256 mergers). From Fig. 3. in \citet{Tafalla:2023}, we can see that the emission from their dense gas tracers appears to extend below their sensitivity limit, suggesting they are indeed missing some emission at low HCN intensity and low column density.  Thus, this discrepancy could be because our models include emission from HCN arising from lower column densities below the detection limit of their study. As we find, this will disproportionately affect clouds with more emission at lower gas surface densities.

\end{appendix}

\end{document}